\documentclass[12pt]{article}

\usepackage{cite}
\usepackage{axodraw}
\usepackage[dvips]{graphicx}
\usepackage{epsfig}
\usepackage{amsmath}
\usepackage{amssymb}

\numberwithin{equation}{section}

\def\g{{\rm I}\hspace{-0.07cm}\Gamma}

\begin{document}
\begin{flushright}
MAN/HEP/2010/10\\[-2pt]
June 2010
\end{flushright}
\bigskip

\begin{center}
{\LARGE {\bf The Minimal Scale Invariant Extension of\\[4mm]
             the Standard Model}}\\[1.5cm] 
{\large Lisa Alexander-Nunneley and Apostolos Pilaftsis }\\[0.5cm]
{\em School of Physics and Astronomy, The University of Manchester, }\\ 
{\em Manchester M13 9PL, United Kingdom }
\end{center}

\vspace{1.5cm}  
\centerline{\bf  ABSTRACT}  

{\small  

\noindent
We perform a systematic analysis of an extension of the Standard Model
that includes a complex singlet  scalar field and is scale invariant at
the  tree level.  We  call such  a model  the Minimal  Scale Invariant
extension  of  the  Standard  Model  (MSISM).   The  tree-level  scale
invariance of  the model is explicitly broken  by quantum corrections,
which  can  trigger  electroweak  symmetry  breaking  and  potentially
provide  a mechanism for  solving the  gauge hierarchy  problem.  Even
though the scale invariant Standard Model is not a realistic scenario,
the  addition  of a  complex  singlet scalar  field  may  result in  a
perturbative  and  phenomenologically  viable  theory.  We  present  a
complete classification of the flat  directions which may occur in the
classical  scalar  potential  of  the MSISM.   After  calculating  the
one-loop effective potential of the  MSISM, we investigate a number of
representative  scenarios  and   determine  their  scalar  boson  mass
spectra,  as  well as  their  perturbatively  allowed parameter  space
compatible   with  electroweak   precision  data.    We   discuss  the
phenomenological  implications  of  these  scenarios,  in  particular,
whether they  realize explicit  or spontaneous CP  violation, neutrino
masses or  provide dark matter  candidates.  In particular, we  find a
new minimal scale-invariant model  of maximal spontaneous CP violation
which can  stay perturbative up to Planck-mass  energy scales, without
introducing  an unnaturally  large hierarchy  in  the scalar-potential
couplings.  }


\newpage
\setcounter{equation}{0}
\section{Introduction}

The Standard  Model (SM) \cite{SM}  is a renormalizable theory  with a
minimal  particle content  which realizes  the famous  Higgs mechanism
\cite{Higgs} to account for the origin of mass of the charged fermions
and the  $W^{\pm}$ and $Z$  bosons.  Despite intense scrutiny,  the SM
remains  resilient to  new physics  and appears  to describe  the data
collected over the years at the LEP collider, TEVATRON and in a number
of low-energy experiments  with remarkable success.  Nevertheless, the
SM predicts the existence of  the Higgs boson which is associated with
the  mechanism of electroweak  spontaneous symmetry  breaking (EWSSB),
but which so  far has remained elusive.  A  natural realization of the
EWSSB mechanism  requires the presence  of a negative  mass parameter,
$-m^{2}$, in the Higgs potential.   The negative mass parameter is the
source  of the  infamous  gauge hierarchy  problem,  in which  quantum
corrections  lead  to quadratically  divergent  terms proportional  to
$\Lambda^{2}$, where $\Lambda$ is  an ultra-violet (UV) cut-off scale.
This  UV cut-off  scale  is usually  associated  with the  scale of  a
possible higher-energy theory in which  the SM might be embedded, such
as Grand Unified  Theory (GUT).  In~the~SM,  with no intermediate mass
scale or theory between the  electroweak (EW) and Planck scale $M_{\rm
  Planck}  \approx  1.2\times 10^{19}$~GeV,  the  cancellation of  the
divergent terms requires excessive fine-tuning.  The avoidance of this
fine-tuning problem  has been the  motivation for many  studies beyond
the  SM, including  supersymmetry  (SUSY).  In  SUSY  this problem  is
naturally  solved,  provided  the  SUSY-breaking mass  scale,  $M_{\rm
  SUSY}$,  stays  close  to  the  EW  scale,  e.g.  $M_{\mathrm{SUSY}}
\lesssim 1$ TeV.

In  this paper we  discuss a  different and  very minimal  approach to
solving the  gauge hierarchy  problem.  It is  remarkable that  the SM
depends on only one mass parameter~$m^2$, whose absence from the Higgs
potential renders  the complete tree-level Lagrangian of  the SM scale
invariant~(SI).   However,  as  first  discussed  by  Coleman  and  E.
Weinberg \cite{CW}  and later by Gildener and  S.  Weinberg \cite{GW},
quantum corrections generate  logarithmic terms which explicitly break
the   scale  invariance  of   the  theory   and  can   trigger  EWSSB.
Unfortunately,  a  perturbative SI  version  of  the  SM is  not  both
theoretically   and   phenomenologically   viable.   Specifically,   a
perturbative SI version of the SM cannot accommodate the LEP2 limit on
the        Higgs-boson       mass,        $m_{H_{\mathrm{SM}}}       >
114.4~\mathrm{GeV}$~\cite{LEP115GeV}, given  the experimental value of
the top-quark  mass.  On  the other hand,  the large  top-quark Yukawa
coupling  gives rise  to an  effective  potential which  is no  longer
bounded  from  below  (BFB).   To overcome  this  difficulty,  several
authors      \cite{Hempfling,Chang,Foot,MNmar2007,MNsept2008}     have
considered  various SI extensions  to the  SI SM  either with  real or
complex singlet scalar fields.

Evidently, one of the main motivations  for a SI theory is the natural
removal of  the $m^{2}$ term  from the Higgs potential.   However, its
absence  alone   does  not  solve  the  gauge   hierarchy  problem  as
$\Lambda^{2}$ terms can still be generated by quantum corrections in a
UV cut-off  scheme of regula\-rization.   This happens because  the UV
cut-off scheme  introduces counter-terms which  explicitly violate the
symmetry  of   classical  scale  invariance  that   governs  the  bare
Lagrangian.  Following the  arguments of \cite{MNmar2007,Bardeen}, one
has to therefore adopt a  regula\-rization scheme which does not break
the classical symmetries  of the local classical action,  in this case
scale invariance.   Dimensional regularization (DR)~\cite{DRscheme} is
such  a SI  scheme within  which the  vanishing of  the $m^2$  term is
maintained to  all orders  in perturbation theory.   Consequently, the
scheme of DR will be used throughout this paper.

An  inherent   field-theoretic  difficulty  of  a  SI   model  is  the
incorporation  of  gravity  which   requires  the  introduction  of  a
dimensionful  parameter, the Planck  mass $M_{\mathrm{Pl}}$,  into the
theory.   The  presence  of  the  Planck mass  explicitly  breaks  the
classical  symmetry  of scale  invariance,  thereby reintroducing  the
issue of quadratic divergences  in the theory.  Even though addressing
this  problem  lies beyond  the  scope of  this  paper,  we note  that
attempts have been  made in the literature to  provide SI descriptions
of quantum gravity \cite{MNmar2007,Footgravity,MNjul2009}.

In this  paper we  study in detail  a minimal  SI extension of  the SM
augmented by a complex singlet  scalar field, $S$.  We call this model
the Minimal  Scale Invariant extension of the  Standard Model (MSISM).
Unlike                        previous                        analyses
\cite{Hempfling,Chang,Foot,MNsept2008,MNmar2007},    we    impose   no
additional constraints on the theory,  such as a U(1) symmetry or some
specific discrete symmetry acting  on $S$.  Hence, the MSISM potential
contains all possible interactions allowed by gauge invariance:
\begin{eqnarray}  
  \label{Vpot}
V(\Phi, S) & = & \frac{\lambda_{1}}{2}\,(\Phi^{\dagger} \Phi)^{2}\ +\
\frac{\lambda_{2}}{2}\, (S^{*} S)^{2}\ +\ \lambda_{3}\,\Phi^{\dagger} \Phi\,
S^{*} S\ +\  \lambda_{4}\,\Phi^{\dagger} \Phi\, S^{2}\
+\  \lambda_{4}^{*}\,\Phi^{\dagger} \Phi\, S^{*2} \nonumber\\  
& & +\  \lambda_{5}\,S^{3}S^{*}\  +\  \lambda_{5}^{*}\,S S^{*3}\ +\
\frac{\lambda_{6}}{2}\,S^{4}\ +\ \frac{\lambda_{6}^{*}}{2}\, S^{*4} \;,\nonumber  
\end{eqnarray}     
where  the  quartic  couplings   $\lambda_{1,2,  \dots,  6}$  are  all
dimensionless  constants and  $\Phi$ is  the usual  SM  Higgs doublet.
Note that the imposition of scale invariance forbids the appearance of
dimensionful   mass   parameters  or   trilinear   couplings  in   the
potential   \footnote{For  recent  studies   of  non-SI   models  with
  dimensionful self-couplings and with  real or complex scalar singlet
  extensions see~\cite{O'Connell,Langsing}.}.

The  tree-level SI  scalar potential  can  possess a  large number  of
different  phenomenologically  viable flat  directions,  which may  be
classified into three major  categories: Type~I, Type~II and Type~III.
Flat directions  of Type  I are characterized  by a singlet  field $S$
with  vanishing  vacuum  expectation  value  (VEV),  whereas  in  flat
directions  of Type  II both  $S$  and $\Phi$  possess non-zero  VEVs.
Finally, in flat directions of Type~III  the SM $\Phi$ has a zero VEV,
which  makes  it somehow  difficult  to  naturally  realize EWSSB  and
therefore we do not study them in detail in this paper.

In  our analysis  of  the  MSISM effective  potential,  we follow  the
perturbative  approach introduced  by  Gildener and  S. Weinberg  (GW)
\cite{GW}.   With  the  aid  of  this  approach  we  can  analytically
calculate  the scalar boson  mass spectrum  and determine  the allowed
range of  parameter space for  which the theory  remains perturbative,
i.e. the theory has perturbative quartic couplings, and which keep the
effective  potential  BFB.   Further  constraints  on  the  MSISM  are
obtained  from  an  analysis  of  the  LEP2  data~\cite{PDG}  and  the
electroweak oblique parameters,  $S$, $T$ and $U$ \cite{Peskin,Guido}.
Of  the  electroweak  oblique  parameters,  $S$ and  $T$  (the  latter
associated with  Veltman's $\rho$ parameter  \cite{Veltman}) yield the
strongest  constraints on  the range  of the  scalar-potential quartic
couplings.

An  interesting feature  of  the MSISM  is  that it  can be  naturally
extended by  right-handed neutrinos in a  SI way, such  that a singlet
Majorana mass scale,  $m_{M}$, can be generated if  the complex scalar
$S$  possesses a VEV  \cite{Footneutrinos, MNsept2008}.   The expected
size of $m_{M}$ is typically of the EW scale.  This can give rise to a
low-scale seesaw  mechanism~\cite{seesaw}, which  in turn can  offer a
natural explanation for  the smallness in mass of  the light neutrinos
as observed in the low-energy neutrino data.  Moreover, unlike the SM,
the MSISM can  realize both explicit and spontaneous  CP violation. Of
particular interest is  a new minimal model of  maximal spontaneous CP
violation along  a maximally  CP-violating flat direction  of Type~II,
which  can stay  perturbative up  to  energy scales  of order  $M_{\rm
Planck}$, without  the need to  introduce a large hierarchy  among the
scalar-potential quartic  couplings or between the VEVs  of the $\Phi$
and  $S$ fields~\cite{Footrecent}.  The  new CP-violating  phase could
act  as a source  for creating  the observed  Baryon Asymmetry  in the
Universe  (BAU),  e.g.~via  a  strong  first-order  electroweak  phase
transition.  Finally, the MSISM  can predict stable scalar states that
could qualify as Dark Matter (DM) candidates.

This paper  is set out as  follows.  In Section  \ref{WI4SI} we review
the  basic properties of  a SI  classical action  and derive  the Ward
identity  which is obeyed  by the  tree-level scalar  potential.  This
Ward identity  for scale  invariance is then  used to define  the flat
direction in the scalar  potential.  In Section \ref{GandW}, we review
the EWSSB mechanism in  multi-scalar SI models following the formalism
outlined in \cite{GW}.  In Section \ref{model}, we present the general
Lagrangian describing  the MSISM.   Furthermore, we present  a general
classification of the flat directions that may occur in the tree-level
scalar potential and then  calculate the one-loop effective potential.
We  also discuss  the  possible phenomenology  of  the different  flat
directions.   Section \ref{TypeI}  investigates models  having  Type I
flat  directions in  both the  U(1)  invariant limit  and the  general
non-invariant  scenario.  Likewise, Section  \ref{TypeII} investigates
models that realize flat directions  of Type II, in the U(1) invariant
limit   and   a  simplified   non-invariant   scenario.   In   Section
\ref{Neutrinos}, we  discuss extensions of the MSISM  that include the
interactions  of  the  complex  singlet  field  $S$  and  its  complex
conjugate $S^{*}$ to right-handed neutrinos.  Technical details of all
our  calculations  have been  relegated  to  a  number of  appendices.
Finally, Section \ref{conclusions} summarizes our conclusions.


\setcounter{equation}{0}
\section{The Ward Identity for Scale Invariance}\label{WI4SI}

In this  section we  derive the Ward  identity (WI) that  results from
imposing the  property of  scale invariance on  a theory.  The  WI for
scale invariance  will then  be used to  consistently define  the flat
directions as local minima of the scalar potential.

To start  with, let us  consider a simple  model with one  real scalar
field, $\Phi(x)$, described by the Lagrangian:
\begin{equation} 
	\label{eqn:Lsimple}
\mathcal{L}\ =\ \frac{1}{2} \partial_{x \mu} \Phi (x) \partial^{\mu}_{x}
\Phi (x)\:  +\:  \frac{1}{2} m^{2} \Phi^{2} (x)\: -\: \lambda \Phi^{4} (x) \; , 
\end{equation}
with the  notation $\partial^{\mu}_{x} \equiv \frac{\partial}{\partial
  x_{\mu}}$.  Under a scale transformation, the scalar field $\Phi(x)$
transforms as
\begin{equation} 
	\label{eqn:PhiunderSI}
\Phi(x)\ \to\ \Phi' (x)\ =\ \sigma \Phi (\sigma x) \; ,
\end{equation}
where  $\sigma =  e^{\epsilon} >  0$.  We  note that  a  general scale
transformation is defined  as $\Phi(x) \to \Phi' (x)  = e^{\epsilon a}
\Phi  (e^{\epsilon} x)$,  where $a$  is the  scaling dimension  of the
field $\Phi(x)$.   At the classical level the  scaling dimension takes
the   value  $a=1$,   if  $\Phi(x)$   is  a   boson,  and   the  value
$a=\frac{3}{2}$, if~$\Phi (x)$~is  a fermion.  The effect of the scale
transformation~(\ref{eqn:PhiunderSI}) of  the scalar field  $\Phi (x)$
on the classical action
\begin{equation} 
 	\label{eqn:Saction}
S[\Phi(x)]\ =\ \int d^{4} x\; \mathcal{L}[\, \partial_{\mu} \Phi(x), \Phi(x)]
\end{equation}
is to give rise to a transformed action given by
\begin{eqnarray} 
	\label{eqn:SIonSM}
S[\sigma \Phi(\sigma x)] & = & \int_{-\infty}^{\infty} d^{4} x\; \bigg[\,
  \sigma^{2}\frac{1}{2} \partial_{x \mu} \Phi (\sigma x)
  \partial^{\mu}_{x} \Phi (\sigma x)\ +\ \frac{1}{2} m^{2} \sigma^{2}
  \Phi^{2} (\sigma x)\ -\ \lambda \sigma^{4} \Phi^{4} (\sigma x) \bigg]
\nonumber\\  
& & \hspace{-2cm} = \int_{\sigma(-\infty)}^{\sigma(\infty)} d^{4}
(\sigma x)\; \bigg[\, \frac{1}{2}\partial_{(\sigma x) \mu} \Phi (\sigma x)
  \partial^{\mu}_{(\sigma x)} \Phi (\sigma x)\ +\ \frac{1}{2}
  \sigma^{-2} m^{2} \Phi^{2} (\sigma x)\ -\ \lambda \Phi^{4} (\sigma x)
  \bigg] \; .\qquad 
\end{eqnarray}
Obviously, the transformed action  $S[\sigma \Phi(\sigma x)]$ is equal
to the original one  $S[\Phi(x)]$, provided the dimensionful parameter
$m^{2}$ vanishes, i.e.~the absence of the $m^{2}$ term results in a SI
theory.

Having  gained  some insight  from  the  above  simple model,  we  now
consider  a general theory,  where $\Phi  (x)$ represents  the generic
field of the theory, which could be a scalar, fermion or vector boson.
The   variation   $\delta   S[\Phi(x)]$   of  the   classical   action
(\ref{eqn:Saction}) under a scale transformation is calculated as
\begin{eqnarray}
\delta S[\Phi(x)] & = & \int d^{4} y\; \bigg[\, \delta \Phi_{i} (y)
  \frac{\delta}{\delta \Phi_{i}(y)}\ +\ \delta \Phi_{i}^{\dagger} (y)
  \frac{\delta}{\delta \Phi_{i}^{\dagger} (y)}\ +\ \delta \big(
  \partial_{\mu} \Phi_{i} (y) \big)\; \frac{\delta}{\delta\big(
    \partial_{\mu} \Phi_{i}(y) \big)} \nonumber \\ 
& & +\ \delta \big( \partial_{\mu} \Phi_{i}^{\dagger} (y) \big)\;
  \frac{\delta}{\delta\big( \partial_{\mu} \Phi_{i}^{\dagger}(y)
    \big)} \bigg] \int d^{4} x\; \mathcal{L} [\Phi(x)]\; ,
\end{eqnarray}
where summation over repeated indices is implied for all the fields in
the  theory.  Given  $\delta \Phi  (x) =  \epsilon \big(  a\Phi  (x) +
x^{\mu}  \partial_{\mu} \Phi  (x)  \big)$ for  an infinitesimal  scale
transformation, the variation $\delta S[\Phi(x)]$ is found to be
\begin{eqnarray}
\delta S[\Phi(x)] & = & \epsilon \int d^{4} x \bigg[ a \frac{\partial
    \mathcal{L} [\Phi(x)] }{\partial \Phi_{i}(x)}  \Phi_{i} (x)  + a
  \Phi_{i}^{\dagger} (x) \frac{\partial \mathcal{L} [\Phi(x)]
  }{\partial \Phi_{i}^{\dagger}(x)}  + (1+a) \frac{\partial
    \mathcal{L}[\Phi(x)]}{\partial \big( \partial_{\mu} \Phi_{i}(x)
    \big)} \big(\partial_{\mu} \Phi_{i}(x) \big) \nonumber\\  
& & + (1+a) \big(\partial_{\mu} \Phi_{i}^{\dagger}(x)
  \big)\frac{\partial \mathcal{L}[\Phi(x)]}{\partial \big(
    \partial_{\mu} \Phi_{i}^{\dagger}(x) \big)} - 4
  \mathcal{L}[\Phi(x)]  \bigg]  + \epsilon x^{\mu} \mathcal
       {L}[\Phi(x)]  \arrowvert_{x^{\mu} \to \pm \infty}\; .  
\end{eqnarray}
In the  above, the last  term is a  surface term which we  assume that
vanishes at infinity.   Requiring that $\delta S[\Phi(x)] =  0$, as it
should for a SI theory, we derive the WI for scale invariance:
\begin{eqnarray} 
	\label{eqn:WIgeneral}
4 \mathcal{L}[\Phi(x)]  & = & a \Bigg[ \frac{\partial
    \mathcal{L}[\Phi(x)] }{\partial \Phi_{i}(x)}\Phi_{i}(x)  +
  \Phi_{i}^{\dagger}(x) \frac{\partial \mathcal{L}[\Phi(x)] }{\partial
    \Phi_{i}^{\dagger}(x)} \Bigg] \nonumber\\ 
& & + (a+1) \Bigg[ \frac{\partial \mathcal{L}[\Phi(x)]}{\partial
    \big(\partial_{\mu} \Phi_{i}(x) \big)}\big( \partial_{\mu}
  \Phi_{i}(x) \big) + \big( \partial_{\mu} \Phi_{i}^{\dagger}(x)
  \big)\frac{\partial \mathcal{L}[\Phi(x)]}{\partial
    \big(\partial_{\mu} \Phi_{i}^{\dagger}(x) \big)} \Bigg]\; . 
\end{eqnarray} 

If the scalar potential $V(\Phi)$ of a theory is SI at tree-level then
the WI (\ref{eqn:WIgeneral}) implies that
\begin{equation} 
	\label{eqn:Wardpotential}
\frac{\partial V^{\mathrm{tree}}(\Phi) }{\partial \Phi_{i}} \Phi_{i}\ +\
\Phi^{\dagger}_{i} \frac{\partial V^{\mathrm{tree}}(\Phi) }{\partial
  \Phi_{i}^{\dagger}}\  =\  4 V^{\mathrm{tree}}(\Phi)\; . 
\end{equation}
For notational simplicity we hereafter suppress the $x$-dependence of
the scalar  field $\Phi$, i.e. $\Phi = \Phi(x)$.  From the  context it
should be clear whether we refer to the  $x$-dependent quantum field
excitation or to its  stationary and $x$-independent background field
value.  If~${\bf \Phi} = (\phi_{1},  \phi_{2}, \dots , \phi_{n})$ is a
vector whose components represent all  the scalar fields of the theory
as   real  degrees   of   freedom,  the   WI~(\ref{eqn:Wardpotential})
straightforwardly generalizes to
\begin{equation} 
	\label{eqn:WIvector}
{\bf \Phi} \cdot \nabla V^{\rm tree} ({\bf \Phi})\  =\  4 V^{\rm tree}
({\bf \Phi}) \; , 
\end{equation}
where   $\nabla  \equiv   \big(   \frac{\partial}{\partial  \phi_{1}},
\frac{\partial}{\partial  \phi_{2}},  \cdots, \frac{\partial}{\partial
  \phi_{n}}  \big)$.  Moreover,  the  dot indicates  the usual  scalar
product of vectors  in an $n$-dimensional vector space  spanned by all
$n$ real scalar fields of the theory.

The WI (\ref{eqn:WIvector}) can be  applied to a specific direction in
the $n$-dimensional field space.  To this end, we may parametrize the
field vector  ${\bf \Phi}$  as ${\bf \Phi}  = \varphi {\bf  N}$, where
${\bf N}$  is a fixed given  $n$-dimensional unit vector  in the field
space  and $\varphi$ is  the radial  distance from  the origin  of the
field space.  In this case, we may rewrite (\ref{eqn:WIvector}) as
\begin{equation}
	\label{eqn:WIN}
\varphi {\bf N} \cdot \nabla V^{\mathrm{tree}}(\varphi {\bf N})\ =\
\varphi \frac{d {\bf \Phi}}{d\varphi} \cdot \nabla
V^{\mathrm{tree}}(\varphi {\bf N})\ =\ \varphi \frac{d
  V^{\mathrm{tree}}(\varphi {\bf N})}{d\varphi}\  =\ 4
V^{\mathrm{tree}}(\varphi {\bf N}) \; .  
\end{equation}
The condition for $V^{\mathrm{tree}}(\varphi  {\bf N})$ to have a flat
direction along a given unit vector ${\bf N} = {\bf n}$ is
\begin{equation}
	\label{eqn:WIphi}
\frac{d V^{\mathrm{tree}}(\varphi {\bf  n})}{d \varphi}\ =\ 0 \; .
\end{equation}
On  account  of  the  WI  (\ref{eqn:WIN}),  the  latter  condition  is
equivalent to $V^{\mathrm{tree}}(\varphi {\bf  n}) = 0$.  In addition,
the condition for this flat  direction to be an extremal or stationary
line is
\begin{equation}
	\label{eqn:Vextremum}
\nabla V^{\mathrm{tree}}({\bf  \Phi})\Big|_{{\bf \Phi} = \varphi {\bf
    n}}\ =\ {\bf 0} \; . 
\end{equation}
In  order  for  this extremal  line  to  be  a  local minimum  of  the
potential, one has to require that
\begin{equation}
	\label{eqn:Vmin}
({\bf v } \cdot \nabla)^2  V^{\mathrm{tree}}({\bf  \Phi})\Big|_{{\bf
            \Phi}\ =\ \varphi {\bf n}}\ \ge\ 0  \; ,  
\end{equation}
for any arbitrary vector ${\bf  v}$ belonging to the $n$-dimensional
field space.  Finally, one has to ensure that the scalar potential is
BFB, i.e. $V^{\rm tree}({\bf  N}) \ge 0$, for all possible directions
${\bf N}$. 


\setcounter{equation}{0}
\section{The Gildener--Weinberg Approach to EWSSB} 
\label{GandW} 

Here we  review the GW  perturbative approach \cite{GW} to  EWSSB that
occurs in generic multi-scalar SI  models.  We also discuss the scalar
mass spectrum  of these models.   The analytic results  presented here
will be used in the next section  to study the EWSSB in the MSISM and
to calculate its scalar mass spectrum.

According to the GW approach,  the minimization of the full potential,
$V = V^{\mathrm{tree}}  + V^{\mathrm{1-loop}}_{\mathrm{eff}} + \dots$,
is performed perturbatively along an extremal (minimal) flat direction
as defined  in the previous section.   This approach is  only valid if
the theory is weakly coupled, which constitutes the regime of validity
for our investigations.

Let us consider a renormalizable  gauge field theory with an arbitrary
set  of $n$  real  scalars ${\phi_{i}}$  (with $i=1,2,\dots,n$)  which
represent the  components of an $n$-dimensional  field multiplet ${\bf
  \Phi}$ (see  also Section \ref{WI4SI}).  We assume that the theory
is SI at tree-level so that its scalar potential is generically given
by
\begin{equation}
 	\label{eqn:VtreeSI}
V^{\mathrm{tree}} ({\bf \Phi})\ =\  \frac{1}{4!}\;  f_{ijkl}\; \phi_{i}
\phi_{j}  \phi_{k}  \phi_{l} \; , 
\end{equation}
where summation over repeated indices is implied and $f_{ijkl}$ stands
for  the  quartic couplings  of  the  potential;  $f_{ijkl}$ is  fully
symmetric in  all its indices.   Notice that (\ref{eqn:VtreeSI}) is a
general solution to the WI for SI given in (\ref{eqn:WIvector}).

As   we   discussed   in   the   previous   section,   the   potential
(\ref{eqn:VtreeSI})  may have a  non-trivial continuous  local minimum
along the  ray ${\bf \Phi}  = \varphi {\bf  N}$, in a  given direction
${\bf  N}  =   {\bf  n}$  of  the  unit  vector   and  at  a  specific
renormalization group (RG) scale $\mu  = \Lambda$.  To find this local
minimum one first needs to identify all the flat directions present in
the potential by solving the equation:
\begin{equation} 
	\label{eqn:3.5}
V^{\mathrm{tree}} ( {\bf N} )\ =\ 
\frac{1}{4!} f_{ijkl}(\mu)  N_{i}  N_{j} N_{k} N_{l}\ =\ 0 \; ,
\end{equation}
where  we have  explicitly  displayed the  dependence  of the  quartic
couplings  $f_{ijkl}$  on  the  RG  scale $\mu$.   Suppose  that  this
condition is met for a particular  unit vector ${\bf N} = {\bf n}$ and
for the specific value of the RG scale, $\mu = \Lambda$.  According to
(\ref{eqn:WIphi}), one  then has $V^{\mathrm{tree}} ({\bf  \Phi}) = 0$
everywhere along  the ray  ${\bf \Phi}^{\mathrm{flat}} =  \varphi {\bf
  n}$, which represents the flat direction.

The  next   step  is   to  ensure  that   the  flat   direction  ${\bf
  \Phi}^{\mathrm{flat}}$, as  determined above, represents  a stationary
line.  This leads to the condition $\partial V^{\mathrm{tree}}({\bf N}
)  / \partial  N_{i} |_{{\bf  N} =  {\bf n}}  = 0$,  and hence  to the
constraint
\begin{equation} 
	\label{eqn:min}
f_{ijkl}(\Lambda)\, n_{j} n_{k} n_{l}\ =\ 0 \; .
\end{equation}
Observe  that   this  constraint   is  equivalent  to   the  condition
(\ref{eqn:Vextremum}).  It  should also be  noted that (\ref{eqn:min})
imposes a single constraint  on the parameters $f_{ijkl}$, independent
of how  many parameters $f_{ijkl}$  contains and specifically  only at
the RG scale $\Lambda$.  Finally, one needs to implement the condition
(\ref{eqn:Vmin}), i.e.   the stationary line is a  local minimum line.
Therefore, one has to require that the Hessian matrix, defined as
\begin{equation} 
	\label{eqn:Hessian}
({\bf P})_{ij}\ \equiv\ \frac{\partial^{2} V^{\mathrm{tree}}({\bf N})}
              {\partial N_{i} \partial N_{j} }  \Bigg\arrowvert_{{\bf
                  N} = {\bf n}}\  =\  \frac{1}{2} f_{ijkl} n_{k} n_{l} \; ,  
\end{equation}
is  non-negative definite,  i.e. the  $n \times  n$-dimensional matrix
${\bf P}$ has either vanishing or positive eigenvalues.

Since $V^{\mathrm{tree}}({\bf  N})$ vanishes along  the flat direction
${\bf  \Phi}^{\rm flat}$,  the full  potential of  the theory  will be
dominated        by       higher-loop        contributions       along
${\bf\Phi}^{\mathrm{flat}}$ and specifically by the one-loop effective
potential,  $V^{\mathrm{1-loop}}_{\mathrm{eff}}({\bf  \Phi})$.  Adding
higher order quantum corrections gives a small curvature in the radial
direction ${\bf \Phi}^{\mathrm{flat}} =  \varphi {\bf n}$, which picks
out a  specific value,  $v_{\varphi}$,
along the ray as the minimum.   In addition, a small shift may also be
produced in a  direction $\delta {\bf \Phi} =  v_{\varphi} \delta {\bf
  n}$ perpendicular  to the  flat direction ${\bf  n}$, i.e.  ${\bf n}
\cdot \delta{\bf n} = 0$.   We may now extend the stationary condition
(\ref{eqn:Vextremum}) to the one-loop corrected scalar potential, i.e.
\begin{equation}
\nabla \left(V^{\mathrm{tree}}({\bf \Phi} )\: +\:
V^{\mathrm{1-loop}}_{\mathrm{eff}}( {\bf \Phi} ) \right) \Big|_{{\bf
    \Phi}\ =\ v_{\varphi} ({\bf n}   + \delta {\bf n})}\ =\ {\bf 0} \; . 
\end{equation}
According  to the GW  perturbative approach,  one has  to consistently
expand this last  expression to the first loop  order, by treating the
perpendicular  shift   $\delta  {\bf  \Phi}$  as   an  one-loop  order
parameter.  In this way, we find
\begin{equation} 
	\label{eqn:Vminpert}
v_{\varphi}^{2} {\bf P} \cdot \delta {\bf \Phi}\: +\: \nabla
V^{\mathrm{1-loop}}_{\mathrm{eff}}( {\bf \Phi}) \Big|_{{\bf \Phi} =
  v_{\varphi} {\bf n}}\ =\ {\bf 0} \; , 
\end{equation}
where the dot indicates the usual matrix multiplication of the Hessian
${\bf P}$ with the vector~$\delta {\bf \Phi}$.

The perturbative  minimization condition (\ref{eqn:Vminpert}) uniquely
determines   $\delta  {\bf   \Phi}$,  except   for   directions  along
eigenvectors  of   ${\bf  P}$  with  zero   eigenvalues.   These  zero
eigenvectors include the flat  direction ${\bf n}$ itself, since ${\bf
  n}  \cdot  {\bf P}  =  {\bf 0}$  by  virtue  of (\ref{eqn:min})  and
(\ref{eqn:Hessian}).  They also  include the Goldstone directions that
may result  from the spontaneous  symmetry breaking of  any continuous
symmetries.    Therefore,  we   may  eliminate   the  first   term  in
(\ref{eqn:Vminpert}) by  contracting the relation (\ref{eqn:Vminpert})
from the left with ${\bf n}$.  Thus, we get the minimization condition
along the radial direction:
\begin{equation}
	\label{eqn:Vminradial}
{\bf n} \cdot \nabla V^{\mathrm{1-loop}}_{\mathrm{eff}} ({\bf \Phi})
\Big|_{{\bf \Phi} =  v_{\varphi} {\bf n}}\ =\ \frac{d
  V^{\mathrm{1-loop}}_{\mathrm{eff}}(\varphi {\bf n})}{d \varphi}
\bigg|_{\varphi = v_{\varphi}}\ =\ 0 \; .  
\end{equation}
Here it is useful to remark that this condition will be used to fully
specify the VEV of $\phi$ to one-loop order in perturbation
theory. 

Along the  flat direction  ${\bf \Phi}^{\mathrm{flat}} =  \varphi {\bf
  n}$,         the         one-loop        effective         potential,
$V^{\mathrm{1-loop}}_{\mathrm{eff}}(\varphi\,  {\bf   n})$,  takes  the
general form:
\begin{equation}
	\label{eqn:general1looppotAB}
V^{\mathrm{1-loop}}_{\mathrm{eff}} (\varphi\, {\bf n})\ =\ 
A({\bf n})\,\varphi^{4}\: +\: B({\bf n})\,\varphi^{4} \ln
\frac{\varphi^{2}}{\Lambda^{2}} \; ,   
\end{equation}
where the ${\bf n}$-dependent  dimensionless constants $A$ and $B$ are
given in the $\overline{\rm {MS}}$ scheme by
\begin{eqnarray}
	\label{eqn:AB}
A & = & \frac{1}{64 \pi^{2} v_{\varphi}^{4}}\; \bigg \{\, \mathrm{Tr}
\bigg[\, m_{S}^{4} \left(-\frac{3}{2} + \ln
  \frac{m_{S}^{2}}{v_{\varphi}^{2}} \right) \bigg]+ 3 \mathrm{Tr}
\bigg[\, m_{V}^{4} \left(-\frac{5}{6} + \ln
  \frac{m_{V}^{2}}{v_{\varphi}^{2}} \right) \bigg]  \nonumber\\
& & -  4 \mathrm{Tr} \bigg[\, m_{F}^{4} \left(- 1 + \ln
  \frac{m_{F}^{2}}{v_{\varphi}^{2}} \right)\, \bigg]\, \bigg \} \; ,
\nonumber\\[3mm]
B & = & \frac{1}{64 \pi^{2} v_{\varphi}^{4}}\; \bigg( \mathrm{Tr}
m_{S}^{4} + 3 \mathrm{Tr} m_{V}^{4} -  4 \mathrm{Tr}  m_{F}^{4} \bigg) \; ,  
\end{eqnarray}
where $m_{S,V,F}$  are the tree-level scalar, vector  and fermion mass
matrices, respectively, which are evaluated at $v_{\varphi}\, {\bf n}$
and the  trace is  taken over  the mass matrix  and over  all internal
degrees of freedom~\footnote{Note that the internal degrees of freedom
  for   Majorana   fermions  are   half   of   those   of  the   Dirac
  fermions. Consequently, if the fermion  $F$ is of the Majorana type,
  the pre-factor  $-4$ in front of  the trace should  be replaced with
  $-2$.}.    Analytic  results  for   the  tree-level   mass  matrices
$m_{S,V,F}$ will be given in the next section, where we will calculate
the  one-loop   effective  potential   of  the  MSISM   following  the
GW~approach.

Minimizing        (\ref{eqn:general1looppotAB})       according       to
(\ref{eqn:Vminradial})  shows  that the  potential  has a  non-trivial
stationary point at a value of the RG scale $\Lambda$, given by
\begin{equation}
	\label{eqn:LambdaAB}
\Lambda\ =\ v_{\varphi}\: \exp \left( \frac{A}{2B}\: +\: \frac{1}{4} \right) \; .
\end{equation}
Note that  since the effective-potential coefficients $A$  and $B$ are
of  the same  loop  order, the  RG  scale $\Lambda$  and the  absolute
minimum $v_\varphi$  are expected to  be of comparable order  as well.
Thus, a natural  implementation of the breaking of  the scale symmetry
can  be  obtained  in  perturbation theory,  where  potentially  large
logarithms of the sort $\ln (\Lambda^2/v^2_\varphi)$ can be kept under
control.
 
The relation (\ref{eqn:LambdaAB}) can now  be used to find the form of
the one-loop effective potential along  the flat direction in terms of
the one-loop VEV $v_\varphi$,
\begin{equation}
 	\label{eqn:minimum}
V^{\mathrm{1-loop}}_{\mathrm{eff}} (\varphi\, {\bf n})\ =\ 
B({\bf n})\: \varphi^4\; \bigg(
  \ln\frac{\varphi^2}{v_{\varphi}^2}\ -\ \frac{1}{2}\,\bigg) \ . 
\end{equation}
Even though the above substitution has made the explicit dependence of
$V^{\mathrm{1-loop}}_{\mathrm{eff}} (\varphi\,  {\bf n})$ on $\Lambda$
to  disappear,  there  still  exists  an implicit  dependence  of  the
kinematic parameters in $B({\bf  n})$ and the flat direction $\varphi$
on  the  RG  scale  $\Lambda$.   On  the  other  hand,  in  order  for
$v_{\varphi}      {\bf       n}$      to      be       a      minimum,
$V^{\mathrm{1-loop}}_{\mathrm{eff}}  (v_{\varphi}  {\bf  n})$ must  be
less than  the value  of the  potential at the  origin $\varphi  = 0$,
hence it must be negative. From~(\ref{eqn:minimum}), it is easy to see
that this can only happen if $B>0$.  Moreover, this constraint ensures
that  the  potential  is  BFB, i.e.~the  one-loop  effective  potential
remains non-negative  for infinitely large values of  $\varphi$ in any
field direction~${\bf N}$.

At the tree-level,  the squared masses of the  scalar bosons are given
by the eigenvalues of the matrix,
\begin{equation}
	\label{eqn:GWmassmatrix}
(m_{S}^{2})_{ij}\ =\ 
\frac{\partial^{2} V^{\mathrm{tree}} ({\bf \Phi})}{\partial \phi_{i}\, 
\partial \phi_{j}} \Bigg|_{{\bf \Phi}\, =\, v_{\varphi}\, {\bf
    n}}\ =\ v_{\varphi}^{2}\, ({\bf P})_{ij} \; .
\end{equation}
From our  discussion above, it is clear  that the Hessian matrix
${\bf P}$ has positive definite  eigenvalues, except for a set of zero
eigenvalues  due   to  the   Goldstone  bosons  associated   with  the
spontaneous symmetry breaking of  compact symmetries of the theory and
one zero eigenvalue due to flat direction.  Hence the model contains a
set  of massive  scalars, a  set of  massless Goldstone  bosons  and a
single massless  scalar, which we  denote as $h$, associated  with the
spontaneous symmetry breaking of scale invariance.

The single  massless scalar does  not remain massless beyond  the tree
approximation.      In     detail,     the     one-loop     correction
$V^{\mathrm{1-loop}}_{\mathrm{eff}}$  to the  scalar  potential shifts
the mass matrix to
\begin{equation}
(m_{S}^{2} + \delta m_{S}^{2})_{ij}\ =\ \frac{\partial^{2}
    \big(V^{\mathrm{tree}}({\bf \Phi})\: +\:
    V^{\mathrm{1-loop}}_{\mathrm{eff}}({\bf \Phi})\big)} {\partial
    \phi_{i} \partial  \phi_{j}} \Bigg \arrowvert_{{\bf \Phi} =
    v_{\varphi} ({\bf n} + \delta {\bf n})} \; .    
\end{equation}
To first order in a perturbative expansion, this becomes
\begin{equation}
	\label{eqn:dmS}
(\delta m_{S}^{2})_{ij}\ =\ \frac{\partial^{2}
          V^{\mathrm{1-loop}}_{\mathrm{eff}}({\bf \Phi})}{\partial
          \phi_{i} \partial  \phi_{j}} \Bigg|_{{\bf \Phi} =
          v_{\varphi}{\bf n}} +\  v_{\varphi}\: f_{ijkl} n_{k}\delta
        \phi_{l} \; .   
\end{equation}
In order  to remove  the second term  in (\ref{eqn:dmS}),  we contract
$(\delta m_{S}^{2})_{ij}$ with $n_{i}$ and $n_{j}$.  Thus, the mass of
the field $h$ is calculated to be
\begin{equation} 
	\label{eqn:mhAB}
m_{h}^{2}\ =\ n_{i} n_{j} (\delta m_{S}^{2})_{ij}\ =\ n_{i} n_{j}
\frac{\partial^{2}  V^{\mathrm{1-loop}}_{\mathrm{eff}}({\bf \Phi})}
     {\partial \phi_{i} \partial \phi_{j}} \Bigg|_{{\bf \Phi} =
       v_{\varphi} {\bf n}} =\ \frac{ d^{2}
       V^{\mathrm{1-loop}}_{\mathrm{eff}}(\varphi {\bf
         n})}{d\varphi^{2}}  \Bigg|_{\varphi = v_{\varphi}} =\ 8 B
     v_{\varphi}^{2} \; ,  
\end{equation}
where     we    have     used     (\ref{eqn:general1looppotAB})    and
(\ref{eqn:LambdaAB})   to    arrive   at   the    last   equality   in
(\ref{eqn:mhAB}).    The   field    $h$   is   commonly   called   the
pseudo-Goldstone  boson of  the anomalously  broken  scale invariance,
since it  is massless at  tree-level when scale invariance  holds, but
acquires a non-zero  mass at the one-loop level  once scale invariance
is broken by quantum corrections.

The  remaining massive  scalar  states  of the  theory  can be easily 
determined provided  $(\delta m_{S}^{2})_{ij}$ remains  a small effect
compared to  the tree-level  mass matrix $(m_{S}^{2})_{ij}$.   In this
case, their masses are determined from the relation:
\begin{equation} 
	\label{eqn:massivescalarmasses}
m_{H}^{2}\ =\ \tilde{n}_{i} \tilde{n}_{j} \frac{\partial^{2}
  V^{\mathrm{tree}} ({\bf \Phi})}{\partial \phi_{i} \partial \phi_{j}}
\Bigg \arrowvert_{{\bf \Phi} = v_{\varphi} {\bf n}} =\ {\bf \tilde{n}}
\cdot {\bf P}\cdot {\bf \tilde{n}} \; ,  
\end{equation}
where  the massive scalar  directions are  defined similarly  to ${\bf
  \Phi}^{\mathrm{flat}}$  as ${\bf  \Phi}^{\mathrm{H}} =  \varphi {\bf
  \tilde{n}}$,  where  ${\bf  \tilde{n}}$  is a  generic  unit  vector
perpendicular  to ${\bf  n}$.  The  Goldstone bosons  remain massless
provided $V^{\mathrm{1-loop}}_{\mathrm{eff}}({\bf \Phi})$ respects the
same global symmetries as $V^{\mathrm{tree}}({\bf \Phi})$. 


\setcounter{equation}{0}
\section{The MSISM}\label{model}

In this section we use  the analytic results presented in the previous
two  sections to study  the mechanism  of EWSSB  in the  Minimal Scale
Invariant extension  of the Standard Model.  First,  we briefly review
the  general Lagrangian  describing the  MSISM.  We  then  discuss the
parameterization  of   the  flat  directions  and   present  a  general
classification of the flat directions that may occur in the tree-level
scalar potential.   We also  present the one-loop  effective potential
for  the  MSISM,  from  which  we derive  its  scalar  mass  spectrum.
Finally, we  briefly discuss the generic  phenomenological features of
the  different  realizations  of  flat  directions in  the  MSISM.   A
detailed investigation of the physically viable flat directions in the
MSISM is deferred to Sections \ref{TypeI} and~\ref{TypeII}.


\subsection{The MSISM Lagrangian}

The Lagrangian defining  the MSISM can be written  as a sum of
five terms:
\begin{equation} 
	\label{eqn:Lfull}
\mathcal{L}_{\mathrm{MSISM}}\ =\ \mathcal{L}_{\mathrm{inv}}\: +\:
\mathcal{L}_{\mathrm{GF}}\:  +\:  \mathcal{L}_{\mathrm{FP}}\: +\:
\mathcal{L}_{\nu}\:  -\: 
V^{\mathrm{tree}}(\Phi, S) \; ,
\end{equation}
where  $\mathcal{L}_{\mathrm{inv}}$,  $\mathcal{L}_{\mathrm{GF}}$  and
$\mathcal{L}_{\mathrm{FP}}$ are  the gauge-invariant, gauge-fixing and
Faddeev--Popov Lagrangians,  respectively, and a  detailed description
of  these Lagrangians  is  given in  Appendix~\ref{App:SM}.  The  term
$\mathcal{L}_{\nu}$ is  the right-handed neutrino  Lagrangian which is
discussed  separately  in  Section  \ref{Neutrinos}.  The  last  term,
$V^{\mathrm{tree}}(\Phi,  S)$,  is  the  tree-level potential  of  the
MSISM, which is given by
\begin{eqnarray}
	\label{eqn:fullpotential}
V^{\mathrm{tree}}(\Phi, S) & = & \frac{\lambda_{1}}{2}\,(\Phi^{\dagger} \Phi)^{2}\ +\
\frac{\lambda_{2}}{2}\, (S^{*} S)^{2}\ +\ \lambda_{3}\,\Phi^{\dagger} \Phi\,
S^{*} S\ +\  \lambda_{4}\,\Phi^{\dagger} \Phi\, S^{2}\
+\  \lambda_{4}^{*}\,\Phi^{\dagger} \Phi\, S^{*2} \nonumber\\  
& & +\  \lambda_{5}\,S^{3}S^{*}\  +\  \lambda_{5}^{*}\,S S^{*3}\ +\
\frac{\lambda_{6}}{2}\,S^{4}\ +\ \frac{\lambda_{6}^{*}}{2}\, S^{*4}\; .
\end{eqnarray}
where  for  simplicity  the  $x$-dependence  of the  fields  has  been
suppressed and  will continue to  be suppressed unless  distinction is
required between the field  $\phi(x)$ and the flat direction component
$\phi$.   As usual,  we may  linearly decompose  the  SU(2)$_L$ scalar
doublet $\Phi$ and the complex singlet field $S$ as follows:
\begin{equation} 
	\label{eqn:PhiS}
\Phi\ =\ \left( \begin{array}{c}
G^{+}\\
\frac{1}{\sqrt{2}}(\phi+iG)
\end{array}\right) \; , \qquad  S\ =\ \frac{1}{\sqrt{2}}(\sigma+iJ) \; ,
\end{equation}
where $\phi$ and $\sigma$ ($G$ and $J$) are CP-even (odd) real scalar
fields and $G^{+}$ is the charged would-be Goldstone boson.

In order to provide a stable minimum for the scalar potential, we must
ensure  that $V^{\mathrm{tree}}$  is  BFB.  This  can  be achieved  by
placing  a set  of constraining  conditions on  the  quartic couplings
$\lambda_{1,2,\dots,6}$.   These  conditions   can  be  determined  by
analyzing  the potential  in terms  of  the two  real and  independent
gauge-invariant field bilinears,  $ \Phi^{\dagger} \Phi$ and $S^{*}S$.
To  convert  (\ref{eqn:fullpotential})  into this  representation,  we
re-express  the  field  $S$   as  $S  =  |S|e^{i  \theta_{S}}$,  where
$\theta_{S}$  is  the  phase  of  the  complex  field  and  $S^{*}S  =
|S|^{2}$. The tree-level scalar potential can then be rewritten in the
form
\begin{equation} 
	\label{eqn:Vbilinears}
V^{\mathrm{tree}} =  \frac{1}{2}\,\Big( \Phi^{\dagger} \Phi\, ,\, S^{*}S\Big)\:
{\bf \Lambda}
\left(\!\! \begin{array}{c}
\Phi^{\dagger} \Phi \\ 
S^{*}S\\
\end{array}\!\right )\; ,
\end{equation}
where ${\bf \Lambda}$ is a real symmetric matrix with the elements:
\begin{eqnarray} 
	\label{eqn:componentsofLambda}
\Lambda_{11} & = & \lambda_{1}\; , \nonumber\\
\Lambda_{12} & = & \Lambda_{21}\ =\ \lambda_{3}\: +\:  \lambda_{4}e^{2 i
  \theta_{S}}\: +\: \lambda^{*}_{4}e^{- 2 i \theta_{S}} \; , \nonumber\\ 
\Lambda_{22} & = & \lambda_{2}\: +\: 2 \lambda_{5}e^{2 i \theta_{S}}\: +\: 2
\lambda^{*}_{5}e^{- 2 i \theta_{S}}\: +\:  \lambda_{6}e^{4 i \theta_{S}}\: +\:
\lambda_{6}^{*}e^{-4 i \theta_{S}} \; . 
\end{eqnarray}
Since  the  two bilinears  $\Phi^\dagger\Phi$  and  $S^*  S$ are  both
positive-definite     by    definition,     the     requirement    for
$V^{\mathrm{tree}}$  to  be  BFB  depends exclusively  on  the  matrix
elements of ${\bf \Lambda}$.   In detail, the following two conditions
are required to keep $V^{\mathrm{tree}}$ BFB:
\begin{equation} 
	\label{eqn:BFBconditions}
\mbox{(i)}\quad \mathrm{Tr} {\bf \Lambda}\  \ge\   0 \; ,\qquad
\mbox{(ii)}\quad  
\left\{\begin{array}{ll}
\Lambda_{12} \ge 0\; ,& \mbox{if $\Lambda_{11} = 0$ or $\Lambda_{22} = 0$}\\
\mathrm{Det} {\bf \Lambda} \ge 0\; ,& \mbox{if $\Lambda_{11} \ne 0$
  and $\Lambda_{22} \ne 0$}
\end{array}\right.\; .
\end{equation}  
The  above conditions  must hold  for all  directions in  the bilinear
vector  space,  including   the  flat  directions.   Obviously,  these
conditions  explicitly depend  on the  phase $\theta_{S}$  through the
matrix      elements     of      ${\bf     \Lambda}$      given     in
(\ref{eqn:componentsofLambda}).   This phase determines  the direction
of a ray in the $\sigma$-$J$ plane within the entire real scalar field
space.       It     is      therefore      essential     that      the
conditions~(\ref{eqn:BFBconditions})  hold  true  for  all  values  of
$\theta_S$,  ensuring  that  $V^{\mathrm{tree}}$  remains BFB  in  all
possible field directions.

It is  now instructive to show  that the angle $\theta_{S}$  is SI. We
can  prove this  by using  the WI~(\ref{eqn:Wardpotential})  for scale
invariance.   We first  note that  the derivatives  of  the tree-level
potential   $V^{\mathrm{tree}}$   with   respect   to   the   different
  representations,  real  fields, complex  fields  and bilinears,  are
  related through:
\begin{eqnarray}
	\label{eqn:fieldandbilinear}
\Re G^{+}  \frac{\partial V^{\mathrm{tree}}}{\partial \Re G^{+}} + \Im
G^{+} \frac{\partial V^{\mathrm{tree}}}{\partial \Im G^{+}} + 
G \frac{\partial V^{\mathrm{tree}}}{\partial G} + \phi \frac{\partial
  V^{\mathrm{tree}}}{\partial \phi} \!\!& = &\!\!  
\frac{\partial V^{\mathrm{tree}}}{\partial \Phi} \Phi + \Phi^{\dagger}
\frac{\partial V^{\mathrm{tree}}}{\partial \Phi^{\dagger}}\ =\ 
2 \Phi^{\dagger} \Phi \frac{\partial V^{\mathrm{tree}}}{\partial
  (\Phi^{\dagger}\Phi)} \; , \nonumber\\
\sigma \frac{\partial V^{\mathrm{tree}}}{\partial \sigma} + J
\frac{\partial V^{\mathrm{tree}}}{\partial J} \!\!& = &\!\! 
S \frac{\partial V^{\mathrm{tree}}}{\partial S} + S^{*} \frac{\partial
  V^{\mathrm{tree}}}{\partial S^{*}}\  =\ 2 S^{*}S \frac{\partial
  V^{\mathrm{tree}}}{\partial (S^{*}S)} \; ,\nonumber\\ 
\end{eqnarray}
with $\Re G^{+} = \frac{1}{\sqrt{2}}(G^{+}  + G^{-})$ and $\Im G^{+} =
\frac{i}{\sqrt{2}}(G^{-}    -    G^{+})$.     The   second    equation
in~(\ref{eqn:fieldandbilinear})  involving the  complex  singlet field
$S$ was derived by employing the relations:
\begin{equation} 
S^{*}S\; \frac{\partial V^{\mathrm{tree}}}{\partial (S^{*}S)}\ +\
\frac{\partial V^{\mathrm{tree}}}{\partial (2i\theta_{S})}\ =\ S
\frac{\partial V^{\mathrm{tree}}}{\partial S} \; ,  \qquad 
S^{*}S\; \frac{\partial V^{\mathrm{tree}}}{\partial (S^{*}S)}\ -\
\frac{\partial V^{\mathrm{tree}}}{\partial (2i\theta_{S})}\ =\ S^{*}
\frac{\partial V^{\mathrm{tree}}}{\partial S^{*}} \; .
\end{equation}
Hence, the  WI~(\ref{eqn:Wardpotential}) can be  re-expressed in terms
of derivatives with respect to bilinears only, i.e.
\begin{equation}
S^{*}S\; \frac{\partial V^{\mathrm{tree}}}{\partial (S^{*}S)}\ +\ 
\Phi^{\dagger} \Phi\; \frac{\partial V^{\mathrm{tree}}}{\partial
  (\Phi^{\dagger}\Phi)}\ =\ 2 V^{\mathrm{tree}} \; . 
\end{equation}
Evidently, the absence of a  derivative term with respect to the phase
$\theta_{S}$ implies that  $\theta_{S}$ is a truly SI  quantity in the
MSISM.

A comment  regarding the predictive power  of the Higgs  sector of the
MSISM  is in  order.   The MSISM  potential  contains several  quartic
couplings  that  would seem  to  imply that  the  MSISM  will be  less
predictive  than  the  SM.    However,  imposing  the  flat  direction
condition~(\ref{eqn:min}) and possible  additional symmetries, such as
a U(1) or  a ${\bf Z}_4$ discrete symmetry  acting on~$S$, reduces the
number of  the independent parameters significantly. In  fact, most of
the generic  cases that  we will  be studying have  only two  or three
independent  quartic  couplings, thereby  making  the  MSISM a  rather
predictive theory.

\subsection{Classification of the Flat Directions}\label{Flatdirections}   

Following   the  approach   presented   in  Sections~\ref{WI4SI}   and
\ref{GandW}, we  parametrize the flat direction  as an $n$-dimensional
vector, whose components represent all  real degrees of freedom of the
scalars fields in the theory.   For the MSISM, the flat direction lies
in the vector space spanned by the real scalar fields,
\begin{displaymath}
\{ \Re G^{+},\ \Im G^{+},\ G,\ \phi,\ \sigma,\ J \} \; .
\end{displaymath}  
Without loss  of generality, we may  exploit the SM  gauge symmetry to
set $\Re G^{+} = \Im G^{+} = G = 0$ and
restrict the  field space to  the neutral fields $\phi$,  $\sigma$ and
$J$, which may develop an  electrically neutral VEV. Thus, the general
flat direction ${\bf \Phi}^{\rm flat}$ can be dimensionally reduced to
\begin{equation} 
	\label{eqn:generalflat}
{\bf \Phi}^{\mathrm{flat}}\ =\ \varphi \left( \begin{array}{c}
n_{\phi} \\ 
n_{\sigma} \\
n_{J}
\end{array}\right )\  
=\ \left( \begin{array}{c}
\phi   \\ 
\sigma \\
J \end{array}\right ) \; ,
\end{equation}
where  the  components  $n_{\phi,\sigma,J}$  satisfy  the  unit-vector
constraint:  $n^2_\phi  +  n^2_\sigma  +  n^2_J =  1$.   Observe  that
$v_{\varphi}  n_{\phi}  \equiv  v_{\phi}  $,  $v_{\varphi}  n_{\sigma}
\equiv  v_{\sigma} $  and  $v_{\varphi}  n_J \equiv  v_{J}  $, at  the
minimum of the one-loop effective potential.

In  order that  the flat  directions  represent minimal  lines of  the
tree-level potential, we  need to require that all  the derivatives of
$V^{\rm tree}$ with respect to the fields $\phi$, $\sigma$ and $J$, or
equivalently with  respect to the  fields $\Phi$ and $S$,  vanish when
evaluated along  the flat direction  [cf.\ (\ref{eqn:Vextremum})].  In
this  way, the  following two  complex tadpole  conditions need  to be
satisfied:
\begin{eqnarray}
  \label{eqn:diffwrtphi}
 \frac{\partial V^{\mathrm{tree}}} {\partial \Phi}  \Bigg|_{{\bf \Phi}^{\mathrm{flat}}}
   & = &
\Phi^{\dagger} \Big[ \lambda_{1}(\Lambda)
  \Phi^{\dagger} \Phi  + \lambda_{3}(\Lambda)
  S^{*}  S  +  
\lambda_{4}(\Lambda)  S^{2} +
  \lambda_{4}^{*}(\Lambda)  S^{*\, 2} \Big]\ =\ 0 \; ,\qquad\\[3mm] 
  \label{eqn:diffwrtS} 
 \frac{\partial V^{\mathrm{tree}}} {\partial S}  \Bigg|_{{\bf \Phi}^{\mathrm{flat}}}
 & = &  S^{*}   \Big[
  \lambda_{2}(\Lambda) S^{*} S +
  \lambda_{3}(\Lambda) \Phi^{\dagger}  \Phi
  + 3 \lambda_{5}(\Lambda)  S^{2} +
  \lambda_{5}^{*}(\Lambda)   S^{*\, 2}  \Big]
\nonumber\\ 
& & +\ S  \Big[ 2 \lambda_{4}(\Lambda) 
  \Phi^{\dagger}  \Phi + 2 \lambda_{6}(\Lambda)
   S^{2}  \Big]\  =\ 0 \; ,
\end{eqnarray}  
where ${\bf \Phi}^{\mathrm{flat}}$ is defined in (\ref{eqn:generalflat}).
As we will discuss in more detail below, there are three distinct ways
to satisfy  the above minimization conditions,  which generically lead
to  three different  types  of flat  directions:  Type~I, Type~II  and
Type~III.

\subsubsection{Flat Direction of Type I} \label{FlatdirectionI}

Along the Type~I flat direction,  the scalar doublet $\Phi$ develops a
VEV, but not the complex field $S$, i.e.~the flat direction components
$\sigma$ and  $J$ in (\ref{eqn:generalflat})  are both zero.  If  $S =
0$,  the minimization condition~(\ref{eqn:diffwrtS})  is automatically
satisfied,  whilst the condition  (\ref{eqn:diffwrtphi}) forces  us to
set  $\lambda_{1}(\Lambda) =  0$.   The values  of  the other  quartic
couplings     are     constrained     by    the     BFB     conditions
(\ref{eqn:BFBconditions}),    such    that    $\Lambda_{22}    >    0$
and~$\Lambda_{12} > 0$.

Since the  complex field $S$ has  a vanishing VEV,  the flat direction
(\ref{eqn:generalflat}) gets dimensionally reduced to
\begin{equation}
	\label{eqn:TypeIflatdir}
{\bf \Phi}^{\mathrm{flat}}\ =\  \varphi \, n_{\phi}\ =\ \phi \; ,
\end{equation}
with  $n_{\phi} =  1$.   This  implies that  the  flat direction  lies
directly  along the  $\phi$ axis  and  that the  quantum field  $\phi$
corresponds exactly  to the  massless scalar field  $h$, which  is the
pseudo-Goldstone  boson  associated  with  broken scale invariance  (see  our
discussion in Section \ref{GandW}).

\subsubsection{Flat Direction of Type II} \label{FlatdirectionII}

Along  the Type~II  flat direction,  both the  doublet $\Phi$  and the
singlet   $S$   fields    develop   non-zero   VEVs.    This   implies
(\ref{eqn:diffwrtphi}) and (\ref{eqn:diffwrtS})  can only be satisfied
if specific relations  among the quartic couplings are  met at some RG
scale  $\Lambda$.   For  instance,  consider a  U(1)-invariant  MSISM
scalar  potential which  is invariant  under U(1)  rephasings  of the
field $S  \to e^{i \alpha} S$,  where $\alpha$ is  an arbitrary phase.
As  a  consequence  of  the  U(1) invariance  the  quartic  couplings
$\lambda_{4,5,6}$  vanish.    Moreover,  the  minimization  conditions
(\ref{eqn:diffwrtphi})   and    (\ref{eqn:diffwrtS})   lead   to   the
constraint:
\begin{equation}
	\label{eqn:Type2flatdircond}
\frac{  \Phi^{\dagger}  \Phi } { 
  S^*  S  }\ =\
\frac{n_{\phi}^{2}}{n_{\sigma}^{2} + n_{J}^{2}}\ =\ -
\frac{\lambda_{3}(\Lambda)}{\lambda_{1}(\Lambda)}\ =\ -
\frac{\lambda_{2}(\Lambda)}{\lambda_{3}(\Lambda)} \; . 
\end{equation}
In  addition, in  order  to satisfy  the  above relation  and the  BFB
condition (\ref{eqn:BFBconditions}), we  must demand that $\lambda_{1}
> 0$, $\lambda_{2} > 0$ and $\lambda_{3} < 0$.

In  a general  Type  II flat  direction,  both $\sigma$  and $J$  will
develop  VEVs, since  $S$  is a  complex  field.  However,  if a  U(1)
symmetry is acting on the  scalar potential, any possible phase of $S$
can be eliminated through a U(1)  rephasing, such that $S$ is real and
$J  = 0$.   Consequently, for  the U(1)  invariant scenario,  the flat
direction  is reduced  to a  two  component vector  and applying  the
constraints    (\ref{eqn:Type2flatdircond})   and    $n_{\phi}^{2}   +
n_{\sigma}^{2} = 1$ yields
\begin{equation}
	\label{eqn:Type2mixingexample}
 {\bf \Phi}^{\mathrm{flat}} \ =\ 
 \varphi \,  \left( \begin{array}{c}
\sqrt{\frac{-\lambda_{3}(\Lambda)}{\lambda_{1}(\Lambda) -
    \lambda_{3}(\Lambda)}} \\  
\sqrt{\frac{\lambda_{1}(\Lambda)}{\lambda_{1}(\Lambda) -
 \lambda_{3}(\Lambda)}}  
\end{array}\right )\  
=\  \phi \, \left( \begin{array}{c}
1 \\ 
\sqrt{\frac{\lambda_{1}(\Lambda)}{- \lambda_{3}(\Lambda)}}
\end{array}\right )  \; .
\end{equation}

Since the  U(1)-invariant Type II  flat direction is composed  of both
the $\phi$ and  $\sigma$ fields, there will be  mixing between the two
CP even states  in the mass basis, where the mass  basis is defined by
the field along  the flat direction and those  fields along directions
perpendicular to it.  Thus, for the U(1) invariant  scenario, the mass
eigenstates are  the massless Goldstone boson $J$  associated with the
spontaneous  breaking of  the  U(1) symmetry  and  the massive  scalar
states $h$ and $H$, given by
\begin{equation}
  \label{hHJ}
h \ =\ \cos  \theta\; \phi\: +\: \sin \theta\; \sigma \; , \qquad 
H \ =\ -\ \sin  \theta\; \phi \: +\:  \cos \theta\; \sigma \; ,
\end{equation}
where $\cos^{2} \theta = -\lambda_{3}(\Lambda)/[\lambda_{1}(\Lambda) -
  \lambda_{3}(\Lambda)]$.

The general U(1) non-invariant scenario is much more involved and will
be            discussed             in            detail            in
Section~\ref{MinmodelmaxspontaneousCPviolation}.     In    the    U(1)
non-invariant  scenario, the  flat  direction is  in  general a  three
component vector.  Hence,  unless $S$ is either real  or imaginary and
so  preserves  the  CP  symmetry,  all three  quantum  fields  $\phi$,
$\sigma$  and $J$  will mix  together  to form  the scalar-boson  mass
eigenstates.

\subsubsection{Flat Direction of Type III} \label{FlatdirectioIII}

The third  type of  flat direction  is characterized by  $ \Phi  = 0$.
However, a zero  VEV for the $\Phi$ doublet  is not phenomenologically
viable,  since  it  is  difficult  to realize  successful  EWSSB.   In
particular,  the  electroweak  gauge  bosons remain  massless  at  the
tree-level.   Beyond the  tree approximation,  there will  be  a small
shift in the direction of the flat direction, but this turns out to be
generically  too  small to  account  for  the  $W^\pm$- and  $Z$-boson
masses, unless  a large hierarchy between  the VEVs of  $\Phi$ and $S$
fields  is introduced~\cite{Footrecent}.  Therefore,  we do  not study
the Type III flat direction in this paper.

It is important  to note here that the three  types of flat directions
described above give a  complete classification of the flat directions
in  the MSISM.   However,  each type  may  contain several  different
variations.  For example, consider the U(1) non-invariant Type~II flat
direction.       It      requires      (\ref{eqn:diffwrtphi})      and
(\ref{eqn:diffwrtS}), but  places no  explicit constraints on  how the
quartic couplings  of the scalar potential satisfy  them.  Each choice
provides a unique  valid flat direction which gives rise  to a vast number
of  possible  variants. We  do  not intend  to  go  through each  such
variant,  but rather  concentrate  on a  few representative  scenarios
which appear to be physically  interesting, in terms of new sources of
CP violation, neutrino masses and DM candidates.


\subsection{The One-Loop Effective Potential}

We now present the general  one-loop effective potential of the MSISM.
This   has   been  computed   in   terms   of   $\Phi$  and   $S$   in
Appendix~\ref{App:EffPot},   where  the  full   one-loop  renormalized
effective  potential $V^{\mathrm{1-loop}}_{\mathrm{eff}}$ is  given in
(\ref{eqn:fullonelooprenormalisedpot}).    Along   the  minimum   flat
direction,  the RG scale  takes the  specific value  $\mu=\Lambda$ and
$V^{\mathrm{1-loop}}_{\mathrm{eff}}$ can  be put in a  form similar to
the one in~(\ref{eqn:general1looppotAB}), i.e.
\begin{equation}
	\label{eqn:1looppotMSISM}
V^{\mathrm{1-loop}}_{\mathrm{eff}} (\phi )\ =\ \alpha\, \phi^{4}\: +\: 
\beta\, \phi^{4}\; \ln \frac{\phi^{2}}{\Lambda^{2}} \; .
\end{equation}
The coefficients $\alpha$ and $\beta$ are dimensionless parameters and
are given in the $\overline{\rm {MS}}$ scheme by
\begin{eqnarray}
	\label{eqn:alphabeta}
\alpha & = & \frac{1}{64 \pi^{2} v_{\phi}^{4}}\ \bigg[\ \sum_{i=1}^{2}
  m_{H_i}^{4}\; \bigg(\! -\frac{3}{2} + \ln \frac{m_{H_i}^{2}}{v_{\phi}^{2}}
  \bigg)\   +\  6m_{W}^{4}\; \bigg(\! -\frac{5}{6} + \ln
  \frac{m_{W}^{2}}{v_{\phi}^{2}} \bigg) \nonumber\\ 
& & +\  3m_{Z}^{4}\; \bigg(\! - \frac{5}{6} + \ln
  \frac{m_{Z}^{2}}{v_{\phi}^{2}} \bigg)\ -\  
  12 m_{t}^{4}\; \bigg(\! - 1 + \ln \frac{m_{t}^{2}}{v_{\phi}^{2}} \bigg)\ 
   -\ 2 \sum_{i=1}^{3} m_{N i}^{4}\; 
\bigg(\! - 1 + \ln \frac{m_{N i}^{2}}{v_{\phi}^{2}} \bigg)
  \bigg] \; ,\nonumber\\[3mm] 
 \beta & = & \frac{1}{64 \pi^{2} v_{\phi}^{4}}\ \bigg(\, \sum_{i=1}^{2}
 m_{H_i}^{4}\ +\ 6m_{W}^{4}\ +\  3m_{Z}^{4}\  -\  12m_{t}^{4}\ -\ 2
 \sum_{i=1}^{3} m_{N i}^{4} \bigg) \; . 
 \end{eqnarray}
In the above, we have neglected  all light fermions, except of the top
quark  and   the  possible   presence  of  heavy   Majorana  neutrinos
$N_{1,2,3}$    [cf.\   (\ref{eqn:fullonelooprenormalisedpot})].    The
parameters  $m_{X}$, with  $X= \{  H_{1,2}, W,  Z, t,  N \}$,  are the
tree-level particle  masses.  These are  given by the  mass parameters
$M_{X}$,  defined  in  Appendix  \ref{App:EffPot},  evaluated  at  the
minimum $\phi = v_{\phi} \equiv v_{\rm SM}$, where $v_{\rm SM} \approx
246$~GeV is the VEV of the SM Higgs doublet~$\Phi$.

Notice      that       the      one-loop      effective      potential
$V^{\mathrm{1-loop}}_{\mathrm{eff}}            (\Lambda)$           in
(\ref{eqn:1looppotMSISM})  can be  written down  entirely in  terms of
$\phi$  and $v_{\phi}$,  without the  need to  involve the  other flat
direction components $\sigma$ and~$J$.  This is possible, since either
$\sigma = J = 0$ along  the Type~I flat direction, or $\sigma$ and $J$
are  related to  $\phi$ along  the  Type~II flat  direction.  In  this
context,  it   can  be  shown  that  the   MSISM  effective  potential
(\ref{eqn:1looppotMSISM})   can  be  written   in  the   general  form
of~(\ref{eqn:general1looppotAB}). To make this explicit, we employ the
fact  that $\phi  = \varphi\,  n_{\phi}$ in~(\ref{eqn:1looppotMSISM}),
which allows us to make  the following obvious identifications for the
parameters $A$ and $B$:
\begin{equation}
	\label{eqn:ABrelatedtoalphabeta}
A\ =\ \alpha\, n_{\phi}^{4}\: +\: \beta\, n_{\phi}^{4}\, \ln n_{\phi}^{2} \; ,
\qquad \qquad B \ =\ \beta\, n_{\phi}^{4}\; . 
\end{equation}
Substituting the above expressions for $A$ and $B$ in~(\ref{eqn:mhAB})
and   (\ref{eqn:LambdaAB}),  we  may   readily  obtain   the  analytic
dependence  of the Higgs-boson  mass $m_{h}$  and the  minimization RG
scale $\Lambda$  on the effective potential  coefficients $\alpha$ and
$\beta$:
\begin{eqnarray}
  \label{eqn:mhalphabeta}
m_{h}^{2} & = &  8\, \beta\, n_{\phi}^{2} v_{\phi}^{2} \; ,\\
  \label{eqn:Lambdaalphabeta} 
\Lambda & = & v_{\phi}\, \exp \left( \frac{\alpha}{2\beta} + \frac{1}{4}
\right) \; .  
\end{eqnarray}   
We   may  now   employ  the   relation~(\ref{eqn:Lambdaalphabeta})  to
eliminate   the  explicit  dependence   of  the   effective  potential
$V^{\mathrm{1-loop}}_{\mathrm{eff}}$  in~(\ref{eqn:1looppotMSISM})  on
the RG scale $\Lambda$,
\begin{equation}
  \label{VpotVEV}
V^{\mathrm{1-loop}}_{\mathrm{eff}} (\phi ) \ =\ \beta\, \phi^{4}\;
\bigg( \ln\frac{\phi^{2}}{v^{2}_\phi}\ -\ \frac{1}{2}\, \bigg)\ ,
\end{equation}
where all kinematic quantities on  the RHS of (\ref{VpotVEV}), such as
$\beta$, $\phi$ and $v_\phi$, are  evaluated at the RG scale $\Lambda$
[cf.~(\ref{eqn:minimum})].    Hence,  the   size   of  the   radiative
corrections  along the  minimum flat  direction is  determined  by the
effective  potential  coefficient  $\beta$  and  is  therefore  highly
model-dependent. In  our analysis of  the specific flat  directions of
Type~I  and Type~II,  we will  use the  two formulae  for  $m_{h}$ and
$\Lambda$         given         in~(\ref{eqn:mhalphabeta})         and
(\ref{eqn:Lambdaalphabeta}), respectively.

Along  the minimum  flat direction,  the scalar  mass spectrum  of the
MSISM generally  consists of two massive states  $H_{1,2}$ with masses
$m_{H_{1,2}}$,  and  one  massless  state  $h$  corresponding  to  the
pseudo-Goldstone  of the  anomalously broken  scale invariance  at the
tree level.   The would-be Goldstone bosons associated  with the EWSSB
of the SM gauge group receive gauge-dependent masses along the minimum
flat   direction,   e.g.~see   (\ref{eqn:gmasses}).   However,   these
gauge-dependent mass terms do not contribute to the one-loop effective
potential  $V^{\mathrm{1-loop}}_{\mathrm{eff}}(\Lambda)$,  since  they
cancel against  the gauge-dependent part of the  gauge-boson and ghost
contributions. More technical details are given in Appendix~B.

Given the analytic form of the effective potential coefficient $\beta$
in~(\ref{eqn:alphabeta}),  it  is now  interesting  to  see  why a  SI
version  of the  SM  cannot  be phenomenologically  viable.   In a  SI
extension of  the SM, we expect  that the Higgs boson  $H_{\rm SM}$ is
massless  at the  tree  level, but  acquires  an one-loop  radiatively
generated  mass given by  (\ref{eqn:mhalphabeta}).  This  implies that
the SM  Higgs-boson mass $m_{H_{\rm  SM}} \equiv m_{h}$  is explicitly
dependent on $\beta$, i.e.
\begin{equation}
\beta\ =\ \frac{1}{64 \pi^{2} v_{\phi}^{4}}\ \bigg( 6m_{W}^{4}\: +\:
3m_{Z}^{4}\:  -\:  12m_{t}^{4}  \bigg) \; . 
\end{equation}
Considering  the  presently  well-known  experimental  values  of  the
top-quark,  $W^{\pm}$- and $Z$-boson  masses, the  coefficient $\beta$
turns out  to be  negative, giving rise  to an  unphysically tachyonic
mass,   in  gross  violation   to  the   LEP2  limit~\cite{LEP115GeV}:
$m_{H_{\rm SM}} > 114.4$~GeV.  Since $\beta$ and $B$ are negative, the
SI limit of  the SM also fails to realize a  scalar potential which is
BFB, according to our discussion in Section \ref{GandW}.


\subsection{Model Taxonomy}\label{Phenomenology}

As was  already mentioned  in the introduction,  the MSISM  provides a
conceptually  very minimal  solution to  the  gauge-hierarchy problem,
with  a minimal  set of  new fields  and new  couplings.   Following a
bottom-up approach, it is  interesting to analyze the phenomenological
features of the different variants of the MSISM. In particular, we are
interested  in scenarios which  include new  sources of  CP violation,
provide massive DM candidates  and can incorporate a natural mechanism
for  generating the small  light-neutrino masses,  such as  the seesaw
mechanism~\cite{seesaw}.

\begin{table}
\begin{center}
\begin{tabular} {|c|c|c|c|c|}
\hline
 & & & & \\[-2mm]
 &  U(1) Invariant & CP Violation & Massive DM  & Seesaw \\[-1.5mm]
 &  &  & Candidate  & Neutrinos \\[3mm]
\hline
\underline{Flat Direction of Type I}  & & & & \\[3mm]
$ S  = 0$ & Yes & None & Yes & No \\[2mm]
$ S = 0$ & No & Explicit & Yes & No \\[3mm]
\hline
\underline{Flat Direction of Type II}  & & & & \\[3mm]
$ S  =$ real & Yes & None & No  & Yes \\[2mm]
$ S  =$ real & No & Explicit & Model  & Yes \\[-1.5mm]
   & & & Dependent & \\[2mm]
$ S  = $ imaginary & No  & Explicit & Model & Yes \\[-1.5mm]
 & &  & Dependent & \\[2mm]
$ S =$ complex & No & Explicit or  & Model & Yes \\[-1mm] 
 & & Spontaneous & Dependent  & \\[3mm]
\hline
\end{tabular}
\bigskip 
\end{center}
\caption[Short  Caption]{\it Taxonomy  of all  possible U(1)-invariant
  and  U(1)  non-invariant  realizations  that may  occur  within  the
  MSISM,  in  terms  of   their  potential  to  realize  explicit  or
  spontaneous  CP violation,  massive DM~candidates  and possible 
  implementation of  the  seesaw  mechanism  for  naturally  explaining
  the small light-neutrino masses. }\label{tab:taxonomy}
\end{table} 

In the MSISM, naturally small Majorana masses for the light neutrinos
can  be generated via  the seesaw  mechanism, only  if there  exist SI
interactions of $S$ with  right-handed neutrinos and the singlet field
$S$ possesses a  non-zero VEV, $ S \neq 0$.  Hence,  as we have listed
in  Table~\ref{tab:taxonomy}, only  Type-II flat  directions  have the
ability  to  realize  the  seesaw  mechanism.   In~addition,  we  have
presented  in Table~\ref{tab:taxonomy}  the scenarios  of  the MSISM,
which can  contain both explicit  or spontaneous CP  violation through
complex quartic  couplings $\lambda_{4,5,6}$ or a complex  VEV for the
field $S$, respectively.  Notice that the Type-II flat direction along
an  imaginary $S$  does not  violate CP  spontaneously, since  one may
redefine  $S$ as  $S'\equiv iS$  to render  this flat  direction real,
without  introducing any  new phase  in the  quartic couplings  of the
scalar   potential.   Finally,   Table~\ref{tab:taxonomy}   shows  the
different variants of the MSISM,  which have the potential to predict
a massive stable scalar particle that could qualify as a DM candidate.
As was pointed  out in~\cite{Higgsportal}, a natural way  to have a
massive  stable scalar boson  is to  impose a  parity symmetry  on the
scalar  potential.   Such  parity  symmetries could  be:  $\sigma  \to
-\sigma$, $J\to  - J$, or $\sigma \leftrightarrow  \pm J$.  Therefore,
as  we comment  in  Table~\ref{tab:taxonomy}, the  existence  of a  DM
candidate is  model-dependent and requires further  constraints on the
theory.

In  the next two  sections, Sections~\ref{TypeI}  and~\ref{TypeII}, we
discuss  in more detail  the phenomeno\-logy  of a  few representative
scenarios  of  the  MSISM,   without  the  inclusion  of  right-handed
neutrinos.  A detailed analysis of the MSISM augmented by right-handed
neutrinos is given in Section~\ref{Neutrinos}.


\setcounter{equation}{0}
\section{The Type-I MSISM}
\label{TypeI} 

In this section, we investigate the MSISM which realizes a Type I flat
direction, i.e.  the VEV of the  complex singlet field $S$  is zero at
the tree  level.  In detail,  we determine the perturbative  values of
the quartic  couplings of the  potential and consider their  effect on
the scalar mass spectrum.  We then further constrain the theoretically
allowed  parameter space by  applying the  experimental limits  on the
electroweak oblique parameters $S$,  $T$ and $U$~\cite{Peskin} and the
LEP2  limit \cite{LEP115GeV}:  $m_{H_{\rm  SM}} >  114.4$  GeV, for  a
SM-like  Higgs boson.  Finally,  we discuss  the phenomenology  of the
Type-I MSISM.

We  individually consider  the  two cases: the  U(1) invariant  and the
general  U(1)  non-invariant  scenarios  of  the  Type-I  MSISM.   As
discussed in Section \ref{FlatdirectionI}, we should bear in mind that
in addition  to $S =  0$, we must  have $\lambda_{1}(\Lambda) =  0$ to
satisfy             the             tree-level            minimization
condition~(\ref{eqn:diffwrtphi}). Moreover, in  the Type-I MSISM, the
flat   direction   lies  along   the   $\phi$   axis,   as  given   in
(\ref{eqn:TypeIflatdir}), with  $n_{\phi} =  1$, so the  quantum field
$\phi$ can  be identified with  the pseudo-Goldstone boson $h$ of the
anomalously broken scale invariance.


\subsection{The U(1) Invariant Limit}\label{TypeIU1invar}

Assuming that the theory is U(1) symmetric and imposing the constraint
$\lambda_{1}(\Lambda)  =  0$  at  a  given  RG  scale  $\Lambda$,  the
tree-level potential  (\ref{eqn:fullpotential}) for the  Type-I MSISM
reduces to
\begin{equation}
	\label{eqn:Type1U1potential}
V^{\mathrm{tree}}(\Lambda)\  =\  
\frac{\lambda_{2}(\Lambda)}{2}\; (S^* S)^2\: +\:
\lambda_{3}(\Lambda)\, \Phi^{\dagger} \Phi\, S^{*} S \; , 
\end{equation}
where $\lambda_{2}(\Lambda)$ and $\lambda_{3}(\Lambda)$ should both be
positive owing to  the BFB conditions~(\ref{eqn:BFBconditions}).  Even
though  the scalar  potential~(\ref{eqn:Type1U1potential})  depends on
the    two   independent    parameters    $\lambda_{2}(\Lambda)$   and
$\lambda_{3}(\Lambda)$,  it   is  not  difficult  to   show  that  the
tree-level scalar  masses and the renormalization  scale $\Lambda$ are
fully  determined  by  one  single  parameter,  the  quartic  coupling
$\lambda_{3}(\Lambda)$.   More  explicitly, by  setting  $S  = 0$  and
$\lambda_{1}(\Lambda) =  0$ in the general squared  scalar mass matrix
${\cal  M}^2_S$  given  in~(\ref{eqn:123}),  we~obtain that  the  only
non-zero  elements  of ${\cal  M}^2_S$  at  $\phi  = v_\phi$  are  the
following entries:
\begin{equation}
  \label{eqn:miTypeIU1invar}
m_{\sigma}^{2}\ =\ m_{J}^{2}\ =\ \frac{\lambda_{3}(\Lambda)}{2}\,
v_\phi^{2}\; .
\end{equation}
Hence,  the scalar  spectrum consists  of the  mass  eigenstates $\phi
\equiv h$,  $\sigma \equiv H_1$ and  $J \equiv H_2$,  where the latter
two states are degenerate, with equal masses $m_{H_{1,2}} = m_{\sigma}
=  m_J$,  proportional  to $\sqrt{\lambda_{3}(\Lambda)}$.   The  first
state~$h$ corresponds to the pseudo-Goldstone boson of the anomalously
broken scale invariance, which receives its mass $m_h$ at the one-loop
level,  by  means  of  (\ref{eqn:mhalphabeta}).   The  $h$-boson  mass
squared  is  directly  proportional  to $\beta$,  since  $n_\phi  =1$.
Consequently,   $m^2_h$   is   fully   specified   by   the   coupling
$\lambda_{3}(\Lambda)$ through the  scalar masses $m_{H_{1,2}}= m_\sigma = m_{J}$.  Likewise, the renormalization scale $\Lambda$, as was evaluated
in  (\ref{eqn:Lambdaalphabeta}),  depends  on  $m_{H_{1,2}}$  through  the
coefficients $\alpha$ and  $\beta$, and hence its exact  value is also
fixed by $\lambda_{3}(\Lambda)$.

From the above discussion, it is now obvious that possible theoretical
constraints  on $\lambda_{3}(\Lambda)$  will  directly translate  into
limits on  the scalar  mass spectrum and  the RG  scale~$\Lambda$.  An
upper  theoretical constraint on  the value  of $\lambda_{3}(\Lambda)$
originates from  the requirement that the  theory remains perturbative
at  the  scale  $\Lambda$.   We may  enforce  this  constraint  by
requiring~that
\begin{equation}
\beta_{\lambda}\ \le\ 1\; , 
\end{equation}
where   $\lambda$  denotes   a   generic  coupling   of  the   MSISM,
i.e.~$\lambda  =  \{  \lambda_{1,2,\dots,6},\  g',\ g,\  g_{s},\  {\bf
  h}^{e,u,d}\}$, and $\beta_\lambda$  is the one-loop RG beta-function
for  the generic  coupling  $\lambda$.   A complete  list  of all  the
one-loop beta functions $\beta_\lambda$  of the MSISM is presented in
Appendix \ref{App:betafns}.   Assuming $\lambda_{2}(\Lambda)$ is small
and setting $\lambda_{1}(\Lambda)=0$, we  find that the most stringent
upper  limit  on  $\lambda_{3}(\Lambda)$  comes  from  demanding  that
$\beta_{\lambda_{3}} \le 1$ at $\mu =\Lambda$. This implies that
\begin{equation}
  \label{lambda3}
 2 \lambda_{3}^{2}(\Lambda)\: +\: 1.86 \lambda_{3}(\Lambda)\; 
   \le\  8 \pi^{2}\; ,
\end{equation}
and an upper limit of $\lambda_{3}(\Lambda)~\le~5.84$  is deduced,  for
$m_{W}   =    80.4$~GeV,   $m_{Z}   =   91.19$~GeV    and   $m_{t}   =
171.3$~GeV~\cite{PDG}.   If $\lambda_{2}(\Lambda)$  is non-negligible,
the  upper  limit  on  $\lambda_{3}(\Lambda)$  decreases.   The  lower
theoretical constraint  is determined by requiring  that the potential
remains  BFB.  This  is  assured  if the  coefficient  $\beta$ of  the
effective  potential  is  positive,   thus  giving  rise  to  a  lower
theoretical bound of $\lambda_{3}(\Lambda)~>~2.32$.

Further constraints on the allowed range of $\lambda_{3}(\Lambda)$ can
be derived  from experimental  data of direct  Higgs searches  and the
electroweak oblique parameters $S$,  $T$ and $U$.  Analytic results of
the $S$, $T$ and $U$ parameters in the MSISM are presented in Appendix
\ref{App:Obparams}.   Using  these results,  we  may place  additional
limits   on   $\lambda_{3}(\Lambda)$    from   experiment.    In   the
U(1)-invariant  Type-I MSISM, only  the $h$  boson interacts  with the
photon  and the  $W^{\pm}$  and  $Z$ bosons.   As  a consequence,  the
shifts,  $\delta S$,  $\delta T$  and $\delta  U$, to  the electroweak
oblique parameters evaluated in the  MSISM with respect to the SM will
result  from  the  $h$  interactions.  Since  these  interactions  are
identical  to those  of the  SM Higgs  boson $H_{\rm  SM}$,  the shift
parameters $\delta  S$, $\delta T$ and  $\delta U$ only  depend on the
difference between the two  masses, $m_{h}$ and $m_{H_{\rm SM}}$.  Assuming that $\delta S$, $\delta T$  and $\delta U$ fall within their 95\% CL interval for a fixed given SM Higgs-boson mass e.g.~$m_{H_{\rm  SM}} =  117$ GeV~\cite{PDG}, we find that the
 limits from  $\delta S$ and $\delta T$  require the respective constraints: $\lambda_{3}(\Lambda)   <  49.12$ and $\lambda_{3}(\Lambda)   <  74.28$, however the prediction  for $\delta U$ lies entirely inside the
range    $\delta    U_{\rm    exp}$,    even    for    large    values
$\lambda_{3}(\Lambda) <  100$, and so provides no constraint.    
Finally,  applying   the  direct
Higgs-boson searches limit \cite{LEP115GeV}, $m_{H_{\rm SM}} = m_{h} >
114.4$~GeV,  on  the SM-like  $h$  boson,  we  obtain the  constraint:
$\lambda_{3}(\Lambda)  >  6.29$,   which  lies  slightly  outside  the
perturbative limit of $\lambda_3(\Lambda ) \le 5.84$. In this context,
we note  that the  highest RG scale  for a  Landau pole to  appear for
$\lambda_{3}(\Lambda) \approx 6.3$  is $\mu_{\rm Landau} \sim 10^4$~GeV,
which is obtained for $\lambda_2 (\Lambda ) = 0$.

\begin{figure}
\centering 
\includegraphics{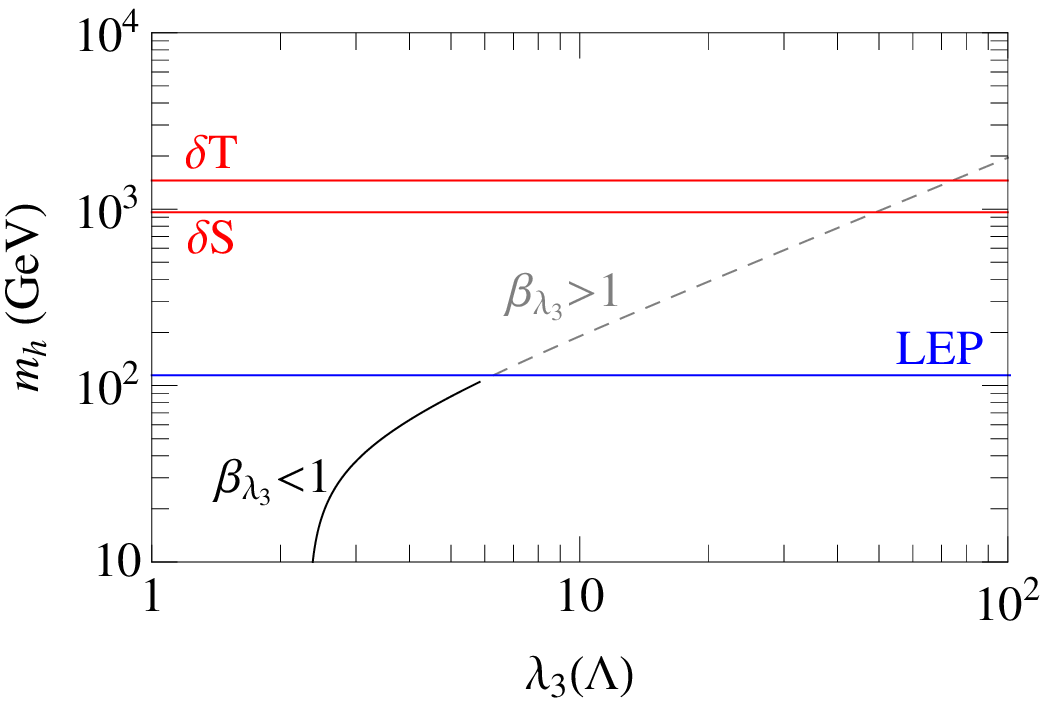}
\includegraphics{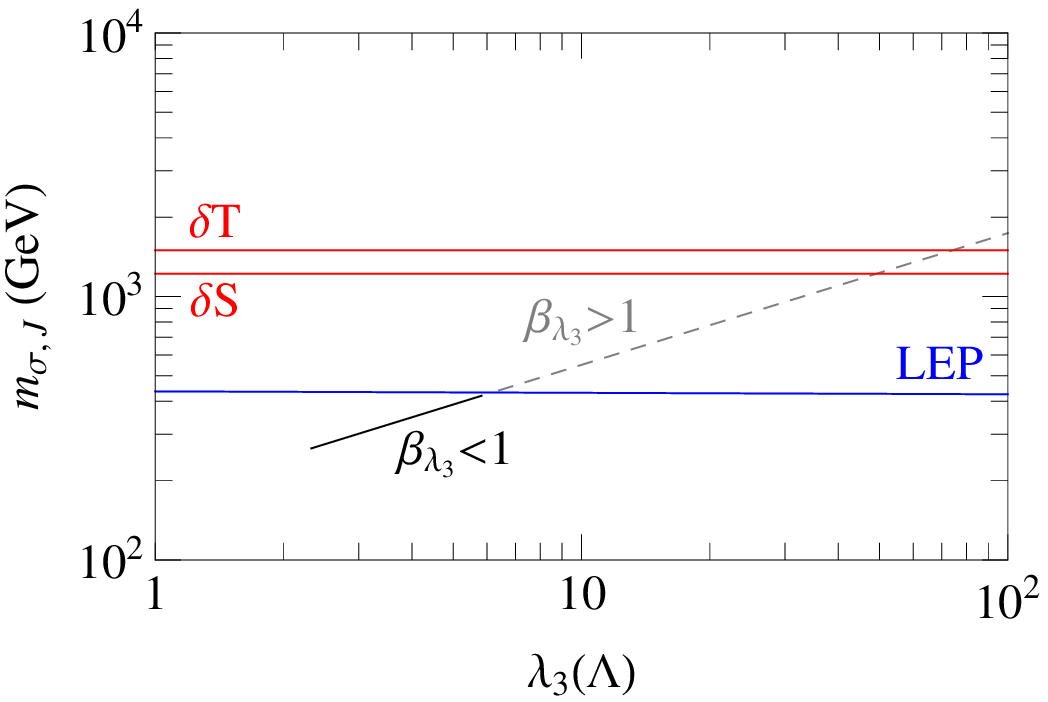}  
\caption{\it Numerical estimates of $m_{h}$ (upper plot) and
  $m_{\sigma, J}$ (lower plot) as functions of $\lambda_{3}(\Lambda)$ in the
  U(1)-symmetric Type-I MSISM.  The solid/black $\beta_{\lambda_{3}}
  <1$ line shows the perturbative values of $\lambda_{3}(\Lambda) \le
  5.84$, whilst the dashed/gray $\beta_{\lambda_{3}} > 1$ line shows
  the non-perturbative values of $\lambda_{3}(\Lambda) > 5.84$.  The
  area between the horizontal blue LEP line and the horizontal red
  $\delta S$ line is allowed by experimental considerations of the
  LEP2 mass limit on the SM-like $h$ boson and the $\delta S$
  parameter respectively.  The area above the horizontal red $\delta
  T$ line is excluded by the $\delta T$ parameter constraint.}
\label{fig:TypeIU1mass}
\end{figure}

\begin{figure}
\centering 
\includegraphics{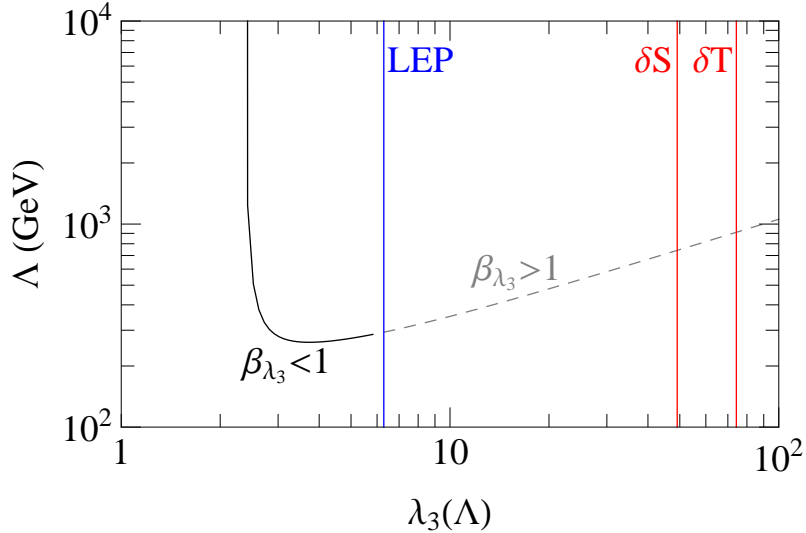}
\caption{\it The RG scale $\Lambda$ as a function of
  $\lambda_{3}(\Lambda)$ in the U(1)-symmetric Type-I MSISM.  The
  solid/black $\beta_{\lambda_{3}} <1$ line shows the perturbative
  values of $\lambda_{3}(\Lambda) \le 5.84$, whilst the dashed/gray
  $\beta_{\lambda_{3}} > 1$ line shows the non-perturbative values.
  The areas lying to the right of the red $\delta S$ and $\delta T$
  lines are excluded, and similarly to the left of the blue LEP line
  is also excluded by the LEP2 Higgs mass limit. }
\label{fig:TypeIU1Lambda}
\end{figure}

In  Fig.   \ref{fig:TypeIU1mass}, we  display  the  dependence of  the
scalar-boson masses $m_h$ and  $m_{\sigma, J}$ on the quartic coupling
$\lambda_{3}(\Lambda)$, for which  the Type-I flat-direction condition
$\lambda_1   (\Lambda)  =   0$   is  realized.    The  solid   (black)
$\beta_{\lambda_{3}}  <1$ lines determine  the perturbative  region of
the scalar-boson masses, which derive from the theoretical constraint,
$2.32  < \lambda_{3}(\Lambda)  \le 5.84$.   The continuation  of these
lines into dashed (grey) $\beta_{\lambda_{3}} > 1$ lines correspond to
the non-perturbative  regime, in which  $\lambda_{3}(\Lambda) > 5.84$.
The area between  the horizontal blue LEP line  and the horizontal red
$\delta  S$   line  indicates  the  combined   experimental  limit  on
$\lambda_{3}(\Lambda)$, i.e.~$6.29  \le \lambda_{3}(\Lambda) < 49.12$.
Similarly,  the region  above the  horizontal red  $\delta T$  line is
excluded by  the $\delta T$ limit.   It is interesting  to remark here
that   unlike  the   well-known   ``chimney  plot"~\cite{Kurt}   which
constrains   the  SM   Higgs-boson  mass   to  an   allowed   band  by
considerations  of   triviality  and  vacuum  stability~\cite{Costas},
Fig.~\ref{fig:TypeIU1mass}  shows  an  exact  value for  the  physical
scalar   masses  $m_{h,\sigma,  J}$   against  the   quartic  coupling
$\lambda_{3}(\Lambda)$ which is related to the RG scale $\Lambda$, see
Fig.~\ref{fig:TypeIU1Lambda}.

Fig.~\ref{fig:TypeIU1Lambda}  shows  the dependence  of  the RG  scale
$\Lambda$  on the quartic  coupling $\lambda_{3}(\Lambda)$.   The same
line colour convention as  in Fig.~\ref{fig:TypeIU1mass} is used, only
now the horizontal LEP, $\delta  S$ and $\delta T$ lines are vertical.
We  observe  that  as  $\lambda_{3}(\Lambda)$ approaches  its  minimum
value, the coefficient $\beta$ gets close to zero, and so the RG scale
$\Lambda$  tends to infinity.   However, this  area is  not physically
viable, as has already been excluded by the LEP limits.

If  we interpret  $\lambda_{3}(\Lambda)  \approx 6.3$  at  the RG  scale
$\Lambda \approx 294$~GeV as  the most experimentally favourable value
of this  quartic coupling within the U(1)-invariant  Type-I MSISM, we
are  then able  to offer  a  sharp prediction  for the  masses of  the
heavier degenerate  scalar bosons $\sigma$ and  $J$.  Specifically, by
virtue  of~(\ref{eqn:miTypeIU1invar}),   we  find  that  $m_{\sigma,J}
\approx 437$~GeV. The fields $\sigma$  and $J$ are both stable and can
qualify   as   DM  candidates   in   the  so-called   ``Higgs-portal''
scenario~\cite{Higgsportal}.   A~detailed   study  of  the   DM  relic
abundances of $\sigma$  and $J$ is beyond the scope  of this paper and
will be given elsewhere.

Since the $h$-boson couplings to fermions and electroweak gauge bosons
have exactly the SM form, its phenomenological distinction from the SM
Higgs boson  itself will  be difficult.  One  possibility would  be to
look for the presence of  large $h \sigma^2$- and $h J^2$-couplings at
the  International $e^+e^-$  Linear  Collider (ILC),  along the  lines
studied  in~\cite{Zerwas}.  Moreover,  even though  the  trilinear and
quadrilinear  $h$ self-couplings  are absent  at the  tree  level, the
large   $h\sigma^2$-   and   $h   J^2$-couplings  can   give   sizable
contributions  at  the one-loop  quantum  level. Therefore,  precision
Higgs experiments at  the ILC might be able  to distinguish the MSISM
from the SM.

From the analysis given above, it is clear that in spite of being very
predictive,  the   U(1)-invariant  Type-I  MSISM  has   a  number  of
weaknesses.   This scenario  satisfies all  experimental limits  for a
large quartic  coupling $\lambda_3 \approx  6.3$, which is close  to the
boundary of non-perturbative  dynamics. Another problematic feature is
that it exhibits  a Landau pole at energy  scales of order $10^4$~GeV,
which is many orders of magnitude below the standard GUT ($M_{\rm GUT}
\approx  2\times  10^{16}$~GeV) and  Planck  ($M_{\rm Planck}  \approx
1.2\times 10^{19}$~GeV) mass scales.   Therefore, in the next section,
we relax the constraint of  U(1) invariance, and investigate whether a
general Type-I  MSISM can  be perturbative up  to the GUT  and Planck
scales.


\subsection{The U(1) Non-Invariant Scenario}\label{Type1U1noninvar}

We now lift  the constraint of U(1) invariance  from the scalar sector
of the  Type-I MSISM.   The tree-level  scalar potential  of the
general Type-I MSISM then reads:
\begin{eqnarray}
V^{\mathrm{tree}}(\Lambda) & = &
\frac{\lambda_{2}(\Lambda)}{2}\,(S^{*} S)^{2}\: +\:
\lambda_{3}(\Lambda)\, \Phi^{\dagger} \Phi\, S^{*} S\: +\:
\lambda_{4}(\Lambda)\, \Phi^{\dagger} \Phi\, S^{2}\: +\:
\lambda_{4}^{*}(\Lambda)\, \Phi^{\dagger} \Phi\, S^{*2} \nonumber \\ &
& +\: \lambda_{5}(\Lambda)\, S^{3} S^{*}\: +\:
\lambda_{5}^{*}(\Lambda)\,S S^{*3}\: +\:
\frac{\lambda_{6}(\Lambda)}{2}\,S^{4}\: +\:
\frac{\lambda_{6}^{*}(\Lambda)}{2}\,S^{*4} \; .
\end{eqnarray}
Exactly as  we did for the  U(1)-invariant scenario, we  can show that
the tree-level scalar-boson  masses and the RG scale  $\Lambda$ do not
depend on all the couplings but only on $\lambda_{3}(\Lambda)$ and the
modulus  $|\lambda_4(\Lambda)|$  of   the  generally  complex  quartic
coupling  $\lambda_{4}(\Lambda)$.   In order  to  show this,  we~first
notice that  by substituting  $S = 0$  and $\lambda_{1}(\Lambda)  = 0$
into  the squared  scalar-boson  mass matrix~${\cal  M}^2_S$ given  in
(\ref{eqn:123}), we obtain only three non-zero matrix elements, i.e.
\begin{eqnarray}
	\label{eqn:miTypeIU1noninvar}
m_{\sigma}^{2} &=&  \frac{1}{2}\; \Big( \lambda_{3}(\Lambda)\: +\:
  \lambda_{4}(\Lambda)\: +\:  \lambda_{4}^{*}(\Lambda) \Big)\, v_{\phi}^{2}
\; , \nonumber\\ 
m_{J}^{2} &=&  \frac{1}{2}\; \Big( \lambda_{3}(\Lambda)\: -\:
  \lambda_{4}(\Lambda)\: -\: \lambda_{4}^{*}(\Lambda) \Big)\, v_{\phi}^{2}
\; , \nonumber\\ 
m_{\sigma J}  &=& \frac{i}{2}\; \Big( \lambda_{4}(\Lambda)\: -\:
  \lambda_{4}^{*}(\Lambda) \Big)\, v_{\phi}^{2} \; . 
\end{eqnarray}
If  $\lambda_{4}(\Lambda)$ is  complex,  the scalar-pseudoscalar  mass
term, $m_{\sigma  J}$, gives rise  to explicit CP violation.   In this
case, the scalar mass spectrum consists of the fields:
\begin{equation}
h\ \equiv\  \phi\; , \qquad 
H_{1}\ =\ \cos \theta\; \sigma\: +\: \sin \theta\, J\; ,\qquad  
H_{2}\ =\ -\: \sin \theta\; \sigma\: +\: \cos \theta\, J\; .
\end{equation}
If the theory preserves CP, we have that $H_{1} = \sigma$ and $H_{2} =
J$ are CP-even and CP-odd scalar fields, respectively.  In the general
case,  however,  the mass  eigenstates  $H_{1,2}$  have indefinite  CP
parities, with their tree-level masses given by
\begin{equation} 
	\label{eqn:TypeIUnot1ms1andms2}
m_{H_1}^2\ =\ \frac{1}{2}\; \Big(\lambda_3 (\Lambda)\: +\: 
  2 |\lambda_{4}(\Lambda)|\, \Big)\, v_{\phi}^{2}\; , \qquad  
m_{H_2}^2\ =\
\frac{1}{2}\; \Big( \lambda_3 (\Lambda)\: -\: 2 |\lambda_{4}(\Lambda)|
  \Big)\, v_{\phi}^{2} \; ,
\end{equation} 
where   $\cos^2  \theta  =   (m^2_\sigma  -   m^2_{H_2})/(m^2_{H_1}  -
m^2_{H_2})$.  Hence,  the scalar-boson masses  $m_{H_{1,2}}$ depend on
only    two   coupling    parameters,   $\lambda_3(\Lambda    )$   and
$|\lambda_{4}(\Lambda)|$.   For  the same  reason,  the two  effective
potential   coefficients   $\alpha$  and   $\beta$   also  depend   on
$\lambda_3(\Lambda   )$  and   $|\lambda_{4}(\Lambda)|$   through  the
scalar-boson masses  $m_{H_{1,2}}$.  It is therefore  not difficult to
see  that  the  one-loop  induced  $h$-boson mass  and  the  RG  scale
$\Lambda$   also   depend    only   on   $\lambda_3(\Lambda   )$   and
$|\lambda_{4}(\Lambda)|$,  by  means  of  (\ref{eqn:mhalphabeta})  and
(\ref{eqn:Lambdaalphabeta}).

The  fact  that  the  scalar-boson  masses $m_{H_{1,2}}$  have  to  be
positive leads to the constraint:
\begin{equation} \label{eqn:TypeIUnot1positivemasses}
\lambda_{3}(\Lambda)\ \ge\ 2|\lambda_{4}(\Lambda)|\ >\ 0\; .
\end{equation}
This  constraint  automatically   enforces  the  second  condition  in
(\ref{eqn:BFBconditions}) for  any value of $\theta_S$,  such that the
potential remains BFB, i.e.~$\Lambda_{12}  \ge 0$, for $\Lambda_{11} =
0$.   The  first BFB  condition  in~(\ref{eqn:BFBconditions}) is  only
fulfilled,  if  $\Lambda_{22}  \ge  0$.  This  restricts  the  allowed
parameter space  of the other  couplings, $\lambda_{2}$, $\lambda_{5}$
and $\lambda_{6}$.  In  order for the first BFB  condition to hold for
any possible value of the phase $\theta_{S}$, we must require that
\begin{equation}
\lambda_2(\Lambda)\ \ge\ 4\,|\lambda_5(\Lambda)|\: +\: 
2\,|\lambda_6 (\Lambda)|\ >\ 0\; .
\end{equation}
As  in   the  U(1)-invariant   scenario,  we  may   derive  additional
theoretical       limits      on       $\lambda_{3}(\Lambda)$      and
$|\lambda_{4}(\Lambda)|$,  by  demanding  that  the  couplings  remain
perturbative at  $\Lambda$ and  that the one-loop  effective potential
$V^{\mathrm{1-loop}}_{\mathrm{eff}}$  is  BFB.   The best  theoretical
upper limit on  $\lambda_{3}(\Lambda)$ and $|\lambda_{4}(\Lambda)|$ is
obtained by  requiring that  $\beta_{\lambda_{3}} \le 1$  at $\Lambda$
and  assuming that $\lambda_{2}(\Lambda)$,  $\lambda_{5}(\Lambda)$ and
$\lambda_{6}(\Lambda)$ are negligible.  This implies that
\begin{equation}
  \label{eqn:TypeINotU1pertlimit}
2\,\lambda_{3}^{2}(\Lambda)\: +\: 8\,|\lambda_{4}(\Lambda)|^{2}\:  +\: 
1.86\,\lambda_{3}(\Lambda)\ \le\ 8 \pi^{2} \; . 
\end{equation}
Correspondingly,  a  lower  theoretical   limit  may  be  obtained  by
requiring that $\beta > 0$, which translates into the constraint:
\begin{equation}
\lambda_{3}^{2}(\Lambda)\: +\: 4\,|\lambda_{4}(\Lambda)|^{2}\ \ge\ 5.39 \; .
\end{equation} 

Experimental data  encoded as  constraints on the  electroweak oblique
parameters  $S$,  $T$ and  $U$  provide  complementary  limits on  the
quartic  couplings  $\lambda_{3}(\Lambda)$ and  $|\lambda_{4}(\Lambda)
|$.  Exactly  as in  the U(1)-invariant scenario,  only the  $h$ boson
interacts  with the  SM  particles,  with couplings  of  the SM  form.
Therefore, as before, useful perturbative constraints on $\lambda_{3}$
and $|  \lambda_{4}(\Lambda) |$ can only  be derived from  the 95\% CL
interval of the electroweak  oblique parameters $\delta S$ and $\delta
T$ for $m_{H_{\rm  SM}} = 117$~GeV.  Since the  $h$ boson has standard
interactions, the  LEP2 lower limit on  the SM Higgs  boson applies in
full, giving rise to the constraint:
\begin{equation}
\lambda_{3}^{2}(\Lambda)\: +\:  4\,|\lambda_{4}(\Lambda)|^{2}\ >\ 39.54\; .
\end{equation}

\begin{figure}
\centering 
\includegraphics[height=0.36\textwidth,width=0.54\textwidth]{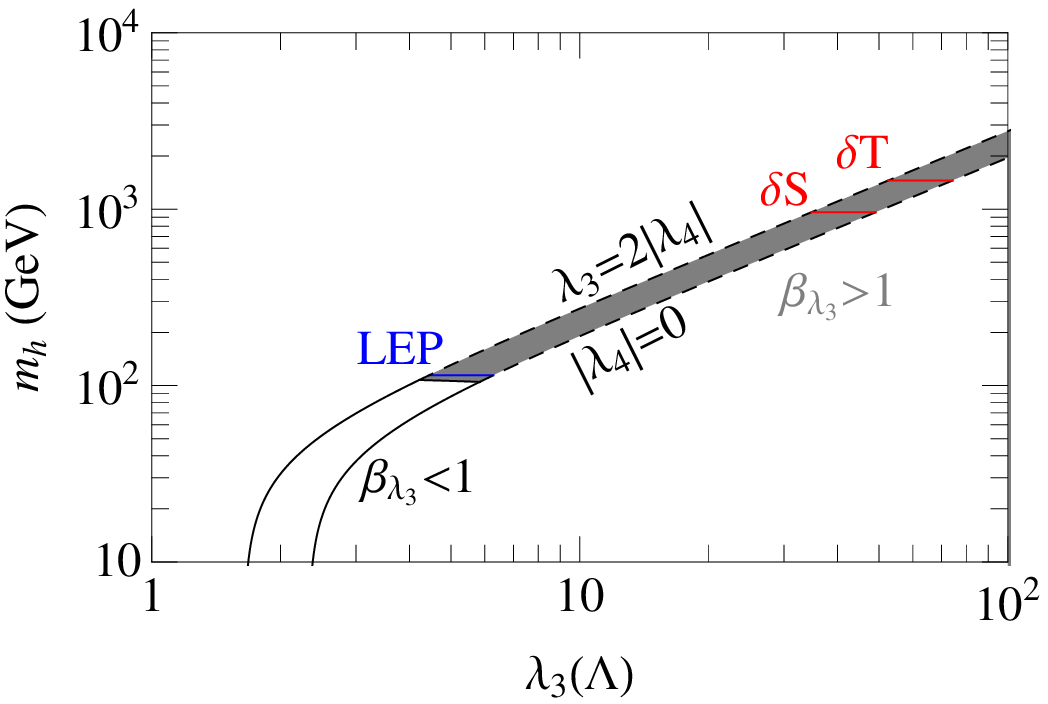}
\qquad
\includegraphics[height=0.36\textwidth,width=0.54\textwidth]{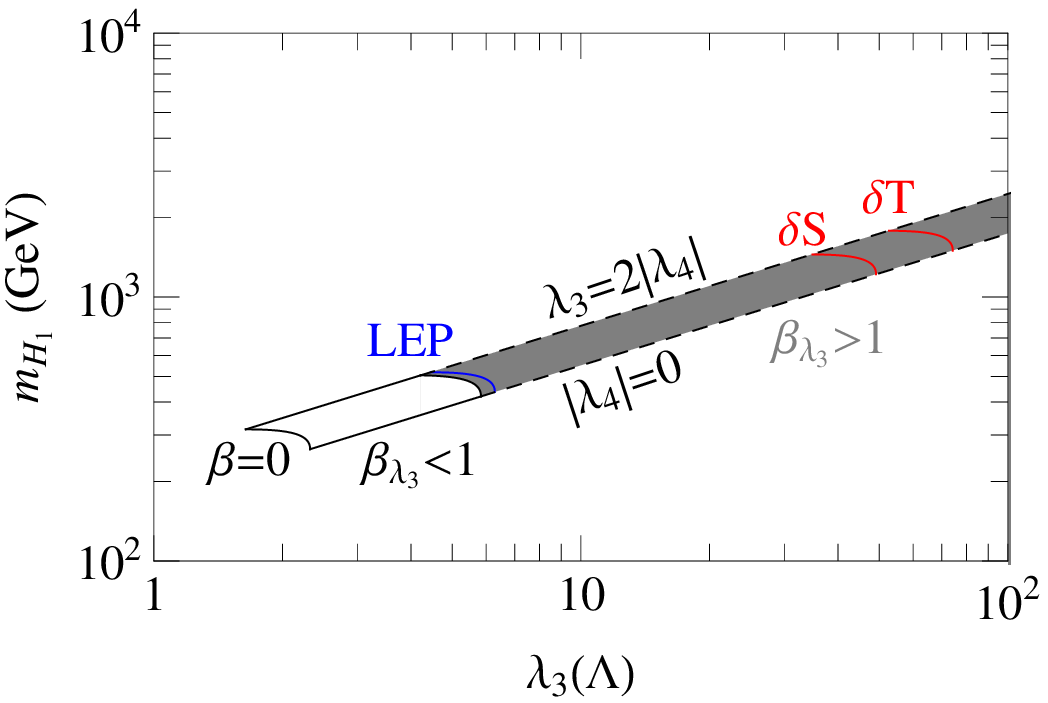}
\qquad
\includegraphics[height=0.36\textwidth,width=0.54\textwidth]{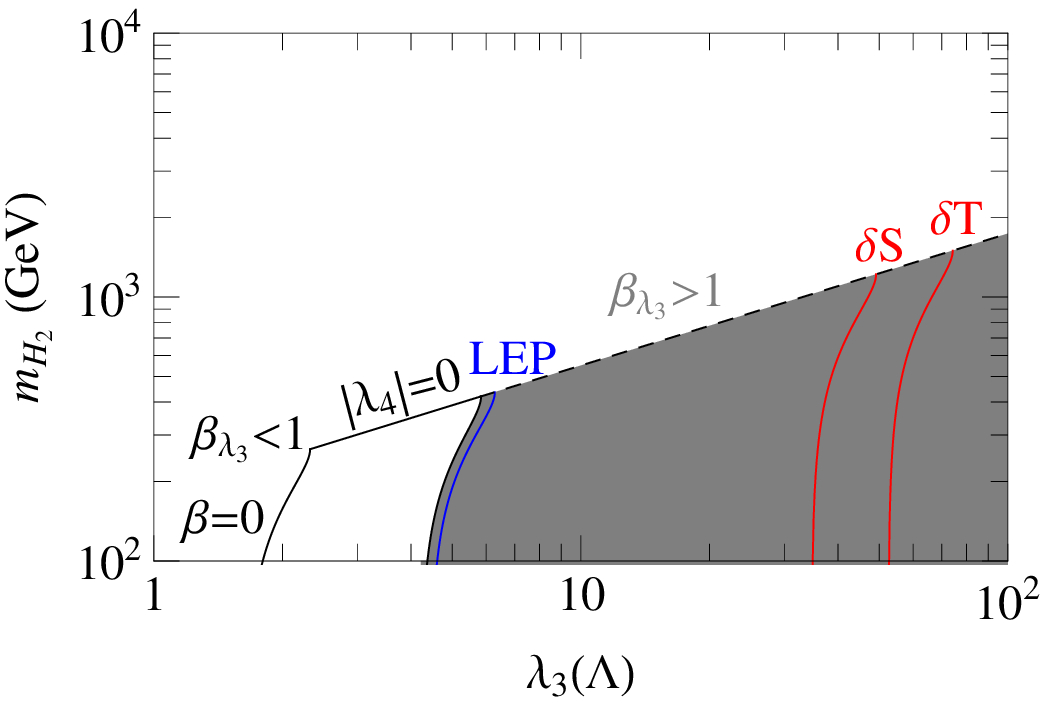} 
\caption{\it Numerical estimates of $m_{h}$~(upper panel),
  $m_{H_{1}}$~(middle panel) and $m_{H_{2}}$~(lower panel) versus
  $\lambda_{3}(\Lambda)$ in the general Type-I MSISM.  The white area
  between the black lines show the regions which correspond to
  perturbative values of $\lambda_{3}(\Lambda)$ and
  $|\lambda_{4}(\Lambda)|$ and positive scalar masses
  (\ref{eqn:TypeIUnot1positivemasses}), whilst the gray-shaded areas
  show their non-perturbative regions.  The areas lying to the right
  of the red lines for $\delta S$ and $\delta T$ are excluded.
  Likewise, the area left of the blue $LEP$ line is ruled out by the
  LEP2 Higgs-mass limit. }
\label{fig:TypeIUnot1mass}
\end{figure}

In  Fig.~\ref{fig:TypeIUnot1mass} we  present numerical  estimates for
the  scalar-boson  masses  $m_{h}$~(upper panel),  $m_{H_{1}}$~(middle
panel)  and $m_{H_{2}}$~(lower  panel),  as functions  of the  quartic
coupling   $\lambda_{3}(\Lambda)$,   after   incorporating   all   the
aforementioned theoretical and  experimental limits.  The perturbative
areas     which     also     contain    positive     scalar     masses
(\ref{eqn:TypeIUnot1positivemasses})  are given  by the  white regions
between the black lines, whereas their non-perturbative extrapolations
are shaded grey with black  dashed border lines.  The LEP2, $\delta T$
and  $\delta S$  limits  are shown  as  the blue  and  two red  lines,
respectively.  The areas to the right of the $\delta S$ and $\delta T$ lines are
excluded by the respective 95\%~CL  limits on $\delta S_{\rm exp}$ and
$\delta  T_{\rm exp}$.   Just  like the  U(1)-invariant scenario,  the
experimentally permitted  regions lie between  the LEP and  $\delta S$
lines, for  quartic couplings which are slightly  outside the boundary
of  perturbative  dynamics.   From  the  middle and  lower  panels  of
Fig.~\ref{fig:TypeIUnot1mass}, we see that the preferred values of the
$H_1$ and $H_2$ masses which  correspond to $m_{h} \sim 114.4$ GeV are
constrained to lie in the intervals:
\begin{equation}
436\ \mathrm{GeV}\ \stackrel{<}{{}_\sim}\ m_{H_{1}}\
\stackrel{<}{{}_\sim}\ 519 \ \mathrm{GeV} \; , \qquad 0 \ \mathrm{GeV}\ \le
m_{H_2}\ \stackrel{<}{{}_\sim}\ 436 \ \mathrm{GeV} \; .
\end{equation}
As in the U(1)-invariant scenario, the distinction of the Higgs sector
of  the  general  Type-I MSISM  from  that  of  the SM  might  require
precision Higgs experiments at the ILC.

\begin{figure}
\centering 
\includegraphics{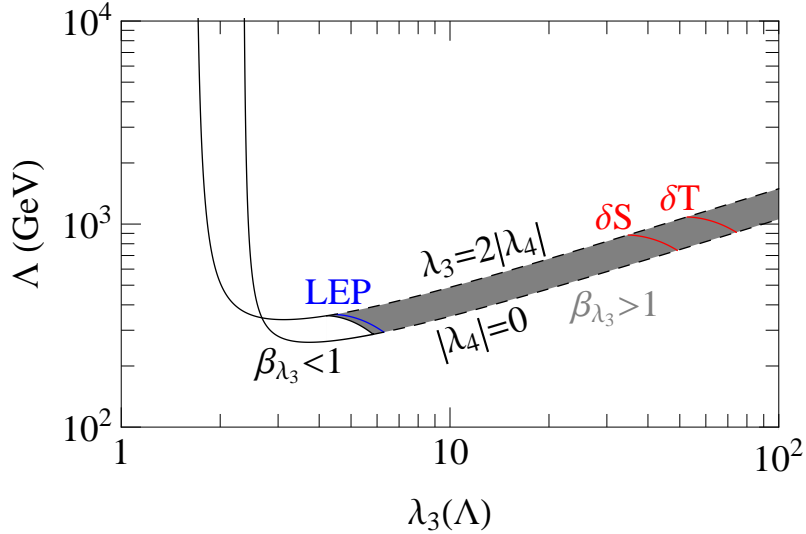}
\caption{\it   The RG scale $\Lambda$ as a function of
  $\lambda_{3}(\Lambda)$ in the general Type-I 
  MSISM.  The white area  between the black lines shows the region 
  that corresponds to  perturbative values  of $\lambda_{3}(\Lambda)$
  and  $|\lambda_{4}(\Lambda)|$,  whilst  the gray-shaded  area  shows
  the non-perturbative  region. The  area between the
  red $\delta S$ and  blue $LEP$ lines is permitted by the oblique
  parameters and the LEP2 Higgs-mass 
  limit. The area to the right of the red $\delta T$ line is excluded
  by the $\delta T$ limit. } 
\label{fig:TypeIUnot1Lambda}
\end{figure}

In Fig.~\ref{fig:TypeIUnot1Lambda} we display the dependence of the RG
scale  $\Lambda$ on  the quartic  coupling  $\lambda_{3}(\Lambda)$ and
include both  the theoretical and experimental limits,  using the same
line  colour  convention  as in  Fig.~\ref{fig:TypeIUnot1mass}.   From
Fig.~\ref{fig:TypeIUnot1Lambda}, we see that the RG scale $\Lambda$ is
of the electroweak order, lying in the range: $293 \ \mathrm{GeV} \, <
\Lambda < 359$~GeV, for perturbative $\lambda_3 (\Lambda )$ couplings,
once the LEP2 Higgs-boson mass limit is taken into account.

In the general  Type-I MSISM, the $H_2$ boson is  a stable particle in
all  the allowed  range of  the quartic  couplings. Therefore,  it can
represent  a viable  cold DM  candidate, provided  the $H_2$  boson is
sufficiently   massive,    e.g.~for   $m_{H_2}   \stackrel{>}{{}_\sim}
30$~GeV. If  $m_{H_{2}}$ is small, then  it opens a  new decay channel
for the  SM-like Higgs boson  $h$ via $h  \to 2H_{2}$ and  for certain
regions of parameter  space can be the dominant mode  of decay over $h
\to b \overline{b}$  when $m_{h} < 2 m_{W}$.  However,  for $m_{h} > 2
m_{W}$, the $h$-boson decay into $W^{+}W^{-}$ or $ZZ$ still dominates.
In addition  to the $H_2$ boson,  the heaviest $H_1$  boson might also
become a stable particle and so a valid DM candidate, if its decay via
the quartic interaction $H_1H^3_2$ is kinematically forbidden, i.e.~as
long as $m_{H_1} < 3 m_{H_2}$.

The general CP-violating Type-I MSISM  shares the same weakness as the
U(1)-invariant Type-I  MSISM. It  turns out that  it also  generates a
Landau  pole at  a maximum of $10^{4}$~GeV,  far below  the  GUT and  Planck
scales.   Unlike  the   U(1)-invariant  scenario,  the  general  model
contains new  sources of  CP violation, which  might be  of particular
importance  for  realizing  electroweak  baryogenesis.   However,  one
serious  drawback of  the Type-I  MSISM is  that it  cannot  provide a
natural implementation of the seesaw  mechanism.  Since the VEV of the
complex singlet  scalar vanishes,  i.e.~$S=0$, no Majorana  mass terms
can be generated in this scenario.  We therefore turn our attention in
the next section to the Type-II MSISM, where $S\neq 0$.

\setcounter{equation}{0}
\section{The Type-II MSISM}  \label{TypeII}

In  this section  we study  the  MSISM that  realizes a  Type II  flat
direction along  which both the  Higgs doublet $\Phi$ and  the complex
singlet scalar $S$ develop  non-zero VEVs.  We investigate the Type-II
MSISM in  two distinct cases: (i)~the U(1)-invariant  limit and (ii)~a
U(1) non-invariant scenario where CP is maximally broken spontaneously
along the  flat direction $\sigma =  J$.  For these  two scenarios, we
determine  the perturbative  values of  the quartic  couplings  of the
potential and the  limits that these set on  the scalar mass spectrum.
Once these limits are considered, we find that the electroweak oblique
parameters $S$, $T$  and $U$ give no further  constraints on the model
parameters. On  the other hand, as  we will see,  the LEP2 Higgs-boson
mass limit does produce useful limits on the quartic couplings and the
scalar-boson  mass   spectrum.   Unlike  in  the   Type-I  MSISM,  the
pseudo-Goldstone $h$ boson in the  Type-II case is in general a linear
composition of all the neutral  fields $\phi$, $\sigma$ and $J$.  As a
consequence, it  is possible for  all the Higgs mass  eigenstates $h$,
$H_{1}$  or $H_{2}$  to  couple to  the  $Z$ boson,  but with  reduced
strength compared to the SM Higgs-boson coupling.


\subsection{The {\boldmath U(1)} Invariant Limit}\label{TypeIIU1invar}

In the U(1) invariant limit, the Type-II MSISM tree-level potential
takes on the simple form:
\begin{equation}
	\label{eqn:Type2U1potential}
V^{\mathrm{tree}}(\Lambda)\  =\
\frac{\lambda_{1}(\Lambda)}{2}\:(\Phi^{\dagger} \Phi )^{2}\ +\
\frac{\lambda_{2}(\Lambda)}{2}\:(S^{*} S)^{2}\ +\
\lambda_{3}(\Lambda)\:\Phi^{\dagger} \Phi\, S^{*}S \; . 
\end{equation}
Imposing   the  minimization  conditions   (\ref{eqn:diffwrtphi})  and
(\ref{eqn:diffwrtS})            on            the           tree-level
potential~(\ref{eqn:Type2U1potential}),   one  gets  a   minimal  flat
direction  at  a given  RG  scale  $\Lambda$,  provided the  following
relations among  the VEVs of  the scalar fields and  quartic couplings
are simultaneously met:
\begin{equation}
  \label{eqn:Type2U1VEVrelation}
\frac{ \phi^{2}}{ \sigma^{2}}\ =\ \frac{ n_{\phi}^{2}}{ n_{\sigma}^{2}}\
=\ -\frac{\lambda_{2}(\Lambda)}{\lambda_{3}(\Lambda)}\ =\
-\frac{\lambda_{3}(\Lambda)}{\lambda_{1}(\Lambda)} \ ,
\end{equation}
where we have made use of the  U(1) symmetry to set the VEV of the $S$
field real. Hence, the  flat direction ${\bf \Phi}^{\rm flat}$ becomes
a two-dimensional vector with components $\phi$ and $\sigma$, given by
(\ref{eqn:Type2mixingexample}).        Moreover,       as       stated
after~(\ref{eqn:Type2flatdircond}),  the quartic couplings  should lie
in the ranges: $\lambda_{1}(\Lambda)  > 0$, $\lambda_{2}(\Lambda) > 0$
and $\lambda_{3}(\Lambda) < 0$.

The flat direction relation (\ref{eqn:Type2U1VEVrelation}) may be used
to reduce the number of  independent quartic couplings at $\Lambda$ to
two,   i.e.~$\lambda_{1}(\Lambda  )$   and   $\lambda_{3}(\Lambda  )$.
Instead, the quartic coupling $\lambda_{2}(\Lambda)$ may eliminated in
favour      of     the      relation:      $\lambda_{2}(\Lambda)     =
[\lambda_{3}(\Lambda)]^{2}/\lambda_{1}(\Lambda)$.   Consequently,  the
scalar masses and the RG  scale $\Lambda$ can be expressed entirely in
terms  of $\lambda_{1}(\Lambda)$  and  $\lambda_{3}(\Lambda)$.  Taking
the relations~(\ref{eqn:Type2U1VEVrelation})  into account, the scalar
mass matrix given  in (\ref{eqn:123}) has the following
non-zero entries:
\begin{equation}
	\label{eqn:massesTypeIIU1invar}
m_{\phi}^{2}\  =\ \lambda_{1}(\Lambda)\,v_{\phi}^{2} \; , \qquad
m_{\sigma}^{2}\ =\ - \lambda_{3}(\Lambda)\, v_{\phi}^{2} \; , \qquad
m_{\phi \sigma}\  =\  - \sqrt{- \lambda_{1}(\Lambda)
  \lambda_{3}(\Lambda)}\: v_{\phi}^{2} \; .
\end{equation}
We  note that  the  U(1)-invariant Type-II  MSISM  cannot realize  CP
violation in the Higgs  sector. Explicitly, the scalar mass spectrum
consists of the mass eigenstates
\begin{equation}
	\label{eqn:Type2U1hH}
h\ =\ \cos \theta \, \phi + \sin \theta \, \sigma \; , \qquad 
H_1\ \equiv\ H\ =\ - \sin \theta \,
\phi + \cos \theta \, \sigma \; , \qquad H_2\ \equiv\ J \; ,
\end{equation}
where     $\cos^{2}    \theta     =    -     \lambda_{3}(\Lambda)    /
[\lambda_{1}(\Lambda) - \lambda_{3}(\Lambda)]$.   The $h$ and $H\equiv
H_1$ bosons are CP even and the $J\equiv H_2$ boson CP odd. The CP-odd
scalar $J$ is the massless Goldstone boson, associated with the 
spontaneous symmetry breaking of
the U(1) symmetry. At the tree-level, the only massive scalar is the
$H$ boson, whose mass squared is given by
\begin{equation}
	\label{eqn:miTypeIIU1invar}
m_{H}^{2}\ =\ \big[ \lambda_{1}(\Lambda) - \lambda_{3}(\Lambda) \big]\:
v_{\phi}^{2} \; .
\end{equation}
Since    $m_{H}^{2}$    depends     solely    on    the    combination
$\lambda_{1}(\Lambda)  -  \lambda_{3}(\Lambda)$, so  do  the  two
effective potential coefficients  $\alpha$ and $\beta$.  Likewise, the
RG    scale    $\Lambda$    also    depends   on    the    combination
$\lambda_{1}(\Lambda)       -      \lambda_{3}(\Lambda)$,      through
(\ref{eqn:Lambdaalphabeta}).   However, the  one-loop  contribution to
$m_{h}$,     given    in    (\ref{eqn:mhalphabeta}),     depends    on
$\lambda_{3}(\Lambda)$ as  well, through the  flat direction component
$n_{\phi}  = \cos  \theta$,  given in  (\ref{eqn:Type2mixingexample}).
Thus, the Higgs sector of the U(1)-invariant Type-II MSISM depends on
$\lambda_{1}(\Lambda)        -        \lambda_{3}(\Lambda)$        and
$\lambda_{3}(\Lambda)$, or  equivalently on $\lambda_{1}(\Lambda)$ and
$\lambda_{3}(\Lambda)$.

\begin{figure}
\centering 
\includegraphics{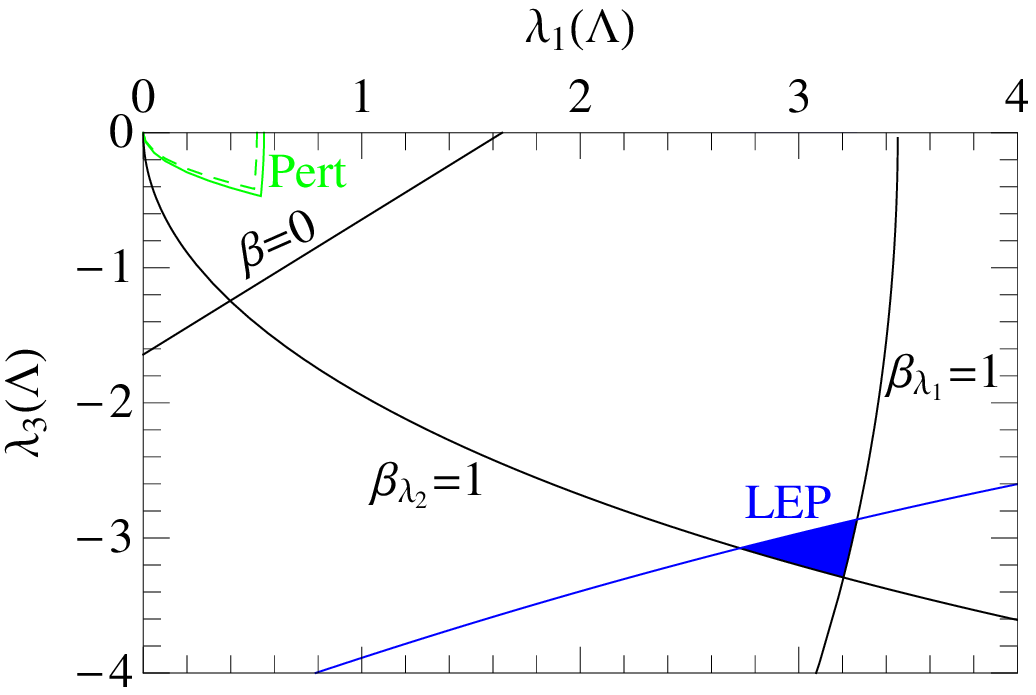}\\[10mm] 
\includegraphics{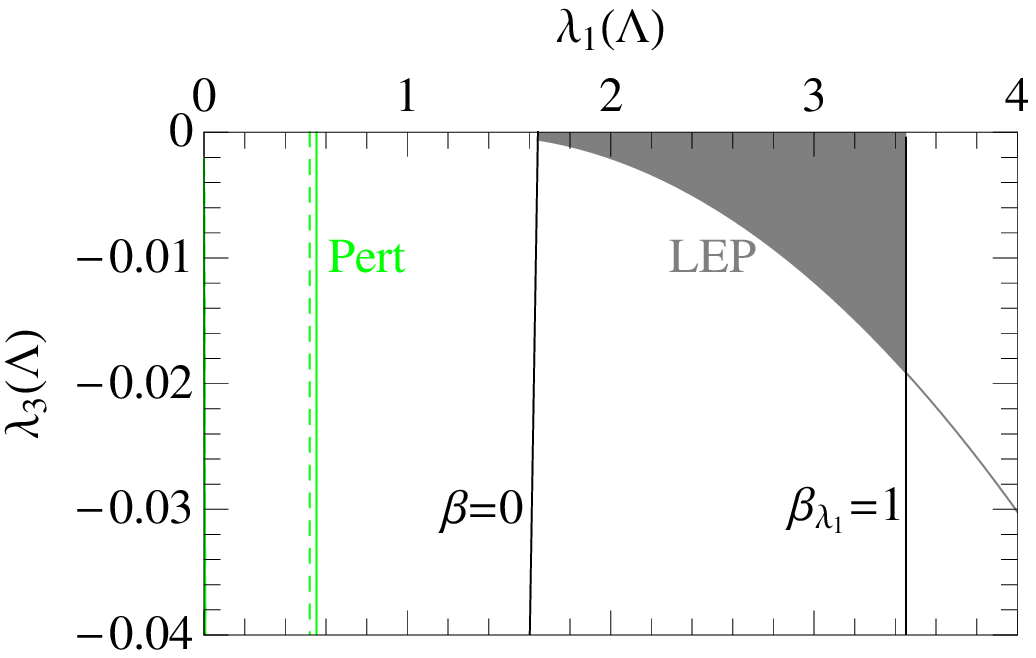}   
\caption{\it Theoretical and experimental exclusion contours in the
  $\lambda_{1}(\Lambda)$-$\lambda_{3}(\Lambda)$ parameter space in the
  U(1)-invariant Type-II MSISM.  The upper panel shows the full
  perturbative parameter space, whilst the lower panel focuses on the
  region with small $\lambda_{3}(\Lambda)$.  The theoretically allowed
  areas are enclosed by the black lines which correspond to keeping
  $\beta_{\lambda_{1,2}} \le 1$, $\beta > 0$ and
  $\lambda_{3}(\Lambda) \le 0$.  The LEP2 limit is given by the blue
  (grey) LEP line and above (below) is excluded for the upper (lower)
  panel.  The blue and grey shaded areas are allowed by the
  theoretical constraints, the LEP2 Higgs-mass limit and the oblique
  parameters.  The region of parameter space which remains
  perturbative to GUT (Planck) scale is enclosed by the solid (dashed)
  green Pert lines.  }
\label{fig:Type2U1paramspace}
\end{figure}

The  full  theoretical and  experimental  limits  on  the two  quartic
couplings   $\lambda_{1}(\Lambda)$   and  $\lambda_{3}(\Lambda)$   are
displayed in Fig.~\ref{fig:Type2U1paramspace}.  The top panel displays
the  full range,  whilst  the lower  panel  focuses on  a very  narrow
region,    which    is   viable    for    very    small   values    of
$\lambda_{3}(\Lambda)$.  As  theoretical constraints, we  require that
the   model   remains   perturbative   at  the   RG   scale~$\Lambda$,
i.e.~$\beta_{\lambda_{1,2}} (\Lambda) \le  1$, which is represented by
the        black        $\beta_{\lambda_{1,2}} = 1$        lines        in
Fig.~\ref{fig:Type2U1paramspace}.  From  these considerations, we find
the    upper     limits    $\lambda_{1}(\Lambda)    <     3.45$    and
$\lambda_{3}(\Lambda)   >  -   3.29$.    Another  useful   theoretical
constraint  is  obtained  by  requiring that  the  one-loop  effective
potential remains BFB ($\beta > 0$):
\begin{equation}
	\label{eqn:Type2U1lowertheor}
\lambda_{1}(\Lambda) - \lambda_{3}(\Lambda)\ >\ 1.64 \; ,
\end{equation}
which    is     indicated    by    the    black    $\beta = 0$     lines    in
Fig.~\ref{fig:Type2U1paramspace}.  Thus,  the theoretically admissible
region  is the one  enclosed by  the $\beta_{\lambda_{1,2}} = 1$,  $\beta = 0$ and
$\lambda_3 (\Lambda)  = 0$ lines  in the upper  panel and by  the $\beta = 0$,
$\beta_{\lambda_{1}} = 1$ and $\lambda_3 (\Lambda) = 0$ lines in the lower
panel.

The $\lambda_{1}(\Lambda)$-$\lambda_{3}(\Lambda)$  parameter space may
be further constrained by  experimental LEP2 limits on the Higgs-boson
mass and  by the electroweak oblique  parameters $S$, $T$  and $U$. We
find that the 95\% CL limits on $S$, $T$ and $U$ parameters provide no
additional    constraints     on    the    theoretically    admissible
region.   Instead,   the  LEP2   Higgs-boson   mass  limits   
significantly  restrict   the    $\lambda_{1}(\Lambda)$-$\lambda_{3}(\Lambda)$
parameter space.   To properly derive  these limits, we  first observe
that the  pseudo-Goldstone boson $h$ and the  heavy $H$-boson interact
with reduced couplings $g_{hVV}$ and $g_{HVV}$ with respect to the SM
coupling of $H_{\rm SM}$ to a pair of vector bosons $V = W^\pm,Z$. The
squared reduced couplings $g^2_{hVV}$ and $g^2_{HVV}$ are given by
\begin{equation} 
	\label{eqn:Type2reducedcouplings}
g^{2}_{hVV}\  =\  \cos^{2} \theta\ =\  \frac{- \lambda_{3}(\Lambda)
}{\lambda_{1}(\Lambda)  - \lambda_{3}(\Lambda) } \; , \qquad
g^2_{HVV}\ =\
\sin^{2} \theta\ =\  
\frac{ \lambda_{1}(\Lambda) }{\lambda_{1}(\Lambda)  -
  \lambda_{3}(\Lambda) } \; ,
\end{equation}  
satisfying  the identity:  $g^2_{hVV}  + g^2_{HVV}  =  1$.  Since  the
reduced $hZZ$-coupling  can be  much smaller than  the SM one,  the SM
Higgs-boson  mass  limit  $m_h  >  114.4$~GeV   no  longer  applies.
Instead,  we  use  the  combined constraints  on  $\xi^{2}_{h}  \equiv
g^2_{hVV}$  and  the  scalar  mass  $m_{h}$, which  are  presented  in
Fig.~10(a) of  Ref.~\cite{LEP115GeV}.  We perform a  polynomial fit up
to  order 10  on the  LEP2  data to  obtain a  reliable constraint  on
$\xi^{2}_{h}    (m_{h})$,     which    in    turn     restricts    the
$\lambda_{1}(\Lambda)$-$\lambda_{3}(\Lambda)$  parameter  space.  This
constraint is  represented by  the blue (grey)  LEP line in  the upper
(lower)  panel of  Fig.  \ref{fig:Type2U1paramspace},  where  the blue
(grey) shaded region respect  both the theoretical constraints and the
LEP2 Higgs-mass limit.  As there  are two distinct shaded regions blue
and grey,  which correspond respectively to higher  and lower values  of $m_{h}$, we
shall consider each scenario separately.

 
\subsubsection{The Electroweak Mass {\boldmath $h$}-Boson
  Scenario}\label{electroweak} 

\begin{figure}
\centering 
\includegraphics{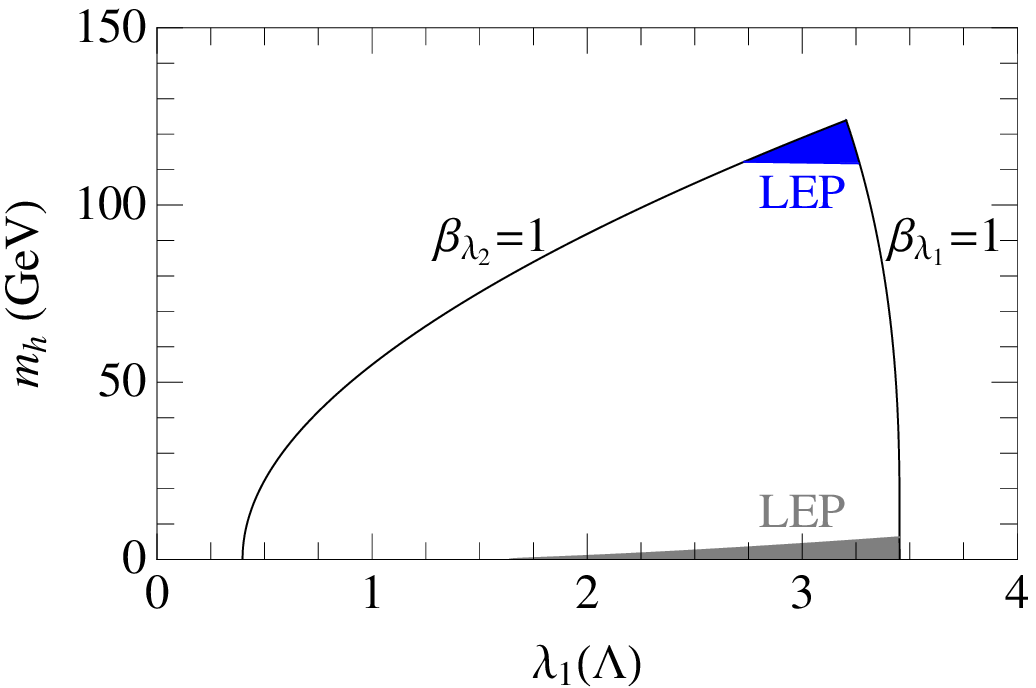}\\[7mm]
\includegraphics{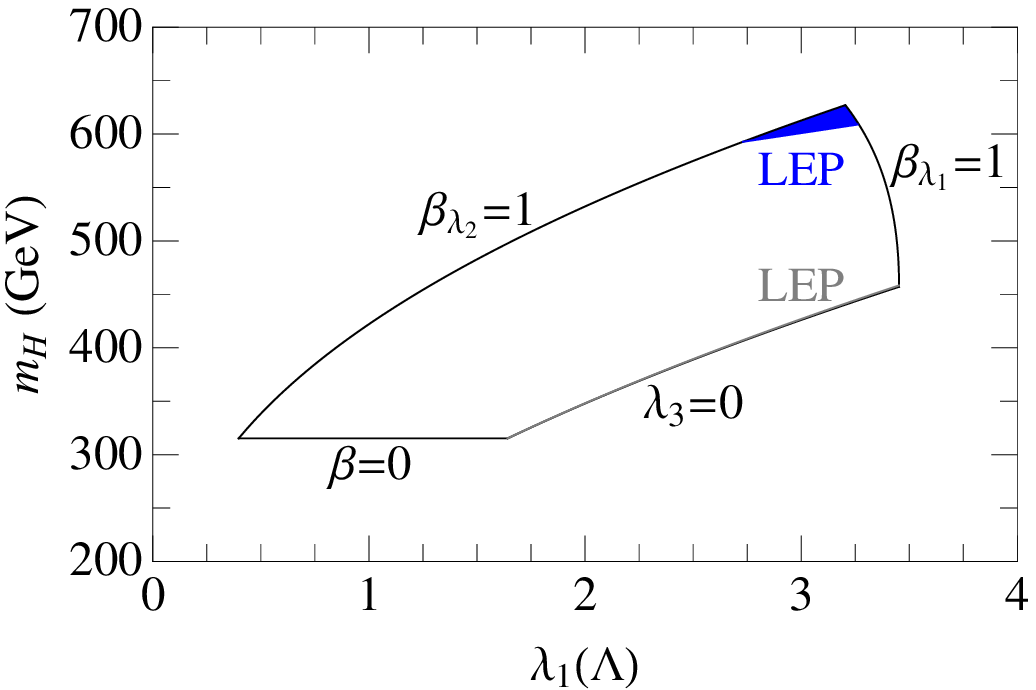}  
\caption{\it Predicted numerical values of $m_{h}$ (upper panel) and
  $m_{H}$ (lower panel) as a function of $\lambda_{1} (\Lambda)$ in
  the U(1)-symmetric Type-II MSISM.  The areas within the black lines
  show the regions which respect the theoretical constraints
  i.e. keeping $\beta_{\lambda_{1,2}} \le 1$, the potential BFB and
  $\lambda_{3}(\Lambda) \le 0$.  The blue (electroweak $m_{h}$) and grey
  (ultra-light $m_{h}$) shaded regions (denoted LEP) are permitted by the
  LEP2 Higgs-mass limit and the theoretical constraints. }
\label{fig:Type2U1mhandmH}
\end{figure}

We first  consider the higher  mass $h$-boson scenario  represented by
the  shaded blue  area in  Fig.~\ref{fig:Type2U1paramspace},  which is
dominated    by   large    values   of    $\lambda_{1}(\Lambda)$   and
$\lambda_{3}(\Lambda)$. In Fig~\ref{fig:Type2U1mhandmH} we show the
dependence  of the  scalar boson  masses  $m_{h}$ and  $m_{H}$ on  the
quartic  coupling $\lambda_{1}(\Lambda)$.  The  areas enclosed  by the
black lines are the  regions which respect the theoretical constraints
i.e.~$\beta_{\lambda_{1,2}}    \le    1$,     $\beta    >    0$    and
$\lambda_{3}(\Lambda) \le 0$. Including  the LEP2 Higgs-mass limit, we
obtain the shaded blue areas, corresponding to an electroweak mass $h$
boson, with mass in the range: $111.7~{\rm GeV} < m_{h} \le 123.9~{\rm
GeV}$,  and a  rather  heavy $H$  boson,  with mass  in the  interval:
$593~{\rm  GeV} <  m_{H} \le  627~{\rm GeV}$.   In addition,  the gray
shaded  areas correspond  to  a very  light  $h$ boson,  which is  not
clearly visible  on the lower  frame of Fig.~\ref{fig:Type2U1mhandmH},
as it follows the $\lambda_{3} =  0$ line. This ultra-light $h$-boson mass
scenario     will     be    discussed     in     more    detail     in
Subsection~\ref{Type2U1lowmh}.

The electroweak  mass $h$  boson could be  detected at the  CERN Large
Hadron  Collider  (LHC),  through  the  decay channel  $h  \to  \gamma
\gamma$.   The  observation of  the  $H$  boson  may proceed  via  the
so-called  ``golden channel,''  $H \to  ZZ \to  4l$.  However,  in the
region $\lambda_{1}(\Lambda) \approx  - \lambda_{3}(\Lambda)$, we have
$g^{2}_{hVV}   \approx   g^{2}_{HVV}    \approx   0.5$,   on   account
of~(\ref{eqn:Type2reducedcouplings}),  which  means  that both  decays
will give reduced signals compared  to the SM Higgs signals. Moreover,
the heavier $H$ boson may predominantly decay invisibly into a pair of
U(1)  Goldstone bosons  $J$~\cite{Majoron}, thanks  to  the relatively
large  quartic couplings.   This  last characteristic  makes the  U(1)
Type-II MSISM distinguishable from the corresponding Type-I one.

\begin{figure}[t]
\centering 
\includegraphics[scale=1.1]{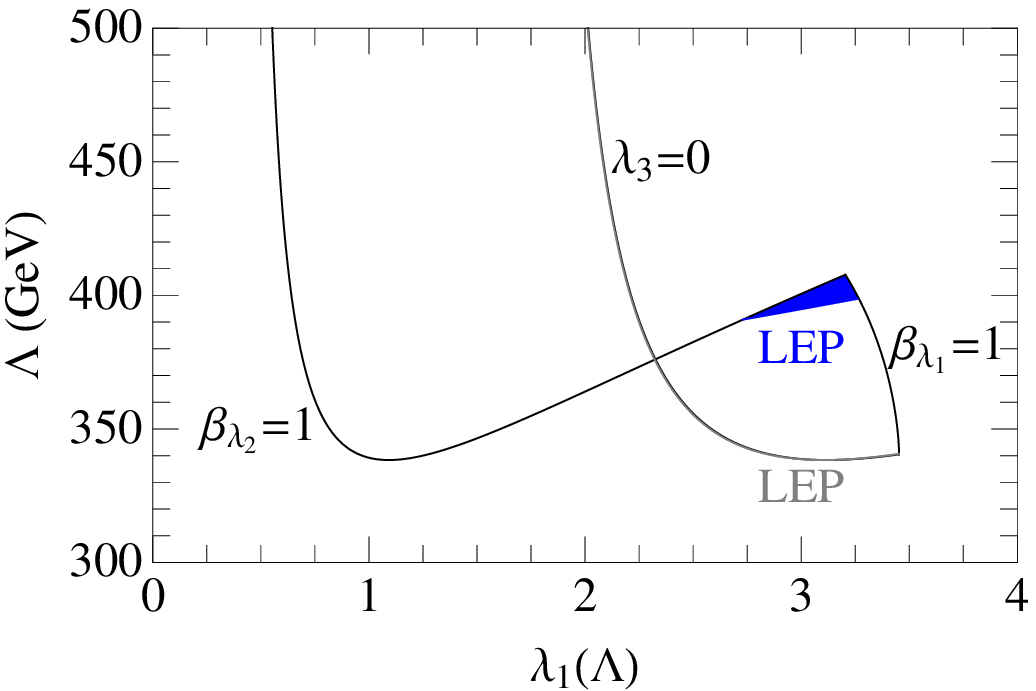}
\caption{\it Predicted numerical values of $\Lambda$ as a function of
  $\lambda_{1} (\Lambda)$ in the U(1)-invariant Type-II MSISM. The
  areas within the black lines show the regions which respect the
  theoretical constraints i.e. keeping $\beta_{\lambda_{1,2}} \le 1$,
  the potential BFB and $\lambda_{3}(\Lambda) \le 0$.  The blue (electroweak $m_{h}$) and grey
  (ultra-light $m_{h}$) shaded regions (denoted LEP) are
  permitted by the LEP2 Higgs-mass limit and the theoretical
  constraints.}
\label{fig:Type2U1Lambda}
\end{figure}

In Fig.~\ref{fig:Type2U1Lambda}  we present  the dependence of  the RG
scale   $\Lambda$   as   a    function   of   the   quartic   coupling
$\lambda_{1}(\Lambda)$.  The area within  the black lines respects the
theoretical constraints,  $\beta_{\lambda_{1,2}}~\le~1$, the effective
potential BFB  condition $\beta >0$ and  $\lambda_{3}(\Lambda) \le 0$.
The areas  which also  respect the LEP2  limit are shaded  blue, which
correspond  to the  electroweak  $h$-boson scenario,  and grey,  which
correspond  to a scenario  with a  very light  $h$ boson.   The latter
region is very narrow and not  clearly visible in the figure, since it
very   closely    follows   the   $\lambda_{3}   =    0$   line.    If
$\lambda_{1}(\Lambda)$  and  $\lambda_{3}(\Lambda)$  are in  the  blue
shaded region and remain perturbative,  then the RG scale $\Lambda$ is
of  the  order of  the  EW  scale and  lies  in  the  range $390.9  \,
\mathrm{GeV} \, \le \Lambda < 407.7$ GeV.  The region of the parameter
space that remains perturbative to GUT or Planck scales lies firmly in
the region excluded by the LEP2 limit. Taking this limit into account,
the theory  becomes non-perturbative at energies  $400$~GeV, with a
Landau pole at around $2 \times 10^{4}$ GeV.


\subsubsection{The Ultra-Light {\boldmath $h$}-Boson
  Scenario}\label{Type2U1lowmh} 

Another  experimentally   and  theoretically  viable   region  of  the
$\lambda_{1}(\Lambda)$-$\lambda_{3}(\Lambda)$      parameter     space
corresponds to  a very  small quartic coupling  $\lambda_3 (\Lambda)$,
giving rise to an ultra-light $h$ boson. The relevant region is shaded
grey in the lower  panel of Fig.~\ref{fig:Type2U1paramspace}. We will
not present a detailed phenomenological analysis of this scenario, but
rather highlight its key features.

As  can  be   seen  from  Fig.~\ref{fig:Type2U1paramspace},  the  LEP2
Higgs-mass  limit  puts   an  upper  bound  on  $-\lambda_{3}(\Lambda)
\stackrel{<}{{}_\sim}  0.019$.   In  view  of this  upper  bound,  the
largest $h$-boson  mass is $m_{h}  \stackrel{<}{{}_\sim} 6.3~{\rm GeV}$,
as     illustrated     by     the     grey    shaded     region     in
Fig.~\ref{fig:Type2U1mhandmH}. In this ultra-light $h$-boson scenario,
the reduced $hZZ$-coupling is rather suppressed, with $g^{2}_{hVV} \le
0.0055$, as  can be determined from~(\ref{eqn:Type2reducedcouplings}).
This fact renders the $h$ boson difficult to detect at the LHC.

The  other CP-even  $H$ boson  has almost  a SM-like  coupling  to the
vector bosons,  with $g^{2}_{hVV} \approx  1$. Its mass may  range for
perturbative     values     of     $\lambda_{1}(\Lambda)$,     between
$315~\mathrm{GeV} < m_{H} < 458~{\rm GeV}$. This range is given by the
$\lambda_{3}(\Lambda)  =  0$  line  in  Fig.~\ref{fig:Type2U1mhandmH}.
Since    the   $HJJ$-coupling   is    proportional   to    the   small
$\lambda_{3}(\Lambda)$ coupling  it is suppressed and  the SM-like $H$
boson would most likely be detected via the ``golden channel,'' $H \to
ZZ \to 4 l$.

An interesting  feature of the  ultra-light $h$-boson scenario  is the
existence of a region that  remains perturbative to higher scales
than the previously  considered models. This is indicated  by the area
enclosed    by    the    solid    and   dashed    green    lines    in
Fig.~\ref{fig:Type2U1paramspace}.   Specifically,  within the  allowed
region,  the  model  becomes  non-perturbative at  energies  of  order
$10^{4}~{\rm    GeV}$    and    develops    a    Landau    pole    at
energies~$10^{6}~{\rm GeV}$,  which is  higher than the
electroweak   mass  $h$-boson  scenario.   In  conclusion,   it  worth
reiterating that the U(1)-invariant Type-II MSISM has no new source of
CP   violation    beyond   the   standard    Kobayashi--Maskawa   (KM)
phase~\cite{KM}    and    predicts    no    massive    DM    candidate
(cf.~Table~\ref{tab:taxonomy}). In spite of these drawbacks, the model
does have the ability to generate Majorana neutrino masses through the
seesaw   mechanism,   as  we   will   discuss   in   more  detail   in
Section~\ref{Neutrinos}.   In  the following  section,  we consider  a
minimal   U(1)-violating   Type-II   MSISM  which   realizes   maximal
spontaneous CP violation~(SCPV).


\subsection{Minimal {\boldmath U(1)} Non-Invariant Model of Maximal
  SCPV}\label{MinmodelmaxspontaneousCPviolation}

Without  the restriction  of  U(1) invariance,  the tree-level  scalar
potential~(\ref{eqn:fullpotential})  of   the  general  Type-II  MSISM
contains  a total  of  9 real  quartic  couplings which  results in  a
multitude  of  valid  solutions  that  all  satisfy  the  minimization
requirements    (\ref{eqn:diffwrtphi})    and    (\ref{eqn:diffwrtS}).
However,  not  all  of  these possible  cases  are  phenomenologically
interesting. Therefore, we have  focused our investigation on a single
U(1)   non-invariant   scenario   that  minimally   realizes   maximal
spontaneous  CP violation,  i.e.~it  has a  flat  direction along  the
$\sigma =  J$ field line.  The  tree-level scalar potential  of such a
scenario is given by
\begin{equation}
	\label{eqn:potminmaxCP}
V^{\mathrm{tree}}\ =\ \frac{\lambda_{1}}{2}\;(\Phi^{\dagger} \Phi)^{2}\: +\:
\frac{\lambda_{2}}{2}\;(S^* S)^{2}\: +\: \lambda_{3}\,\Phi^{\dagger} \Phi\, 
S^* S\: +\: \frac{\lambda_{6}}{2}\;(S^{4} + S^{* 4}) \; ,
\end{equation}
where  $\lambda_4  = \lambda_5  =  0$ and  $\lambda_6$  is  real as  a
consequence  of  CP  invariance.   In  addition to  CP  symmetry,  the
tree-level scalar potential~(\ref{eqn:potminmaxCP}) is invariant under
the ${\bf Z}_4$ discrete symmetry: $S \to S' = \omega S$ and $\Phi \to
\Phi' =  \Phi$, with $\omega^4  = 1$.  The  CP symmetry and  the ${\bf
  Z}_4$ discrete symmetry  are sufficient to uniquely fix  the form of
the      tree-level      scalar     potential      $V^{\mathrm{tree}}$
given~in~(\ref{eqn:potminmaxCP}).

Minimizing  the tree-level  potential at  the RG  scale  $\Lambda$, by
means of (\ref{eqn:diffwrtphi})  and (\ref{eqn:diffwrtS}), we find the
following relations for the flat direction:
\begin{equation}
	\label{eqn:minconditionsminmaxCP}
\frac{ \phi^{2}}{ \sigma^{2}}\ =\ \frac{ n_{\phi}^{2}}{ n_{\sigma}^{2}}\
=\ - \frac{2 \lambda_{3}(\Lambda)}{\lambda_{1}(\Lambda)}\ =\ - \frac{2
  \big[ \lambda_{2}(\Lambda) - 2 \lambda_{6}(\Lambda)
    \big]}{\lambda_{3}(\Lambda)} \; , \qquad  \sigma\  =\ J  \; , \qquad
n_{\sigma}\  =\ n_{J}\; .  
\end{equation} 
Note that  the second  (or third) condition  implies a  flat direction
that  triggers maximal  spontaneous  CP violation  with $\theta_S  =
\pi/2$.   Combining  (\ref{eqn:minconditionsminmaxCP})  with  the  BFB
condition        (\ref{eqn:BFBconditions})        requires        that
$\lambda_{1}(\Lambda)   >   0$,   $\lambda_{3}(\Lambda)   <   0$   and
$\lambda_{2}(\Lambda) - 2\lambda_{6}(\Lambda) > 0$, where the signs of
$\lambda_{2}(\Lambda)$ and  $\lambda_{6}(\Lambda)$ individually remain 
undetermined.   Another  choice   of  a  maximally  CP-violating  flat
direction would be to have $\theta_{S} = 3 \pi/4$, i.e.~$\sigma = -J$.
However,  such a  choice does  not affect  the scalar  masses  or the
phenomeno\-logy of the model in an essential manner.

Other solutions to the minimization conditions, (\ref{eqn:diffwrtphi})
and  (\ref{eqn:diffwrtS}), are  possible  but they  either reduce  the
potential to the U(1)  invariant scenario ($\lambda_{6}(\Lambda) = 0$)
or modify it  to a Type I flat  direction ($\sigma = J =  0$), both of
which  have  been  previously  investigated  in Sections  6.1  and  5,
respectively.

The flat direction for this  model can be expressed as a 3-dimensional
vector, with non-zero $\phi$, $\sigma$ and $J$ components, i.e.
\begin{equation}
	\label{eqn:flatdirminmixCP}
{\bf \Phi}^{\mathrm{flat}}\ =\ 
\frac{\varphi}{\sqrt{ 2 \big[ \lambda_{1}(\Lambda) -
      \lambda_{3}(\Lambda) \big] }} \left( \begin{array}{c} 
\sqrt{- 2 \lambda_{3}(\Lambda)} \\ 
\sqrt{\lambda_{1}(\Lambda)} \\
\sqrt{\lambda_{1}(\Lambda)} 
\end{array}\right )\  
=\  
\frac{\phi}{ \sqrt{- 2 \lambda_{3}(\Lambda)}} \left( \begin{array}{c}
 \sqrt{- 2 \lambda_{3}(\Lambda)} \\ 
\sqrt{\lambda_{1}(\Lambda)} \\
\sqrt{\lambda_{1}(\Lambda)}
\end{array}\right )  \; .
\end{equation}
Considering   the   relations~(\ref{eqn:minconditionsminmaxCP}),   the
scalar mass matrix elements in (\ref{eqn:scalarmassesfor123}) become
\begin{eqnarray}
m_{\phi}^{2}\ =\ \lambda_{1}(\Lambda)\: v_{\phi}^{2} \; , & \quad &
m_{\sigma}^{2}\ =\ m_{J}^{2}\ =\ \big[ \lambda_{2}(\Lambda) +
  2\lambda_{6}(\Lambda)\big]\: v_{\sigma}^{2} \; , \nonumber \\ 
m_{\sigma J}\ =\ \big[ \lambda_{2}(\Lambda) - 6
  \lambda_{6}(\Lambda)\big]\: v_{\sigma}^{2} \; ,  & \quad & m_{\phi
  \sigma}\ =\ m_{\phi J}\ =\ \lambda_{3}(\Lambda)\: v_{\phi}
v_{\sigma}\; . 
\end{eqnarray}
Note  that   the  elements  $m_{\phi   J}$  and  $m_{\sigma   J}$  are
CP-violating.   In terms of  the quantum  fields $\phi$,  $\sigma$ and
$J$, the mass eigenstates $h$, $H_{1}$ and $H_{2}$ are given by
\begin{eqnarray}
	\label{eqn:Type2Unot1masseigenstates}
h & = &  \sqrt{\frac{- \lambda_{3}}{\lambda_{1} - \lambda_{3}}}\; \phi\ +\
\sqrt{\frac{\lambda_{1}}{2(\lambda_{1} - \lambda_{3})}}\;(\sigma + J)
\; ,\nonumber\\ 
H_{1} & = & \sqrt{\frac{\lambda_{1}}{\lambda_{1} - \lambda_{3}}}\; \phi\
-\ \sqrt{\frac{- \lambda_{3}}{2(\lambda_{1} - \lambda_{3})}}\;(\sigma +
J) \;, \nonumber \\ 
H_{2} & = & \frac{1}{\sqrt{2}}\;(- \sigma + J)\; .  
\end{eqnarray}
Correspondingly, their tree-level masses squared are given by
\begin{equation}
	\label{eqn:Type2Unot1mass}
m_{h}^{2} = 0 \; , \qquad 
m_{H_{1}}^{2} =  \big[ \lambda_{1}(\Lambda)
  - \lambda_{3}(\Lambda) \big]\: v_{\phi}^{2} \; , \qquad 
m_{H_{2}}^{2} = 4 \frac{\lambda_{1}(\Lambda) \lambda_{6}(\Lambda)}{ -
  \lambda_{3}(\Lambda)}\: v_{\phi}^{2}\; ,  
\end{equation}
where    we   employed    the    relation   $\lambda_{2}(\Lambda)    =
[\lambda_{3}^{2}(\Lambda)/\lambda_{1}(\Lambda)]                       +
2\lambda_{6}(\Lambda)$,      as     can     easily      be     derived
from~(\ref{eqn:minconditionsminmaxCP}).  In order for  the $H_2$-boson
mass   squared  $m^2_{H_{2}}$   to  be   positive,  we   require  that
$\lambda_{6}(\Lambda) > 0$, implying $\lambda_{2}(\Lambda) > 0$.

The Higgs sector of the Type-II MSISM of maximal SCPV depends on the
three         quartic        couplings:        $\lambda_{1}(\Lambda)$,
$\lambda_{2}(\Lambda)$   and   $\lambda_{3}(\Lambda)$.   The   quartic
coupling  $\lambda_{6}(\Lambda)$  can   be  eliminated  in  favour  of
$\lambda_2 (\Lambda)$,  by means of (\ref{eqn:minconditionsminmaxCP}).
Explicitly, the  scalar masses  $m_{H_{1}}$ and $m_{H_{2}}$  depend on
the three quartic couplings $\lambda_{1,2,3}(\Lambda)$, as can be seen
from  (\ref{eqn:Type2Unot1mass}).  Likewise,  the  RG scale~$\Lambda$
determined  in~(\ref{eqn:Lambdaalphabeta})  depends  on the  effective
potential coefficients $\alpha$ and $\beta$ that are both functions of
$m_{H_{1}}$       and      $m_{H_{2}}$.        Finally,      according
to~(\ref{eqn:mhalphabeta}), the pseudo-Goldstone  $h$ boson depends on
$\beta$            and            $n_{\phi}$.            Nevertheless,
from~(\ref{eqn:flatdirminmixCP}),    we    see    that    $n_\phi    =
\sqrt{-\lambda_{3}(\Lambda)/[\lambda_{1}(\Lambda)                     -
    \lambda_{3}(\Lambda)]}$.   Consequently,  the entire  scalar-boson
mass spectrum of the model only depends on the three quartic couplings
$\lambda_{1,2,3}(\Lambda)$.

We may now exploit the  extra freedom of the three independent quartic
couplings to identify  theoretically and experimentally viable regions
of the  parameter space which remain  perturbatively renormalizable up
to  Planck-mass  energy  scales.   To  be  precise,  we  require  that
$\beta_{\lambda_{1,2,3,6}}(M_{\rm  Planck})  \le  1$  and  impose  the
tree-level BFB conditions up  to the Planck scale: $\lambda_{1}(M_{\rm
  Planck}) >  0$, $\lambda_{2}(M_{\rm Planck})  - 2 \lambda_{6}(M_{\rm
  Planck}) >  0$ and $\lambda_{3}(M_{\rm Planck}) <  0$.  Moreover, if
we  assume  $\lambda_3(\Lambda )  \ll  \lambda_1(\Lambda)$, such  that
$\lambda_{6}(\Lambda)\approx  \frac{1}2 \lambda_{2}(\Lambda)$  we find
that the quartic  couplings $\lambda_{1,2}(\Lambda)$ are restricted to
the intervals,
\begin{equation}
  \label{Planckquartic}
0.39\ \stackrel{<}{{}_\sim}\ \lambda_{1}(\Lambda)\ \stackrel{<}{{}_\sim}\ 0.52\;,
\qquad
0\ <\ \lambda_{2}(\Lambda)\ \stackrel{<}{{}_\sim}\ 0.20\;,
\end{equation}
for   $-   0.1  \stackrel{<}{{}_\sim}   \lambda_3   (\Lambda)  <   0$.
From~(\ref{eqn:Type2Unot1mass}),   we  observe   that  in   the  limit
$\lambda_3  (\Lambda) \to 0$,  the $H_2$-boson  mass~$m_{H_2}$ becomes
infinite.   Therefore,   to  obtain  an  upper   limit  on  $\lambda_3
(\Lambda)$, we  require that the coefficients $\alpha$  and $\beta$ of
the     one-loop     effective    $V^{\mathrm{1-loop}}_{\mathrm{eff}}$
in~(\ref{eqn:1looppotMSISM}) are small,  e.g.~$\alpha ,\ \beta \le 1$,
such   that  perturbative   unitarity  in   the  Higgs   sector  holds
true~\cite{Dicus}.  In our numerical analysis, we apply the constraint
$\alpha \le 1$,  which is comparable to the  constraint $\beta \le 1$.
For   definiteness,   we   choose   two   representative   values   of
$\lambda_{2}(\Lambda)$:    $\lambda_{2}(\Lambda)     =    0.02$    and
$\lambda_{2}(\Lambda) = 0.2$.

\begin{figure}
\centering 
\includegraphics[height=0.36\textwidth,width=0.54\textwidth]{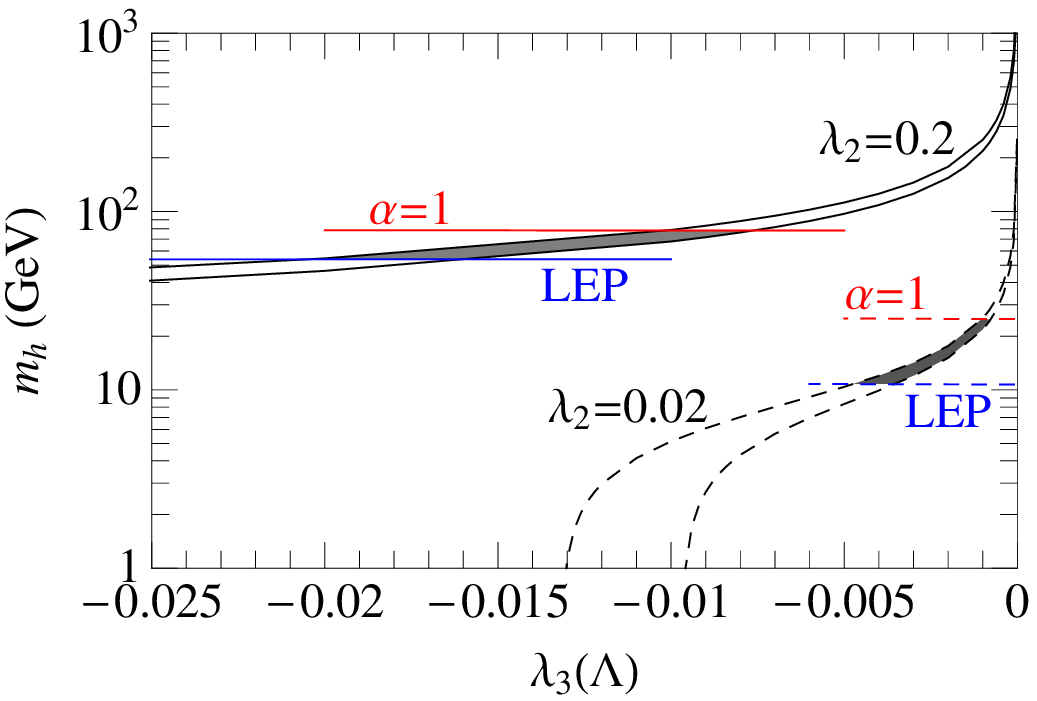}
\includegraphics[height=0.36\textwidth,width=0.54\textwidth]{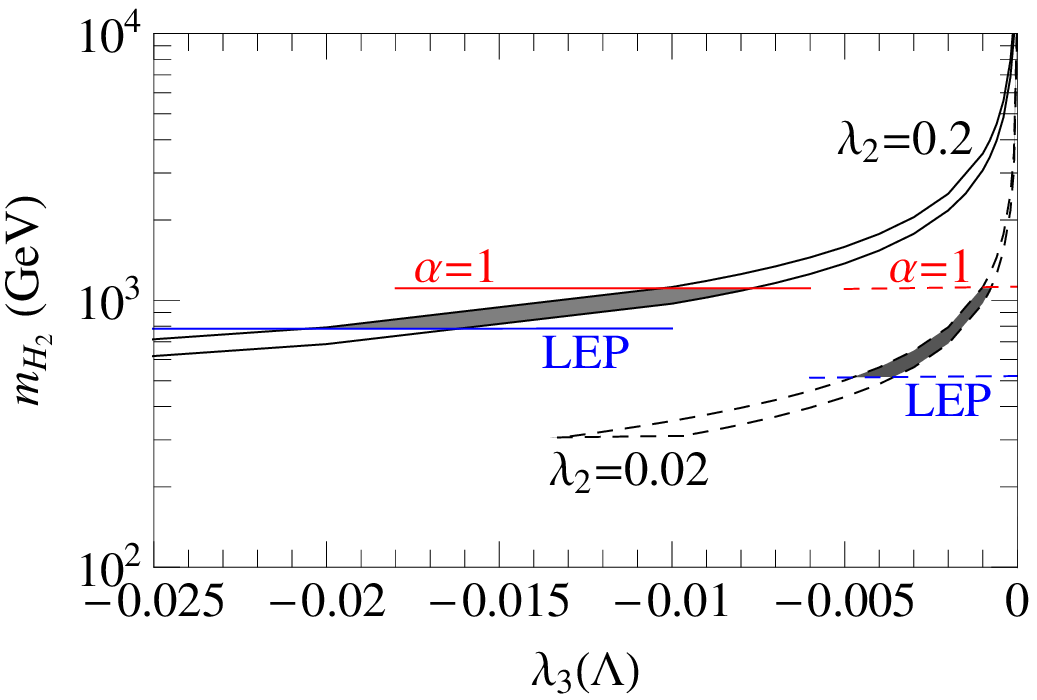}
\includegraphics[height=0.36\textwidth,width=0.54\textwidth]{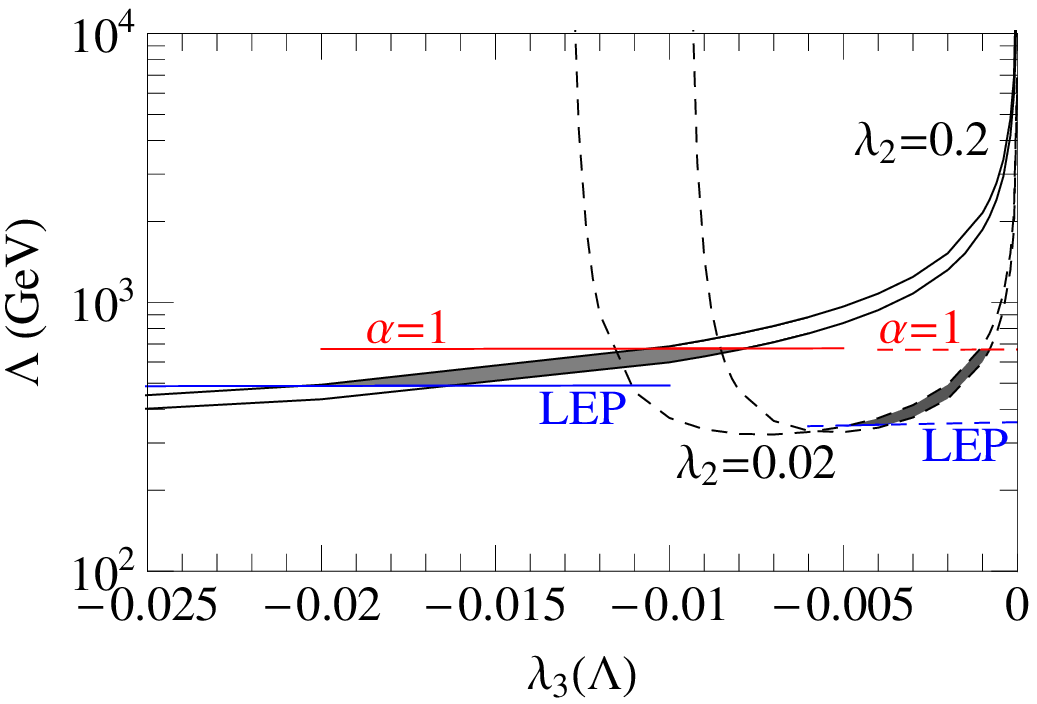}
\caption{\it Numerical estimates of $m_{h}$ (top panel), $m_{H_{2}}$
  (middle panel) and the RG scale $\Lambda$ (lower panel) as functions
  of $\lambda_{3}(\Lambda)$ in a Type-II MSISM of maximal SCPV.
  The areas between the solid and dashed black lines correspond to the masses, for which
  $\beta_{\lambda_{1}}(M_{\rm Planck}) \le 1$, $\lambda_{1}(M_{\rm
    Planck}) \ge 0$ and $\beta>0$ with $\lambda_{2}(\Lambda) = 0.02$ and $0.2$ respectively.  The
  solid and dashed blue lines represent the LEP2 Higgs-mass limit
  below which are excluded.  The solid and dashed red lines represent
  the constraint $\alpha \le 1$ and above each of the lines is
  excluded.  The grey regions correspond to areas that respect the theoretical and LEP2 limits.  The solid lines correspond to
  $\lambda_{2}(\Lambda) = 0.2$ whilst the dashed lines
  correspond to $\lambda_{2}(\Lambda) = 0.02$.}
\label{fig:Type2Unot1mhmH2Lambda}
\end{figure}


\begin{table}
\begin{center}
\begin{tabular} {|c|c|c|c|c|c|c|c|c|}
\hline
$\lambda_{2}(\Lambda)$ & \multicolumn{2}{|c|}{ $m_{h}$ } &
\multicolumn{2}{|c|}{ $m_{H_{1}}$} & \multicolumn{2}{|c|}{
  $m_{H_{2}}$} & \multicolumn{2}{|c|}{ $\Lambda$} \\ 
 & min & max & min & max & min & max & min & max \\
\hline
0.2  & 54 & 78 & 155 & 181 & 783 & 1110 &  490 & 675 \\ 
0.1  & 34 & 56 & 155 & 180 & 703 & 1110 &  444 & 674 \\
0.05 & 21& 39 & 154 & 179 & 607 & 1110 &  395 & 674  \\
0.02 & 11& 25 & 154 & 178 & 515 & 1110 &  350 & 675 \\
\hline
\end{tabular} 
\end{center}
\caption[Short Caption]{\it Minimum and maximum values of $m_{h}$,
  $m_{H_{1}}$, $m_{H_{2}}$ and $\Lambda$ as determined by the LEP2
  Higgs-mass limit and the theoretical constraint $\alpha \le 1$ for a
  range of $\lambda_{2}(\Lambda)$.}
  \label{table:masses}
\end{table}

In Fig.~\ref{fig:Type2Unot1mhmH2Lambda} we present numerical estimates
of  the scalar-boson masses,  $m_{h}$, $m_{H_{2}}$,  and the  RG scale
$\Lambda$,    as    functions   of    the    the   quartic    coupling
$\lambda_{3}(\Lambda)$.  The  solid and dashed black  line enclose the
regions  permitted by  considering the  theoretical bounds  which most
tightly  constrain   the  values  of   $\lambda_{1}(\Lambda)$  namely,
$\beta_{\lambda_{1}}(M_{\rm   Planck})  \le   1$,  $\lambda_{1}(M_{\rm
  Planck}) \ge  0$ and $\beta  > 0$, for $\lambda_{2}(\Lambda)  = 0.2$
and 0.02,  respectively.  The solid (dashed) blue  lines represent the
LEP2 Higgs-boson  mass limit, which  has been applied directly  to the
$h$-boson mass  $m_{h}$ for $\lambda_{2}(\Lambda) =  0.2$ (0.02).  The
regions below the blue LEP  lines are excluded for the specific values
of    $\lambda_{2}(\Lambda)$   considered.     As   a    result,   the
$\lambda_{3}(\Lambda)$  coupling has  to take  small  absolute values,
with  $\lambda_{3}(\Lambda) \stackrel{>}{{}_\sim}  -0.02$.   The solid
(dashed) red lines represent the  theoretical limit $\alpha \le 1$ for
$\lambda_{2}(\Lambda) = 0.2~(0.02)$, where  the area above the $\alpha
= 1$  lines is excluded. The  grey shaded regions are  the areas which
respect all  the theoretical constraints and the  LEP2 limit. Finally,
the electroweak oblique parameters offer no useful constraints, within
the      theoretically      allowed      parameter     space.       In
Table~\ref{table:masses}, we present the upper and lower limits on the
masses of the $h$ and $H_{2}$ bosons and on the RG scale $\Lambda$ for
different  values  of $\lambda_{2}(\Lambda)$.   The  lower bounds  are
determined  from the LEP2  Higgs-mass limit,  whilst the  upper bounds
come from the theoretical constraint $\alpha \le 1$.

\begin{figure}[t]
\centering 
\includegraphics{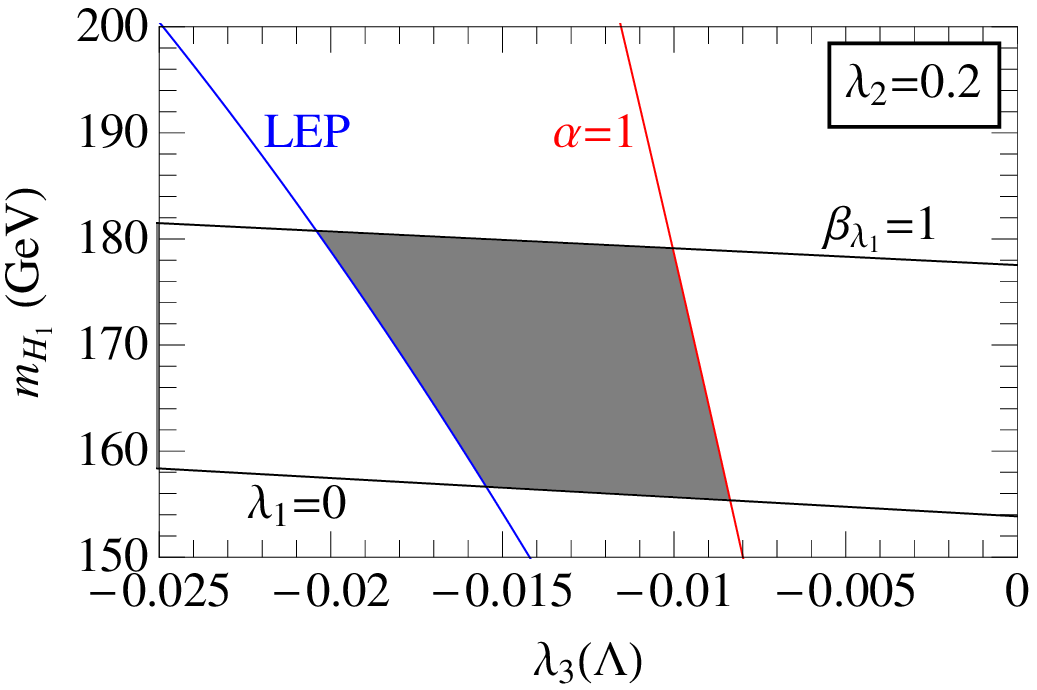}
\includegraphics{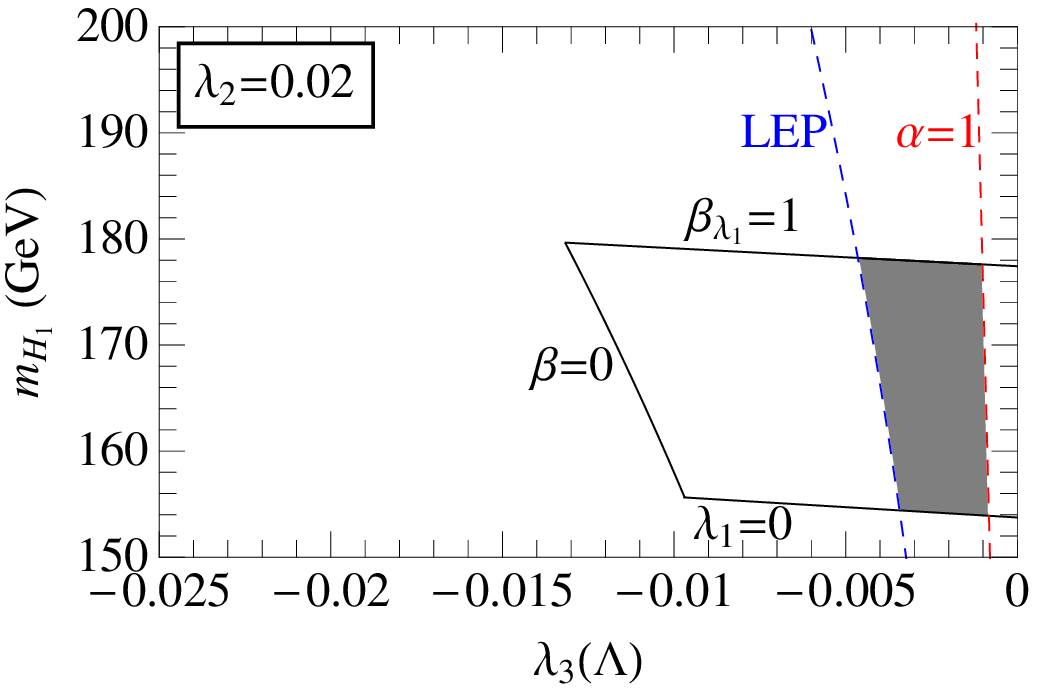}  
\caption{\it Predicted  numerical values of $m_{H_{1}}$  as a function
  of  $\lambda_{3}(\Lambda)$ for  $\lambda_{2}(\Lambda) =  0.2$ (upper
  panel) and $\lambda_{2}(\Lambda) = 0.02$ (lower panel), in a Type-II
  MSISM  of  maximal SCPV.   The  black  lines  corresponds to  masses
  restricted  by the conditions:  $\beta_{\lambda_{1}}(M_{\rm Planck})
  \le  1$, $\lambda_{1}(M_{\rm  Planck})~\ge~0$  and $\beta>0$  (lower
  panel  only).   The grey  shaded  regions  correspond  to the  areas
  permitted by  the LEP2 limit (solid  and dashed blue  lines) and the
  $\alpha \le 1$ limit (solid and dashed red lines).}
\label{fig:Type2Unot1mH1}
\end{figure}

In Fig. \ref{fig:Type2Unot1mH1} we  display numerical estimates of the
$H_{1}$-boson    mass    $m_{H_1}$    as    a    function    of    the
$\lambda_{3}(\Lambda)$ coupling.  The black lines correspond to values
of the quartic couplings  which respect the limits $\lambda_{1}(M_{\rm
  Planck}) > 0$,  $\beta_{\lambda_{1}}(M_{\rm Planck})~<~1$ and $\beta
>0$.   Even  though  the  $H_1$-boson mass  $m_{H_{1}}$  evaluated  in
(\ref{eqn:Type2Unot1mass})    does    not    explicitly   depend    on
$\lambda_{2}(\Lambda)$,  the LEP2  limit  applied to  $m_{h}$ and  the
theoretical  constraint  $\alpha  \le  1$  do, as  can  be  seen  from
Table~\ref{table:masses}.   The grey  shaded areas  between  the solid
(dashed)  blue  LEP and  red  $\alpha =1$  lines  are  allowed by  the
respective constraints  for $\lambda_{2}(\Lambda) =  0.2$ (0.02).  The
LEP2 limit provides an upper limit on the value of $m_{H_{1}}$, whilst
the  $\alpha = 1$  constraint gives  a lower  limit.  These  upper and
lower   limits   on   the   $H_{1}$-boson  mass   are   exhibited   in
Table~\ref{table:masses},     for     various     values    of     the
$\lambda_{2}(\Lambda)$ coupling.

In  spite of the  additional quartic  coupling $\lambda_{6}(\Lambda)$,
the  interactions of  the  $h$ and  $H_1$  scalars to  a  pair of  $V=
W^\pm,Z$ bosons are very  similar to the U(1)-invariant scenario.  The
reduced $hVV$- and $H_1VV$-couplings are given by
\begin{equation} 
	\label{eqn:Type2Unot1reducedcouplings}
g^{2}_{hVV}\  = \  \frac{- \lambda_{3}(\Lambda)
}{\lambda_{1}(\Lambda)  - \lambda_{3}(\Lambda) } \; , \qquad
g^2_{H_{1}VV}\ =\  
\frac{ \lambda_{1}(\Lambda) }{\lambda_{1}(\Lambda)  - \lambda_{3}(\Lambda) } \; .
\end{equation}
In the Type-II MSISM of maximal  SCPV under study, the $h$ boson has a
large  component  from  the  heavy   $H_{2}$  scalar  and  so  it  can
generically be  heavier than the respective $h$  in the U(1)-invariant
model, this allows  it to comfortably evade detection  at the LEP2. On
the  other  hand,  the $H_1$  boson  has  a  SM-like coupling  to  the
electroweak  vector bosons  and would  again most  likely  be detected
through  the  standard discovery  channel  $H_1  \to  ZZ \to  4l$.  In
addition to the  standard discovery channel, $H_1 \to  ZZ \to 4l$, the
$H_1$  boson  may  now decay  favourably  to  a  pair of  $h$  bosons,
i.e.~$H_1 \to  hh$, if  kinematically allowed. Then,  each of  the $h$
bosons  may decay  into a  pair of  $\tau$ leptons  or $b$  quarks.  A
detailed phenomenological study of  this detection channel for the LHC
is beyond the scope of this paper.

The  minimal  Type-II  MSISM  of  maximal  SCPV  gives  rise  to  rich
phenomeno\-logy.  As mentioned previously, the model spontaneously and
maximally violates the CP symmetry.  Since the complex singlet $S$ has
a non-zero VEV,  the model can also generate  naturally small neutrino
masses  through the  seesaw mechanism.   Moreover, the  presence  of a
permutation parity symmetry, $\sigma \leftrightarrow J$, which remains
intact  after EWSSB, renders  the massive  $H_{2}$ boson  stable, with
vanishing  VEV.   Hence, the  $H_2$  boson could  act  as  a cold  DM,
according   to  the   Higgs-portal   scenario~\cite{Higgsportal}.   In
general, there  are two parity symmetries  that could be  imposed on a
general Type-II MSISM with  SCPV, they are: $\sigma \leftrightarrow J$
and $\sigma \leftrightarrow -J$.  Both symmetries lead to similar mass
spectra, so we do not discuss them separately. Also, both symmetries
trigger {\em  maximal} SCPV, which  might open up the  possibility for
successful electroweak baryogenesis in this scenario.

In summary, the  Type-II MSISM of maximal SCPV  is a theoretically and
experimentally viable scenario. The quartic couplings of the model can
remain perturbative  up to Planck  energy scales and  its scalar-boson
spectrum is  compatible with limits  from LEP2 Higgs searches  and the
$S$, $T$ and $U$ oblique parameters.  Most importantly, the model does
not require additional theory to stay perturbatively renormalizable up
to the  standard quantum gravity scale,  i.e.~$M_{\rm Planck}$.  Since
the addition  of right-handed neutrinos can have  a significant impact
on                the                one-loop                effective
potential~$V^{\mathrm{1-loop}}_{\mathrm{eff}}$      and     on     the
phenomenology of  the model  in general, we  analyze in detail  such a
scenario in the next section.

\setcounter{equation}{0}
\section{The MSISM with Right-Handed Neutrinos}\label{Neutrinos}

In  order to  account for  the observed  non-zero neutrino  masses, we
extend the MSISM  with three right-handed neutrinos, $\nu^0_{1,2,3R}$.
As  was already  mentioned in  Section  4.4, the  Type-I MSISM  cannot
realize the seesaw  mechanism since the VEV of the  $S$ field is zero
along the minimal flat direction. The only way of introducing neutrino
masses  in a  SI fashion  into the  Lagrangian is  through  the hugely
suppressed neutrino  Yukawa couplings  of order $10^{-12}$,  which are
about 6 orders of magnitude smaller than the electron Yukawa coupling.
Obviously, such a scenario  has the difficulty of naturally explaining
the  smallness of  the light  neutrino masses.   Moreover,  the Type-I
MSISM with right-handed neutrinos  is a highly uninteresting scenario
as the  actual effect of the  very small neutrino  Yukawa couplings on
the scalar potential is negligible.

We therefore turn our attention to the Type-II MSISM.  The Lagrangian
term ${\cal L}_\nu$ in~(\ref{eqn:Lfull}), which describes the dynamics
of the right-handed neutrinos, is given by
\begin{eqnarray} 
  \label{eqn:neutrinoL}
\mathcal{L}_{\nu} \!& = &\! \bar{\nu}^{0}_{iR} i \gamma^{\mu}\partial_{\mu}
\nu^{0}_{iR}\: -\: {\bf h}^{\nu}_{ij} \bar{L}_{iL} \tilde{\Phi}
\nu^{0}_{jR}\: -\: {\bf h}^{\nu \dagger}_{ij} \bar{\nu}^{0}_{iR}
\tilde{\Phi}^{\dagger} L_{jL}\: -\:  \frac{1}{2}\; {\bf h}^{N}_{ij}
\bar{\nu}^{0C}_{iR} S \nu^{0}_{jR}\: -\: \frac{1}{2}\; {\bf h}^{N \dagger}_{ij}
\bar{\nu}^{0}_{iR} S^{*} \nu^{0C}_{jR}     \nonumber\\ 
\!& &\!  -\: \frac{1}{2}\; {\bf \tilde{h}}^N_{ij} 
\bar{\nu}^{0}_{iR} S \nu^{0 C}_{jR} \:
-\:  \frac{1}{2}\; {\bf \tilde{h}}^{N \dagger}_{ij} \bar{\nu}^{0 C}_{iR}
S^{*} \nu^{0}_{jR} \; . 
\end{eqnarray}
where the usual summation convention over repeated indices is implied,
with $i,j  = 1,2,3$  labelling the three  generations, $e$,  $\mu$ and
$\tau$,  respectively.   In~(\ref{eqn:neutrinoL}), ${\bf  h}^\nu_{ij}$
are the Dirac-neutrino Yukawa couplings of the SM Higgs doublet~$\Phi$
to  the  lepton  doublets  $L_{iL}$,  as defined  in  Appendix~A.   In
addition, ${\bf  h}^N_{ij}$ and  ${\bf \tilde{h}}^N_{ij}$ are  the two
possible Majorana-neutrino  Yukawa couplings of the  singlet field $S$
to the right-handed neutrinos~$\nu^0_{1,2,3R}$.  Note that ${\bf h}^N$
and ${\bf \tilde{h}}^N$ are symmetric $3\times 3$ matrices, i.e.~${\bf
  h}^N =  {\bf h}^{N\,T}$,  ${\bf \tilde{h}}^N =  {\bf \tilde{h}}^{N\,
  T}$.  Since the  Majorana-neutrino Yukawa couplings ${\bf h}^N_{ij}$
and  ${\bf \tilde{h}}^N_{ij}$ can  be sizeable,  we need  to calculate
their  effect   on  the  flat-directions  and   the  one-loop  $\beta$
functions.   Technical  details  of  such calculations  are  given  in
Appendices~B and~C.

Since  $S \neq  0$ along  the  Type-II flat  direction, the  following
neutrino mass terms are generated:
\begin{equation} 
	\label{eqn:neutrinoonlyL}
\mathcal{L}^{\rm Mass}_\nu\ =\ 
-\ \frac{1}{2} (\bar{\nu}^0_{iL} , \bar{\nu}^{0C}_{iR}) \;
\left( \begin{array}{c c}
{\bf 0} & {\bf m}_{Dij}\\
{\bf m}_{Dij}^{T} & {\bf m}_{Mij}
\end{array}\right) \;
\left( \begin{array}{c}
\nu^{0C}_{jL} \\
\nu^0_{jR}
\end{array}\right)
\ +\ {\rm H.c.}
\end{equation}
with
\begin{equation} 
	\label{eqn:mDmM}
{\bf m}_D\ =\ \frac{\phi}{\sqrt{2}}\ {\bf h}^{\nu}\ , \qquad 
{\bf m}_M\ =\ \frac{1}{\sqrt{2}} \Big[\, \sigma ( {\bf h}^N + 
{\bf \tilde{h}}^{N \dagger}) + i J ({\bf h}^N  - {\bf \tilde{h}}^{N
  \dagger})\, \Big]  \; . 
\end{equation}
Without loss of generality, we can assume a weak basis, in which ${\bf
m}_M$  is  diagonal, real  and  positive,  whilst  ${\bf h}^N$,  ${\bf
\tilde{h}}^N$ and ${\bf m}_D$  are in general $3\times 3$ non-diagonal
complex matrices.
  
The  $6\times 6$ mass  matrix in  $\mathcal{L}^{\rm Mass}_\nu$  can be
block-diagonalized via a unitary matrix $U$ as follows:
\begin{equation}
U^{T}  \left( \begin{array}{c c}
{\bf 0} & {\bf m}_{D}\\
{\bf m}_{D}^{T} & {\bf m}_{M}
\end{array}\right) U\ =\ \left( \begin{array}{c c}
{\bf m}_{\nu} & {\bf 0} \\
{\bf 0} & {\bf m}_{N}
\end{array}\right)\ .
\end{equation}
To  leading  order  in an  expansion  in  powers  of ${\bf  m}_D  {\bf
  m}_M^{-1}$, we obtain the standard seesaw formulae:
\begin{equation} 
	\label{eqn:mnmN}
{\bf m}_{\nu}\ =\ -\; {\bf m}_{D} {\bf m}_{M}^{-1} {\bf m}_{D}^{T} \; ,  \qquad 
{\bf m}_{N}\ =\ {\bf m}_{M} \; .
\end{equation}
where  ${\bf m}_\nu$  is  a  $3\times 3$  light  neutrino mass  matrix
pertinent to the masses  of the observed light neutrinos $\nu_{1,2,3}$
and  ${\bf m}_N$  is the  heavy neutrino  mass matrix,  predicting new
heavy Majorana  neutrinos, which  we denote hereafter  as $N_{1,2,3}$.

As  we  will  see  in  this  section,  the  heavy  Majorana  neutrinos
$N_{1,2,3}$ in the  Type-II MSISM are typically not  much heavier than
the  EW scale.   In the  standard seesaw  framework~\cite{seesaw}, all
Dirac-neutrino  Yukawa couplings  ${\bf h}^\nu_{ij}$  have to  be less
than  about $10^{-6}$,  e.g.~of  order the  electron Yukawa  coupling.
However, the  possible presence  of approximate flavour  symmetries in
${\bf m}_D$ and/or  ${\bf m}_M$~\cite{WW/MV,AZPC,KS} are sufficient to
relax this constraint for  some of the Dirac-neutrino Yukawa couplings
${\bf    h}^\nu_{ij}$   and    render   them    sizeable    of   order
$10^{-2}$--1~\cite{KPS,IP}.    Even  though   we  keep   the  analytic
dependence of our  results on ${\bf h}^\nu$, we  assume that all ${\bf
  h}^\nu_{ij} \stackrel{<}{{}_\sim}  0.01$, such that  their numerical
impact on the one-loop effective potential and the electroweak oblique
parameters can be safely ignored.

In the following, we study several representative scenarios within the
framework of the Type-II MSISM with right-handed neutrinos.  First, we
consider a U(1)-symmetric theory that preserves the lepton number.  We
then consider a benchmark scenario  of Type-II MSISM with maximal SCPV
and  analyze two  variants  of  such a  scenario.   The first  variant
assumes a CP-symmetric neutrino Yukawa sector, where the CP invariance
is only violated spontaneously by the ground state of the theory.  The
second  variant  promotes a  parity  symmetry  present  in the  scalar
potential of the model to the neutrino Yukawa sector, thus giving rise
to a massive stable scalar particle. This stable scalar particle could
act as a potential candidate to solve the cold DM problem.


\subsection{Neutrinos in the U(1) Invariant Type-II MSISM}

We now consider the effect  of including right-handed neutrinos in the
the U(1)-invariant Type-II MSISM.  The imposition of U(1) symmetry on
the   neutrino  Yukawa   sector  is   equivalent   to  lepton-number
conservation, where the  right-handed neutrinos $\nu^0_{1,2,3R}$ carry
the lepton  number $+1$  and the singlet  field $S$ the  lepton number
$-2$.  As a  consequence of  lepton-number conservation,  the Majorana
Yukawa  coupling ${\bf \tilde{h}}^N$  vanishes and  the heavy-neutrino
mass matrix along the Type-II flat direction is given by
\begin{equation}
{\bf m}_{N}\ =\ \frac{\sigma}{\sqrt{2}} \; {\bf h}^N\; , 
\end{equation}
where we have set $J = 0$  by virtue of a U(1) rotation.  

With the aid of (6.2), we may now express the
light-  and heavy  neutrino  mass matrices,  ${\bf  m}_\nu$ and  ${\bf
m}_N$, in terms of the SM VEV $v_\phi$:
\begin{equation}
	\label{eqn:U1neutrinomasses}
{\bf m_{\nu}}\ =\ -\; \sqrt{\frac{ - \lambda_{3}(\Lambda)}{ 2
    \lambda_{1}(\Lambda)}}\ 
v_{\phi}\,  {\bf h^{\nu}}({\bf h}^N)^{-1} {\bf h}^{\nu T}  \; ,\qquad
{\bf m}_{N}\ =\ \sqrt{\frac{\lambda_{1}(\Lambda)}{- 2 \lambda_{3}(\Lambda)}}\
v_{\phi}\, {\bf h}^N\; .
\end{equation}
where ${\bf h}^N$  is a real and diagonal  matrix.  For simplicity, we
assume  that three  heavy  Majorana neutrinos  $N_{1,2,3}$ are  nearly
degenerate, specifically by assuming that ${\bf h}^N = h^{N} {\bf 1}_3$
is  SO(3)  symmetric.  The  perturbativity  constraint  on the  Yukawa
couplings  ${\bf h}^N$ may  be translated  into the  inequality, ${\rm
Tr}\big(\mbox{\boldmath  $\beta$}^\dagger_{{\bf  h}^N} \mbox{\boldmath
$\beta$}_{{\bf  h}^N}\big) \le 3$,  at the  RG scale  $\Lambda$.  This
constraint  leads to  the  upper bound,  $h^N(\Lambda  ) <  4.0$, for  a
perturbative  theory  up to  the  EW scale.   If  we  insist that  the
Majorana Yukawa couplings ${\bf h}^N_{ij}$ stay perturbative up to the
Planck  scale, we find  the tighter  upper limit:  $h^{N}(\Lambda) \le
0.89$.  Finally,  the condition that  the one-loop scalar  potential be
BFB,  i.e.~$\beta  > 0$,  along  with  the perturbativity  conditions,
$\beta_{\lambda_{1,2,3}}  \le  1$, yield  the  upper  limit on  $h^N$,
$h^N(\Lambda) < 2.5$, at the EW scale.

\begin{figure}
\centering \includegraphics{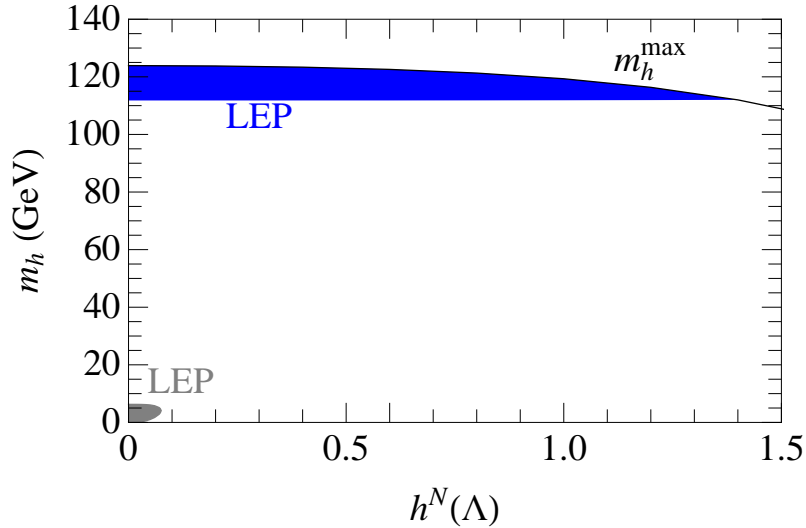}
\caption{\it Predicted numerical values of the LEP2 Higgs-mass limit
  allowed range of $m_{h}$ as a function of $h^{N}$ in the Type-II
  U(1)-invariant MSISM with right-handed neutrinos.  The blue and grey
  shaded areas correspond to those regions allowed by the LEP2 limit,
  for the electroweak and ultra-light $h$-boson scenarios,
  respectively.  The black $m_{h}^{\mathrm{max}}$ line represents the
  maximum perturbatively attainable values of $m_{h}$.}
\label{fig:Type2U1mhvshm2}
\end{figure}

Fig. \ref{fig:Type2U1mhvshm2} shows the allowed parameter space of the
$h$-boson  mass  and  the  Majorana-neutrino  Yukawa  coupling  $h^N$,
compatible with  the LEP2 Higgs-mass limit.   The maximum perturbative
value is  represented by  the black $m_{h}^{\mathrm{max}}$  line, such
that  the  area between  the  black  line and  the  $m_{h}  = 0$  line
corresponds  to perturbative masses.   The maximum  perturbative value
for   $m_h$  depends   on  the   perturbatively  allowed   values  for
$\lambda_{1}(\Lambda)$,   $\lambda_{3}(\Lambda)$    and   $h^N$,   i.e
$\beta_{\lambda_{1,2}}(\Lambda) \le 1$ and $\beta_{h^N}(\Lambda) \le
1$.  Since  right-handed neutrinos  induce a negative  contribution to
the  coefficient $\beta$  defined in~(\ref{eqn:alphabeta})  and  so to
$m_h$ in~(\ref{eqn:mhalphabeta}),  $m_{h}^{\mathrm{max}}$ decreases as
the  right-handed  neutrino   Yukawa  coupling  $h^N$  increases.   In
Fig.~\ref{fig:Type2U1mhvshm2},  the areas which  are permitted  by the
LEP2 Higgs-mass  limit are shaded  blue and grey, for  the electroweak
mass   and   the  ultra-light   $h$-boson   scenarios,  discussed   in
Subsections~\ref{electroweak}  and  \ref{Type2U1lowmh},  respectively.
In     the    electroweak     mass    $h$-boson     scenario,    where
$\lambda_{3}(\Lambda)   \approx  -3$,  the   Majorana-neutrino  Yukawa
coupling $h^N$ is  restricted to be: $h^{N} <  1.40$. Instead, for the
ultra-light  $h$ boson  scenario  (with $\lambda_{3}(\Lambda)  \approx
-0.02$), we  get the upper limit:  $h^{N} < 0.074$.   In this context,
the influence of the Majorana-neutrino Yukawa coupling $h^N$ on the RG
scale  $\Lambda$  is not  significant,  as  we  find $\Lambda  \approx
464$~GeV  for $h^{N}_{\rm  max} =  1.40$.  We  also verified  that all
values  of  $\lambda_{1}(\Lambda)$  and  $\lambda_{3}(\Lambda)$  which
respect $\beta_{\lambda_{1,2}}(\Lambda) \le 1$  lie within the 95\% CL
interval  of $\delta S_{\rm  exp}$, $\delta  T_{\rm exp}$  and $\delta
U_{\rm exp}$, using the limits for $m_{H_{\rm SM}} = 117$~GeV.

\begin{figure}
\centering \includegraphics{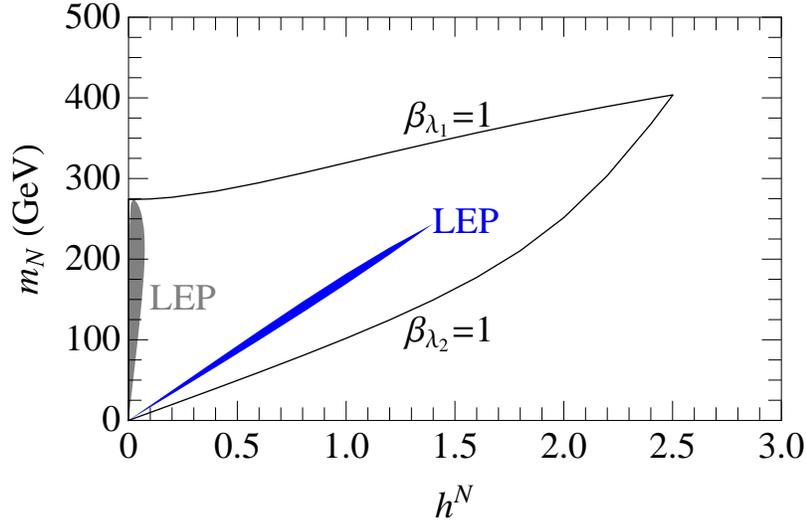}
\caption{\it Perturbatively allowed values of $m_{N}$ against $h^{N}$
  in the Type-II U(1) symmetric MSISM with right-handed neutrinos.
  The perturbatively allowed parameter space of $(h^N, m_N)$ is given
  by the area between the black $\beta_{\lambda_{1,2}}=1$ lines.  The
  internal blue and grey shaded areas represents the regions allowed
  by the LEP2 Higgs-mass limit, for the electroweak and ultra-light
  $h$-boson scenarios, respectively.}
\label{fig:Type2U1mNvshm2}
\end{figure}

In  Fig.~\ref{fig:Type2U1mNvshm2}  we  display the  allowed  parameter
space  of  $m_{N}$  and   $h^{N}$,  for  all  perturbative  values  of
$\lambda_{1}(\Lambda)$    and   $\lambda_{3}(\Lambda)$,    under   the
constraint: $|\beta_{\lambda_{1,2}}| < 1$.  The allowed space is given
by  the area  enclosed by  the two  black $\beta_{\lambda_{1,2}}  = 1$
lines.  The  blue and grey  shaded areas indicate the  parameter space
which is allowed  by the LEP2 Higgs-boson mass limit.   As can be seen
from~Fig.~\ref{fig:Type2U1mNvshm2},  the resulting  allowed  areas set
upper limits on the heavy  Majorana neutrino masses, $m_{N} < 244$~GeV
and  $m_{N}  < 274$  GeV,  for  the  electroweak and  the  ultra-light
$h$-boson scenarios,  respectively.  Depending on the  strength of the
neutrino  Yukawa  couplings ${\bf  h}^\nu_{ij}$,  such heavy  Majorana
neutrinos can be produced at the LHC~\cite{LHCN}, leading to like-sign
dilepton signatures without missing energy.

As  was discussed  in  Section~6.1, the  U(1)-invariant Type-II  MSISM
predicts no massive stable scalar particle that could play the role of
the  cold  DM.  In  fact,  the  presence  of the  Majorana  neutrinos,
$\nu_{1,2,3}$  and $N_{1,2,3}$, leads  to new  decay channels  for the
scalar particles $h$ and $H$, such as $h \to (\nu_i N_j,\ N_iN_j)$ and
$H \to (\nu_i N_j,\  N_iN_j)$~\cite{AZPC}.  Moreover, the inclusion of
right-handed neutrinos does  not change the UV behaviour  of the model
which  becomes non-perturbative and  develop a  Landau pole  far below
$M_{\rm  GUT}$ and  $M_{\rm Planck}$.  For  this reason,  we turn  our
attention to the Type-II MSISM of maximal SCPV, which does not exhibit
this weakness.


\subsection{Neutrinos in a Minimal Model of Maximal SCPV} 

We  now consider  an  extension  of the  Type-II  MSISM presented  in
Section     \ref{MinmodelmaxspontaneousCPviolation},     by     adding
right-handed neutrinos. The Type-II flat direction of this scenario is
given by  $\sigma = J$,  which leads to  maximal SCPV in  the one-loop
scalar  potential.   Along this  flat  direction,  the heavy  Majorana
neutrino mass matrix~${\bf m}_M$ takes on the form:
\begin{equation}
{\bf m}_{M}\ =\ 
\frac{\sigma}{\sqrt{2}}\; \Big[\, (1 + i)\,{\bf h}^N\: +\: 
(1 - i)\, {\bf \tilde{h}}^{N\dagger} \, \Big]\; .
\end{equation}
Since   the  Majorana   Yukawa  couplings,   ${\bf  h}^N$   and  ${\bf
 \tilde{h}}^N$, may contain large number of independent parameters, we
 will  investigate two  simple variants  of the  model.  In  the first
 variant, we assume that both ${\bf h}^N$ and ${\bf \tilde{h}}^N$ are
 real, i.e.~there is  no sources of explicit CP  violation in neutrino
 Yukawa sector.   The second variant  makes use of a  parity symmetry,
 which  gives rise  to a  massive  stable scalar  particle that  could
 qualify as DM.


\subsubsection{The CP Symmetric Limit}\label{CPsymneutrino}

In the  CP symmetric limit of  the theory, the  Yukawa couplings ${\bf
  h}^N_{ij}$ and  ${\bf \tilde{h}}^N_{ij}$ are all  real.  In~the weak
basis,  where ${\bf  m}_M$ is  real and  diagonal, one  then  gets the
constraint: ${\bf  h}^N = {\bf \tilde{h}}^N$.   Implementing this last
constraint along the Type-II flat direction $\sigma = J$, the neutrino
mass matrices read:
\begin{equation}
	\label{eqn:Type2Unot1CPsymneutrinosmasses}
{\bf m}_{\nu}\ =\ -\ \frac{1}{2}\, 
\sqrt{\frac{ - \lambda_{3}(\Lambda)}{ \lambda_{1}(\Lambda)}}\
v_{\phi}\ {\bf h^{\nu}}\,({\bf h}^N)^{-1}\, {\bf h^{\nu}}^{T}  \; ,\qquad
{\bf m}_{N}\ =\ \sqrt{\frac{\lambda_{1}(\Lambda)}{- \lambda_{3}(\Lambda)}}\
v_{\phi}\, {\bf h}^N\; .
\end{equation}
Assuming a  universal scenario with three  degenerate heavy neutrinos,
with  ${\bf h}^N = h^{N}  {\bf 1}_3$, the  coupling parameter $h^N$
has  to  be less  than  $2.6$  to be  perturbative  at  the RG  scale
$\Lambda$.  This perturbativity constraint becomes stronger at the GUT
and Planck  scales, where we obtain  the upper limits,  $h^N \le 0.52$
and $h^{N} \le 0.47$, respectively.

This model depends on  four independent theoretical parameters, namely
$\lambda_{1}(\Lambda)$,           $\lambda_{2}(\Lambda)$           (or
$\lambda_{6}(\Lambda)$),   $\lambda_{3}(\Lambda)$  and   $h^{N}$.   As
particular viable  benchmark models,  we consider the  following three
cases:
 \begin{eqnarray}
	\label{eqn:Type2Unot1CPsymneutrinoscases}
\mathrm{Case \,  A:}\quad & \lambda_{2}(\Lambda)\ =\ 0.1 \, ,  &\quad
\lambda_{3}(\Lambda)\ =\ -0.01 \, ,\nonumber \\
\mathrm{Case \, B:}\quad & \lambda_{2}(\Lambda)\ =\ 0.1\, ,&\quad
\lambda_{3}(\Lambda)\ =\ -0.005 \, ,\nonumber \\ 
\mathrm{Case \, C:}\quad & \ \, \lambda_{2}(\Lambda)\ =\ 0.05 \, ,&\quad
\lambda_{3}(\Lambda)\ =\ -0.005 \, .
\end{eqnarray} 

\begin{figure}
\centering 
\includegraphics[height=0.36\textwidth,width=0.54\textwidth]{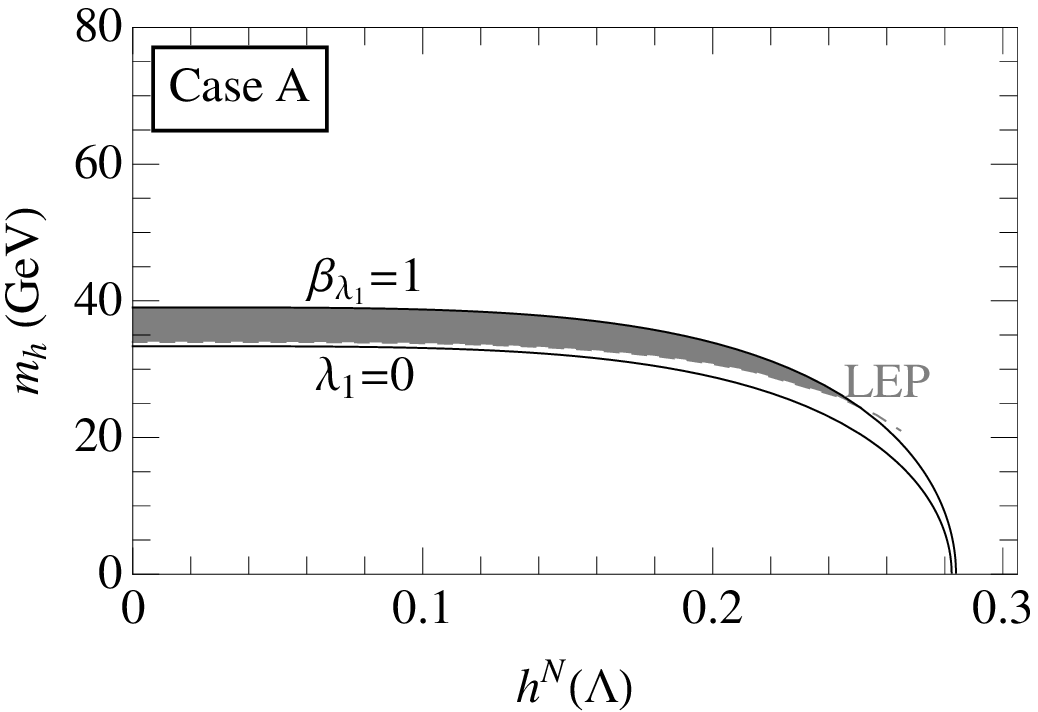}
\qquad
\includegraphics[height=0.36\textwidth,width=0.54\textwidth]{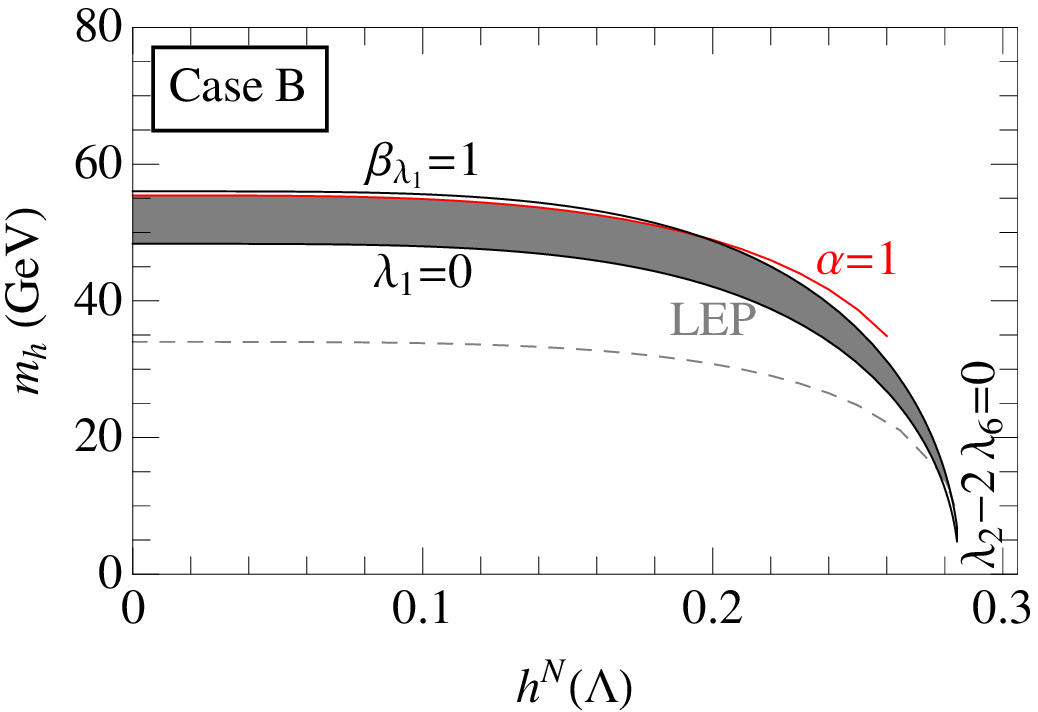}
\qquad
\includegraphics[height=0.36\textwidth,width=0.54\textwidth]{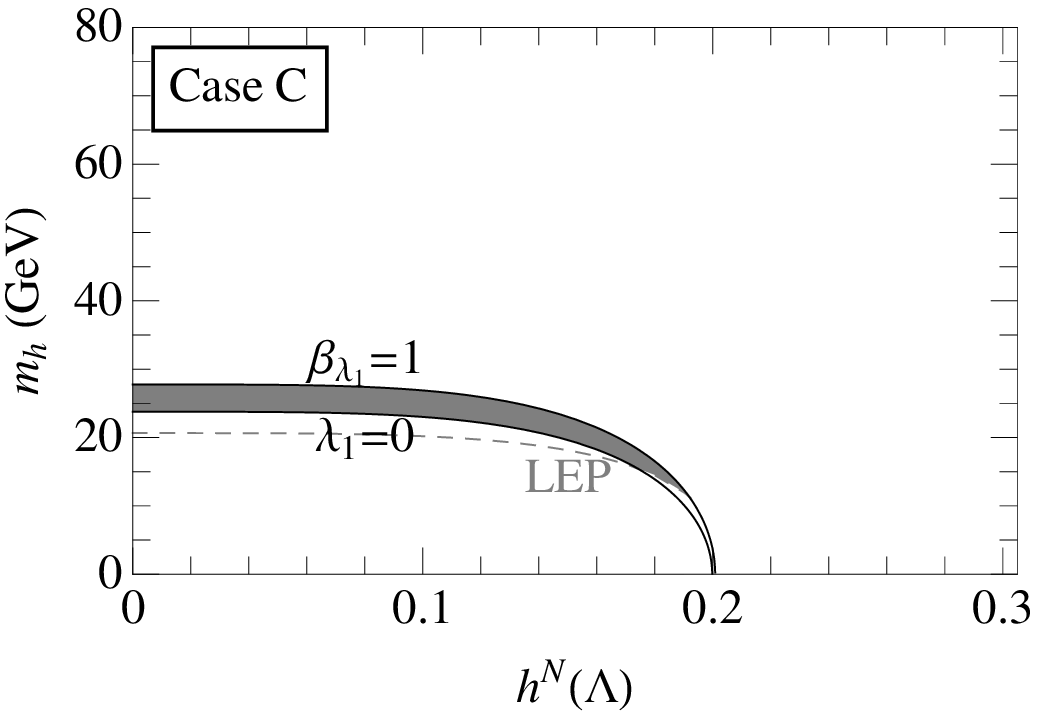} 
\caption{\it Numerical estimates of $m_{h}$ as a function of
  $h^{N}(\Lambda)$ in the minimal Type-II MSISM with maximal SCPV and
  massive Majorana neutrinos for Cases A, B and C defined in
  (\ref{eqn:Type2Unot1CPsymneutrinoscases}).  The area between the
  black lines show the regions which correspond to imposing
  $\beta_{\lambda_{1}}(M_{\mathrm{Planck}}) < 1$,
  $\lambda_{1}(M_{\mathrm{Planck}}) > 0$ and
  $\lambda_{2}(M_{\mathrm{Planck}}) -
  2\lambda_{6}(M_{\mathrm{Planck}}) > 0$ in Case B or $\beta >0$ in
  Cases A and C.  The area above the red $\alpha = 1$ line is
  excluded.  The area below the grey dashed LEP line is excluded by
  LEP2 Higgs-mass limit.  The grey shaded areas correspond to the
  regions allowed by theory and experiment. }
\label{fig:TypeIIUnot1mh}
\end{figure}

In Fig.~\ref{fig:TypeIIUnot1mh} we present the allowed parameter space
in  the  $h^N$-$m_h$  plane,  for  the  Cases A,  B  and  C  given  in
(\ref{eqn:Type2Unot1CPsymneutrinoscases}).  The area between the black
lines is  allowed by the  considerations: $\beta_{\lambda_{1}} (M_{\rm
  Planck})   <    1$,   $\lambda_{1}(M_{\rm   Planck})    >   0$   and
$\lambda_{2}(M_{\mathrm{Planck}})  - 2\lambda_{6}(M_{\mathrm{Planck}})
> 0$ in Case B or $\beta >0$ in Cases A and C, which give the tightest
theoretical constraints  for the model  to remain perturbative  to the
Planck scale.  Furthermore, the area above the red $\alpha =1$ line is
excluded,  because it  violates  perturbative unitarity  in the  MSISM
Higgs sector~\cite{Dicus}.  For Case A  and C, the $\alpha =1$ line is
above the  allowed region and has  not been displayed.   We find that,
within  the  theoretically  allowed  areas, the  predictions  for  the
electroweak oblique parameters  $S$, $T$ and $U$ fall  within the 95\%
CL intervals for the three  scenarios considered. The region below the
grey dashed line is excluded  by the LEP-2 Higgs-mass limit applied to
the $h$-boson  mass $m_{h}$. As  a consequence, the grey  shaded areas
correspond  to  the  regions  which  are allowed  by  our  theoretical
considerations and  the LEP2 and  oblique paameters.  The  presence of
the  right-handed neutrinos  does not  greatly affect  $m_{h}$, except
when $h^{N}$  approaches its maximum  allowed value which  reduces the
prediction for  $m_h$, as shown  in Fig.~\ref{fig:TypeIIUnot1mh}.  The
other scalar masses, $m_{H_{1,2}}$,  are not affected by the inclusion
of neutrinos, since they are independent of $h^{N}$ at the tree level.

\begin{figure}
\centering 
\includegraphics[height=0.36\textwidth,width=0.54\textwidth]{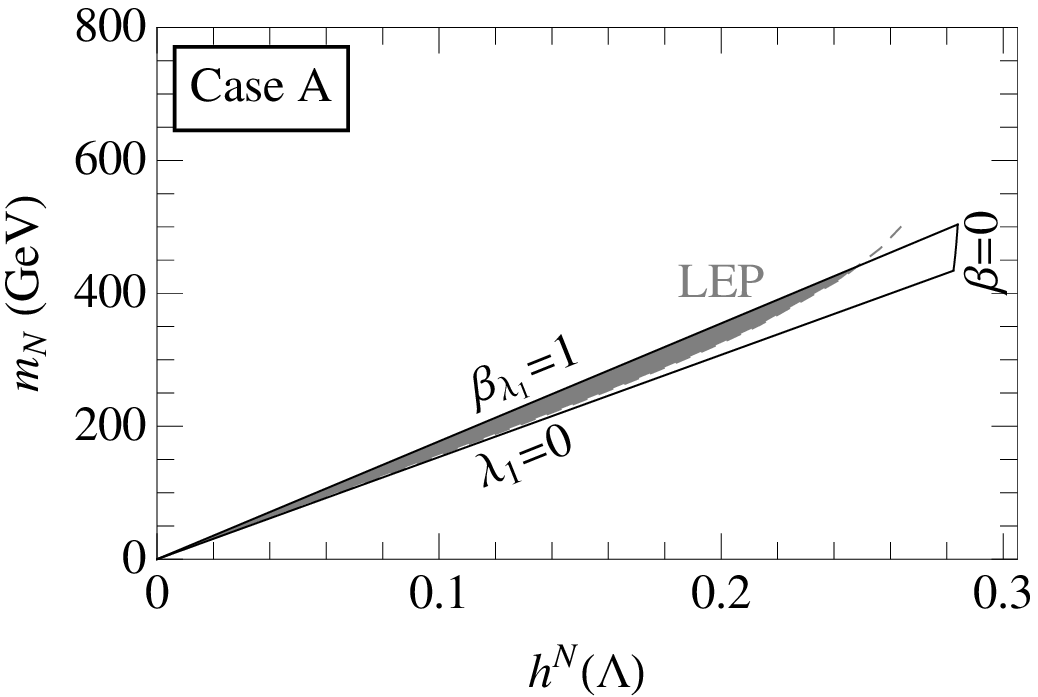}
\qquad
\includegraphics[height=0.36\textwidth,width=0.54\textwidth]{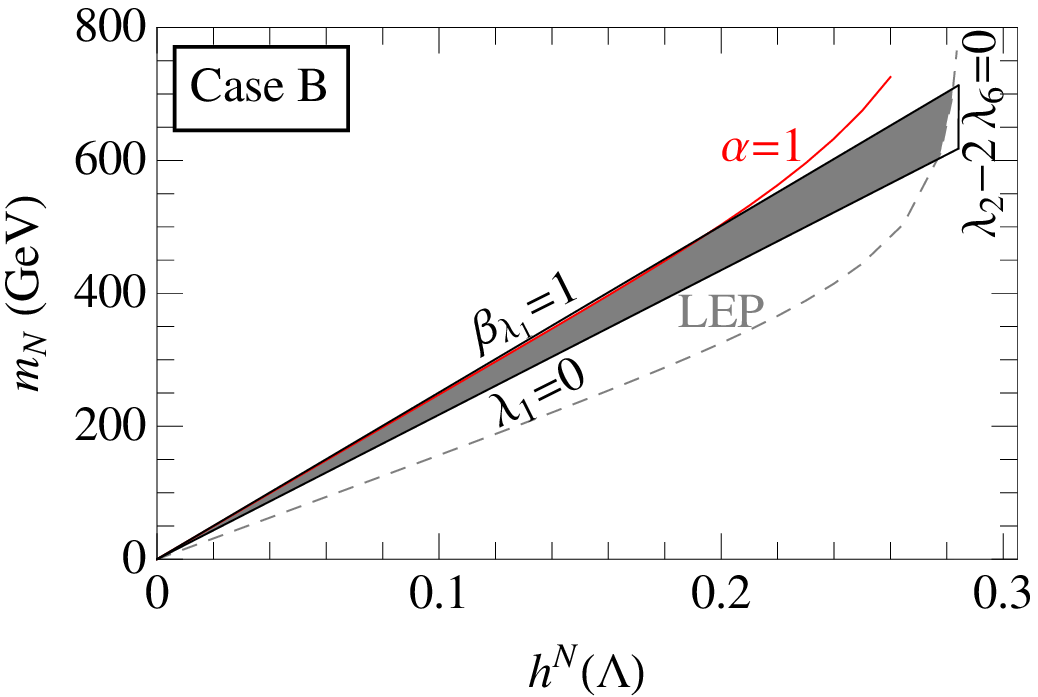}
\qquad
\includegraphics[height=0.36\textwidth,width=0.54\textwidth]{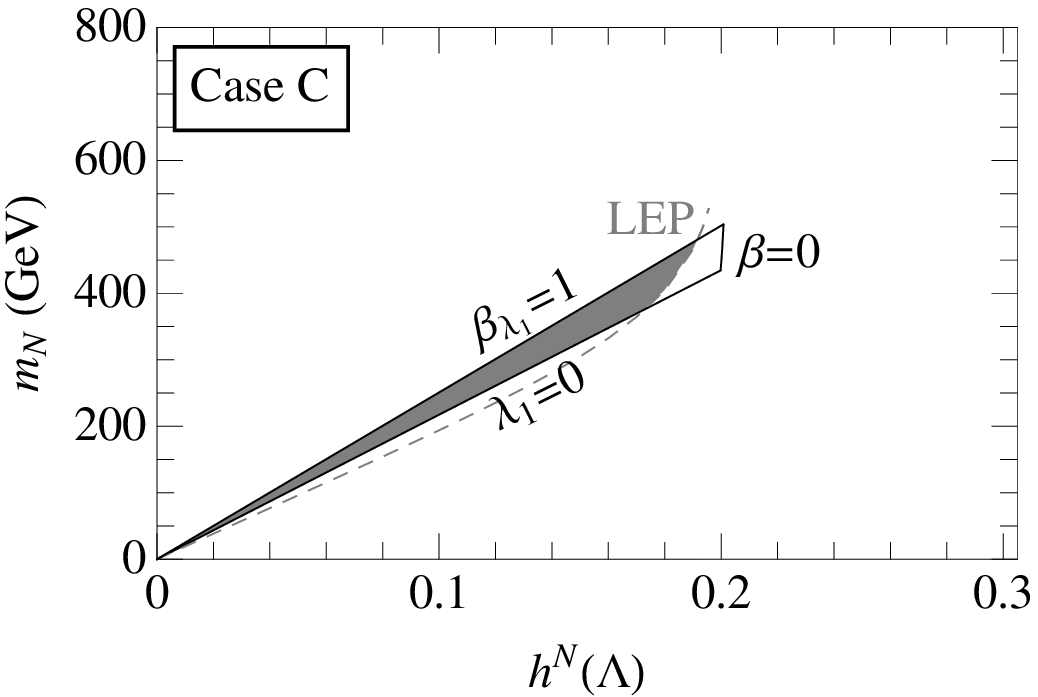} 
\caption{\it Numerical estimates of $m_{N}$ as a function of
  $h^{N}(\Lambda)$ in the minimal Type-II MSISM with maximal SCPV and
  massive Majorana neutrinos for Cases A, B and C defined in
  (\ref{eqn:Type2Unot1CPsymneutrinoscases}).  The area between the
  black lines show the regions corresponding to the constraints:
  $\beta_{\lambda_{1}}(M_{\mathrm{Planck}}) < 1$,
  $\lambda_{1}(M_{\mathrm{Planck}}) > 0$ and
  $\lambda_{2}(M_{\mathrm{Planck}}) -
  2\lambda_{6}(M_{\mathrm{Planck}}) > 0$ in Case B or $\beta >0$ in
  Cases A and C.  The region above the red $\alpha = 1$ line is
  excluded.  The area below the grey dashed LEP line is excluded by
  LEP2 Higgs-mass limit.  The grey shaded areas correspond to the
  regions allowed by both theory and experiment. }
\label{fig:TypeIIUnot1mN}
\end{figure}

Fig.~\ref{fig:TypeIIUnot1mN}  displays  the  allowed  parameter  space
spanned  by  the  Majorana-neutrino  Yukawa  coupling  $h^N$  and  the
universal right-handed  neutrino mass $m_{N}$ for  the three benchmark
scenarios  listed  in  (\ref{eqn:Type2Unot1CPsymneutrinoscases}).   As
before,  the  area  between  the  black  lines  is  permitted  by  the
considerations:   $\beta_{\lambda_{1}}    (M_{\rm   Planck})   <   1$,
$\lambda_{1}(M_{\rm         Planck})          >         0$         and
$\lambda_{2}(M_{\mathrm{Planck}})  - 2\lambda_{6}(M_{\mathrm{Planck}})
> 0$ in Case B or $\beta >0$ in  Cases A and C, and the area above the
red $\alpha  =1$ line  violates perturbative unitarity,  and so  it is
theoretically inadmissible.   The area below the grey  dashed LEP line
is excluded  by LEP2 Higgs-mass  limit applied the $m_{h}$.   The grey
shaded region  is permitted by  theory and the LEP2  limit.  Comparing
the three  cases, we observe that  if $\lambda_{3}(\Lambda)$ decreases
or $\lambda_{2}(\Lambda)$ increases, both  the upper limits on $m_{N}$
and  $h^{N}$ increase.  From  Fig.~\ref{fig:Type2Unot1mhmH2Lambda}, we
see  that if  $\lambda_{2}(\Lambda)$  increases $\lambda_{3}(\Lambda)$
also  needs to  increase to  remain  within the  theoretical and  LEP2
limits and so the two effects  cancel and we assume the maximal values
of $m_{N}$ and $h^{N}$ do not vary significantly from the values given
in  Case  B.  Within  this  benchmark  scenario,  we can  then  derive
approximate upper limits on the  values of $m_{N}$ and $h^{N}$.  Thus,
from the middle panel of Fig.~\ref{fig:TypeIIUnot1mN}, we observe that
the heavy  Majorana neutrinos  can generically have  masses up  to TeV
scale,  i.e.~$m_{N} \stackrel{<}{{}_\sim}  1~{\rm  TeV}$, and  $h^{N}$
must  remain relatively  small  in order  for  the one-loop  effective
potential to be BFB, i.e.~$h^{N} \stackrel{<}{{}_\sim} 0.3$.

The  only  weakness of  the  present model  under  study  is that  the
would-be DM candidate,  the $H_{2}$ boson, is no  longer stable, since
it can  decay to  $\nu_i N_j^*$, where  $N^*_j$ is an  off-shell heavy
Majorana neutrino, which can subsequently decay into off-shell $W^\pm$
and $Z$ bosons and charged  leptons and light neutrinos.  The decay of
the  $H_2$ boson  is  a consequence  of  the violation  of the  parity
symmetry, $\sigma \leftrightarrow  J$, in the Majorana-neutrino Yukawa
sector.  In the following, we  consider a minimal Type-II MSISM, where
the  parity symmetry  is elevated  to an  {\em exact}  global symmetry
acting on the complete Lagrangian of the theory.


\subsubsection{The {\boldmath $H_2$} Boson as a Cold DM Candidate}

As  mentioned above,  in the  absence of  right-handed  neutrinos, the
scalar potential of the Type-II  MSISM with maximal SCPV possesses the
permutation symmetry: $\sigma \leftrightarrow  J$. Under the action of
this symmetry, the scalar field  $H_2 = (J - \sigma)/\sqrt{2}$ is odd:
$H_2  \to -  H_2$. This  parity  symmetry remains  unbroken after  the
EWSSB, leading to  a massive stable scalar particle,  which could play
the role of the cold DM in the Universe.

We may now extend the above permutation or parity symmetry to neutrino
Yukawa sector of  the model, which implies that ${\bf h}^N  = - i {\bf
\tilde{h}^{N \dagger}}$.  As a consequence,  the $H_2$ boson  will
not interact  with the neutrinos, so  it will remain  a massive stable
particle which can potentially act as DM particle.  Given the relation
${\bf  h}^N  =  -  i  {\bf \tilde{h}}^{N  \dagger}$,  the  light-  and
heavy-neutrino mass matrices become
\begin{equation}
{\bf m_{\nu} }\ =\ - \frac{1}{4}\,
\sqrt{\frac{-\lambda_{3}(\Lambda)}{ \lambda_{1}(\Lambda)}}\
v_{\phi}\; {\bf h^{\nu} } ({\rm Re}\,{\bf h}^{N})^{-1}\, {\bf h^{\nu} }^T\;,\qquad
{\bf m}_{N}\ =\
  2\;\sqrt{\frac{ \lambda_{1}(\Lambda)}{-\lambda_{3}(\Lambda)}}\
v_{\phi}\; {\rm Re}\,{\bf h}^N\;,
\end{equation}
where ${\rm Re}\,{\bf h}^N = - {\rm Im}\,{\bf h}^N$ in the weak basis,
in which ${\bf m}_{M}$ is real.  Assuming a universal Majorana flavour
structure with ${\bf  h}^N = h^{N} {\bf 1}_3$, we  find that ${\rm Re}\,
h^{N}$ must  be less than  2.1 in order  to be perturbative at  the RG
scale $\Lambda$ and less than  0.37 and 0.33 to remain perturbative at
the GUT and Planck scales, respectively.

\begin{figure}
\centering 
\includegraphics[height=0.36\textwidth,width=0.54\textwidth]{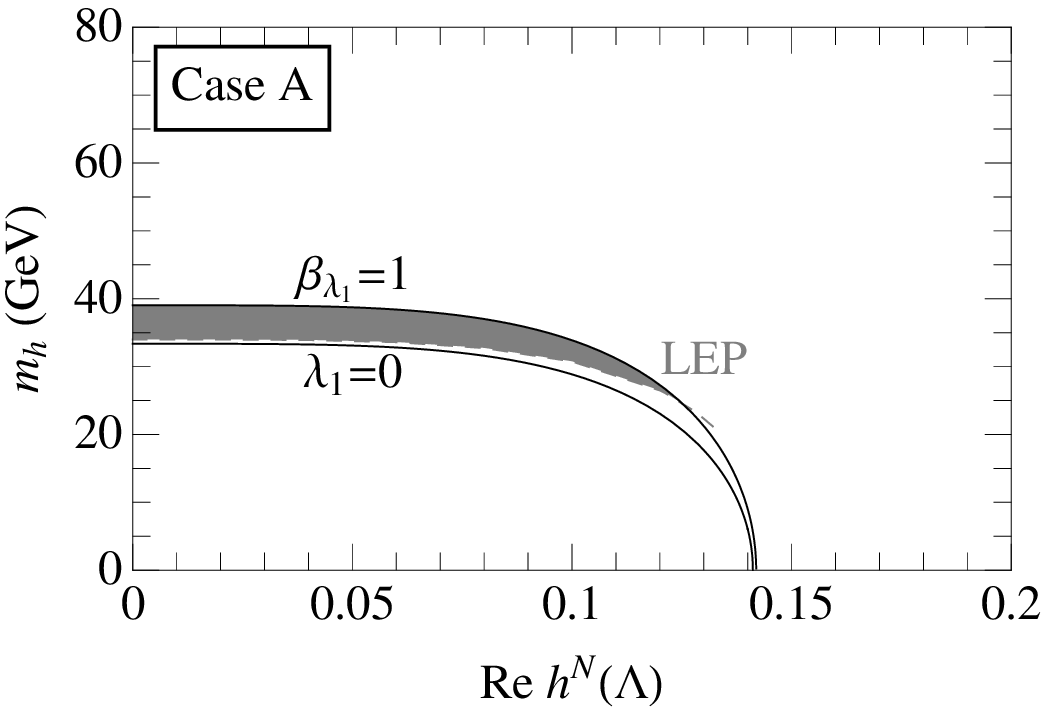}
\qquad
\includegraphics[height=0.36\textwidth,width=0.54\textwidth]{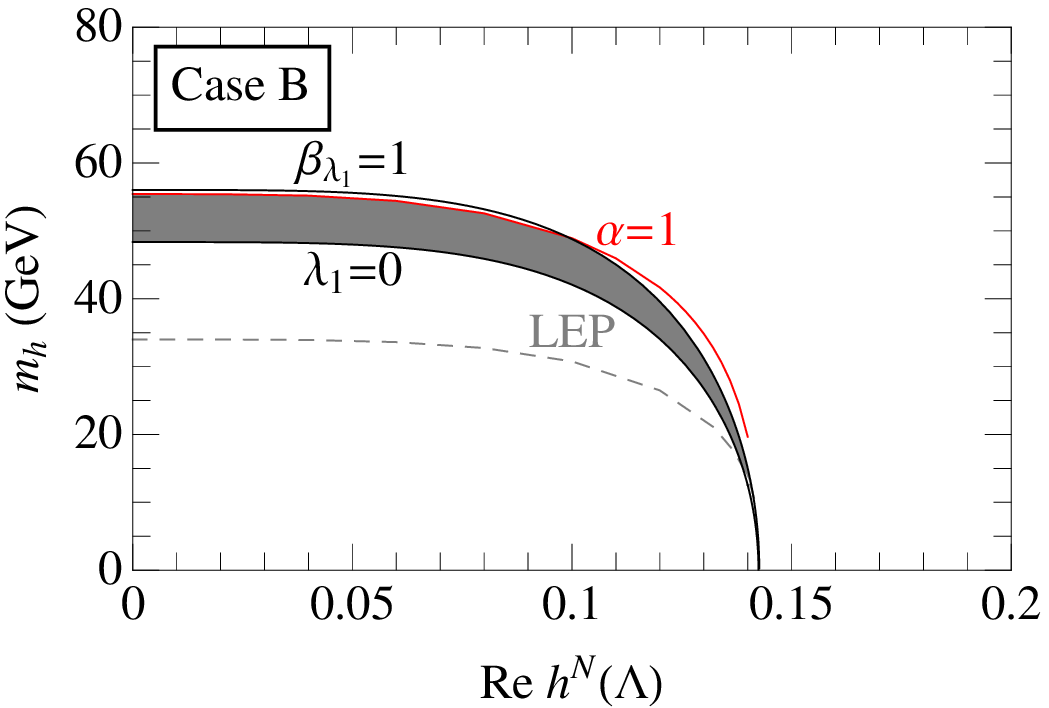}
\qquad
\includegraphics[height=0.36\textwidth,width=0.54\textwidth]{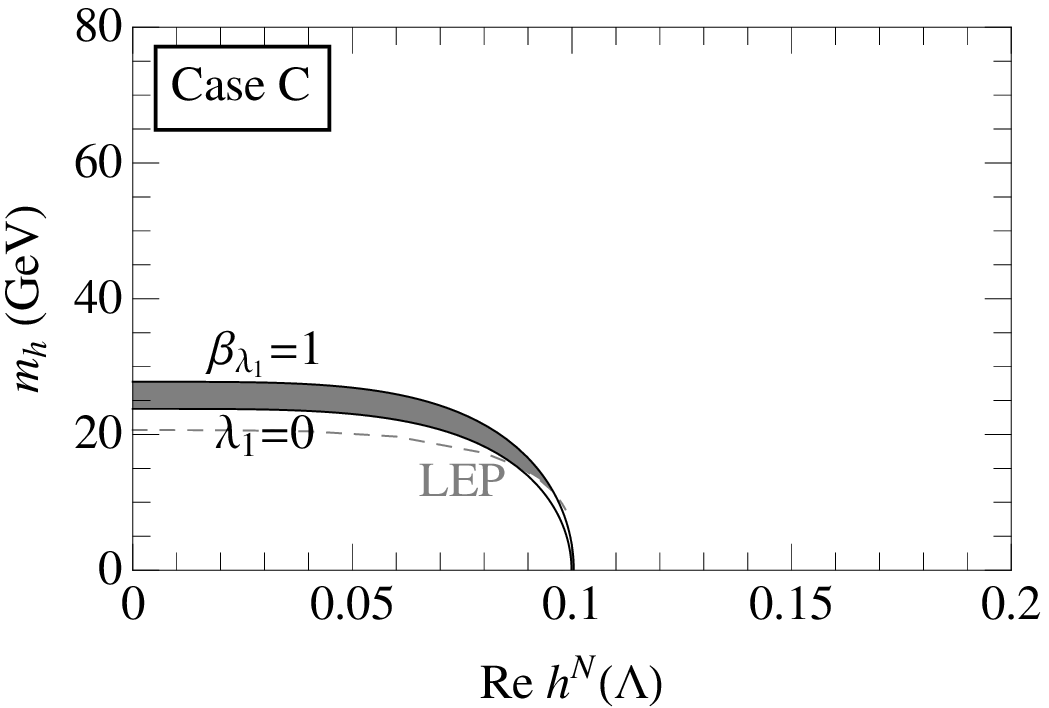} 
\caption{\it Numerical estimates of $m_{h}$ as a function of ${\rm
    Re}\,h^{N}(\Lambda)$ in the minimal Type-II MSISM with maximal
  SCPV, massive Majorana neutrinos and a scalar DM, for Cases A, B and
  C defined in (\ref{eqn:Type2Unot1CPsymneutrinoscases}).  The area
  between the black lines correspond to regions allowed by
  $\beta_{\lambda_{1}}(M_{\mathrm{Planck}}) < 1$,
  $\lambda_{1}(M_{\mathrm{Planck}}) > 0$ and the potential BFB ($\beta
  > 0$). The region above the red $\alpha = 1$ line is excluded.  The
  area below the grey dashed LEP line is excluded by LEP2 Higgs-mass
  limit.  The grey shaded areas correspond to the regions allowed by
  both theory and the LEP2 limit. }
\label{fig:TypeIIUnot1H2mh}
\end{figure}

\begin{figure}
\centering 
\includegraphics[height=0.36\textwidth,width=0.54\textwidth]{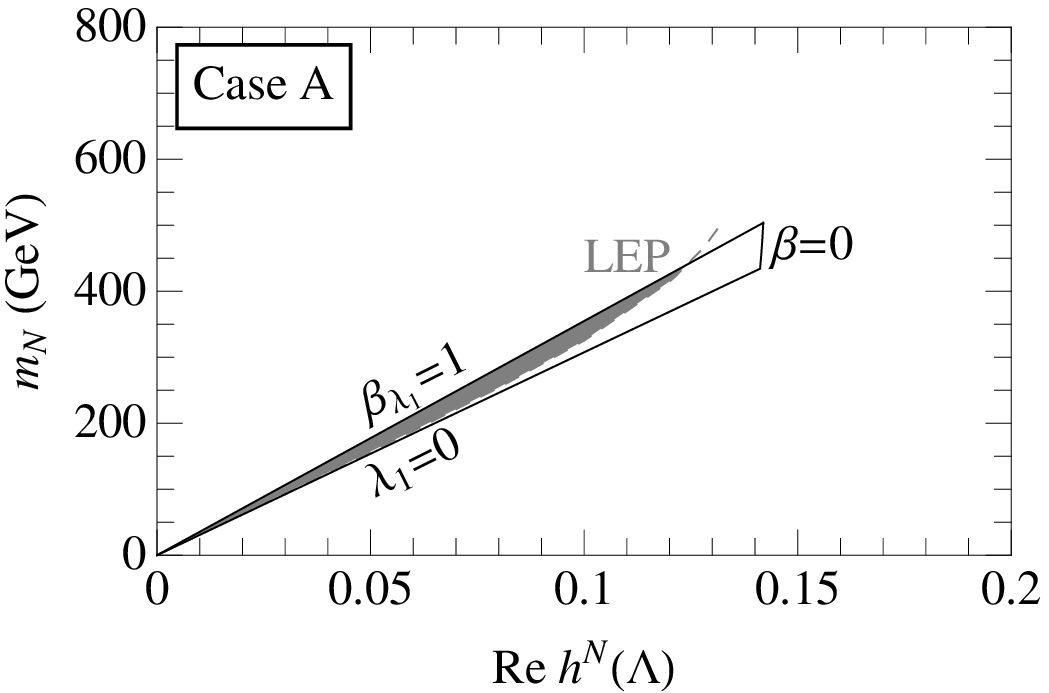}
\qquad
\includegraphics[height=0.36\textwidth,width=0.54\textwidth]{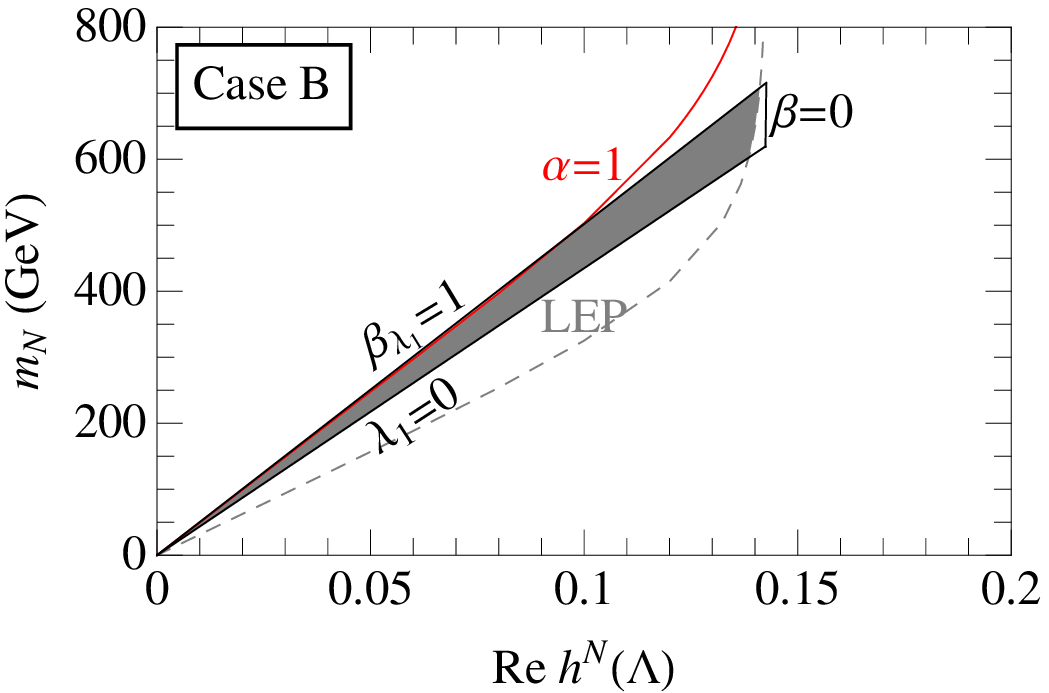}
\qquad
\includegraphics[height=0.36\textwidth,width=0.54\textwidth]{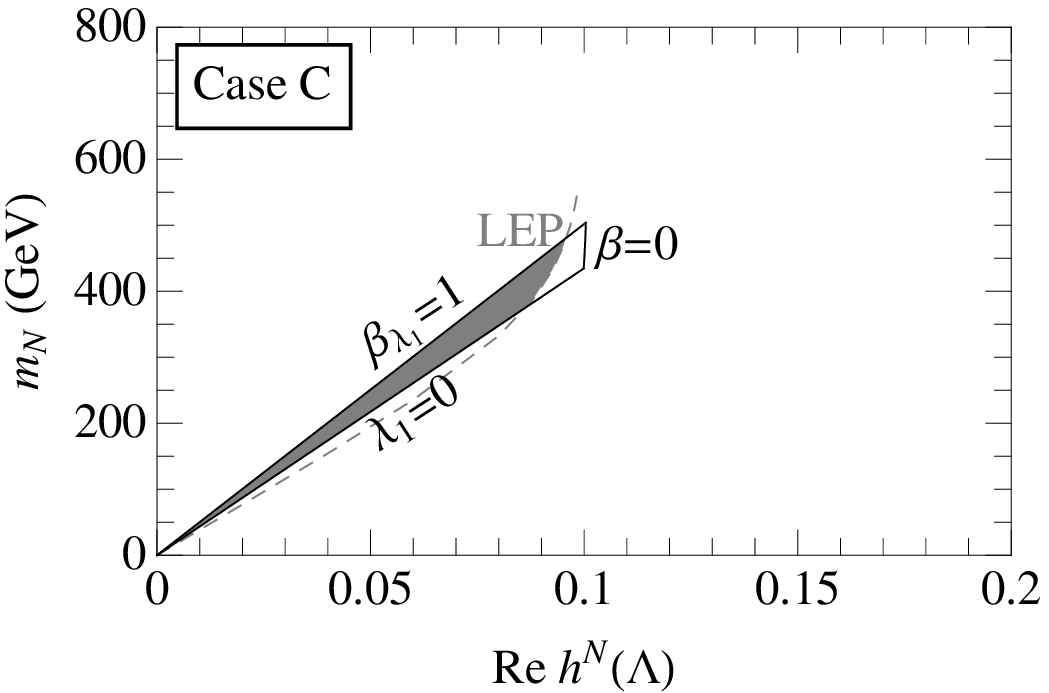} 
\caption{\it Numerical estimates of $m_{N}$ as a function of ${\rm
    Re}\,h^{N}(\Lambda)$ in the minimal Type-II MSISM with maximal
  SCPV, massive Majorana neutrinos and a scalar DM, for Cases A, B and
  C defined in (\ref{eqn:Type2Unot1CPsymneutrinoscases}).  The area
  between the black lines show the regions which satisfy:
  $\beta_{\lambda_{1}}(M_{\mathrm{Planck}}) < 1$,
  $\lambda_{1}(M_{\mathrm{Planck}}) > 0$ and $\beta > 0$.  The red
  $\alpha = 1$ line excludes the region above this line.  The area
  below the grey dashed LEP line is excluded by LEP2 Higgs-mass limit.
  The grey shaded areas correspond to the regions allowed by both
  theory and the LEP2 limit.  }
\label{fig:TypeIIUnot1H2mN}
\end{figure}

Fig.  \ref{fig:TypeIIUnot1H2mh}  shows the allowed  parameter space of
the $h$-boson masses and the real part of the Majorana Yukawa coupling
${\rm Re}\,h^{N}  (\Lambda)$, for the three  Cases A, B  and C defined
in~(\ref{eqn:Type2Unot1CPsymneutrinoscases}).   The  area enclosed  by
the black lines is theoretically  favoured by the perturbative and BFB
conditions:      $\beta_{\lambda_{1,2}}(M_{\rm      Planck})<      1$,
$\lambda_{1}(M_{\rm  Planck}) >  0$ and  $\beta >  0$ which  offer the
tightest  theoretical constraints.   Instead, the  area above  the red
$\alpha  =1$ line  is  disfavoured, because  it violates  perturbative
unitarity  in the  Higgs  sector.   Above the  grey  dashed LEP  lines
correspond  to  the regions  which  are  also  permitted by  the  LEP2
Higgs-mass limit applied to $m_{h}$, whereas constraints from the $S$,
$T$  and $U$  parameters play  no  role in  the theoretically  allowed
parameter  space.   The  grey  shaded regions  are  theoretically  and
experimentally   permitted.   From~Fig.~\ref{fig:TypeIIUnot1H2mh},  we
observe that the  $h$-boson mass has a similar range  of values as the
CP-symmetric MSISM discussed in the previous subsection.

In  Fig.~\ref{fig:TypeIIUnot1H2mN} we  display  the allowed  parameter
space of the universal right-handed neutrino Majorana mass $m_{N}$ and
${\rm Re}\,h^{N}(\Lambda)$, for the three  different Cases A, B and C.
As   before,  we  consider   the  following   theoretical  conditions:
$\beta_{\lambda_{1}}(M_{\mathrm{Planck}})             <            1$,
$\lambda_{1}(M_{\mathrm{Planck}}) >  0$, $\beta  > 0$ and  $\alpha \le
1$.  The theoretically favoured  regions are those, which are enclosed
by  the black  $\beta_{\lambda_{1}}<  1$, $\lambda_{1}  >  0$ and  BFB
($\beta > 0$) lines.  The  grey shaded areas correspond to the regions
which  are also  permitted by  the  LEP2 Higgs-mass  limit applied  to
$m_{h}$. In  all the three  benchmark scenarios considered,  the heavy
Majorana neutrino mass scale $m_N$  stays below the TeV scale and
the value  of ${\rm Re}\,h^{N}(\Lambda)$  is constrained to  be: ${\rm
  Re}\,h^{N} \stackrel{<}{{}_\sim} 0.15$.

In summary,  the variant  of the Type-II  MSISM with maximal  SCPV and
right-handed neutrinos we discussed in this subsection has a number of
physically  interesting properties.   First, it  can realize  a parity
symmetry in  the theory,  such that the  $H_2$ boson becomes  a stable
particle and  so could play the role  of the cold DM  in the Universe.
Second,  the  present  model   can  implement  an  electroweak  seesaw
mechanism to  provide naturally small neutrino masses.   It contains a
new source of spontaneous CP violation, thereby enabling us to address
the  problem of  the  baryon  asymmetry in  the  Universe.  The  model
successfully  passes all  obvious experimental  constraints  from LEP2
Higgs and  other electroweak  precision data.  Finally,  of particular
interest is the  existence of a significant region  of the theoretical
parameter space,  within which the  model can stay perturbative  up to
Planck-mass energy scales.

\setcounter{equation}{0}
\section{Conclusions}\label{conclusions}

We  have  performed a  systematic  analysis  of  an extension  of  the
Standard Model that includes a complex singlet scalar field $S$ and is
scale invariant  at the tree level.   We have called such  a model the
Minimal  Scale  Invariant extension  of  the  Standard Model  (MSISM).
Quantum  corrections  explicitly break  the  scale  invariance of  the
classical Lagrangian of the model  and may trigger EWSSB.  Even though
the  scale invariant SM  is not  a realistic  scenario, the  MSISM may
result in a perturbative and phenomenologically viable theory that may
potentially solve the gauge hierarchy problem.

We  have presented a  complete classification  of the  flat directions
which  may occur  in the  classical  scalar potential  of the  MSISM.
Employing the  perturbative GW approach  to EWSSB, we  have calculated
the one-loop  effective potential along the  different flat directions
and  derived the necessary  and sufficient  conditions for  the scalar
potential to be  BFB [cf.~(\ref{eqn:BFBconditions})].  In addition, we
have   computed  the   scalar-boson   masses,  including   theoretical
constraints  from the  validity  of perturbation  theory,  as well  as
phenomenological  limits from  electroweak precision  data  and direct
Higgs-boson searches at LEP2.

The different flat directions in  the MSISM can be classified in three
major categories: Type~I, Type~II  and Type~III.  In the Type-I MSISM,
the singlet  scalar $S$ has a zero  VEV at the tree  level, whereas in
the Type-II MSISM both the VEVs of $S$ and the SM Higgs doublet $\Phi$
are  non-zero.  In  Type-III MSISM,  the  Higgs doublet  $\Phi$ has  a
vanishing VEV at the tree-level,  which makes it somewhat difficult to
naturally realize EWSSB.  Therefore,  our analysis has focused only on
scenarios realizing Type-I and Type-II flat directions.  We have found
that the general Type-I MSISM is  perturbative only up to the EW scale
and  exhibits  a  Landau   pole  at  energy  scales  $\sim  10^4$~GeV.
Likewise,  we have  found  that the  U(1)-invariant  Type-II MSISM  is
perturbative up to energies $\sim 10^4$~GeV and develops a Landau pole
at energy scales $\sim 10^5$~GeV.  In this respect, our results are in
qualitative  agreement with~\cite{Foot}.  As  we have  shown, however,
this  is not  an indispensable  property of  a general  Type-II MSISM.
Moving away from the  model-building constraint of U(1) invariance, we
have explicitly  demonstrated that a minimal Type-II  MSISM of maximal
SCPV can stay perturbative up to the Planck scale, without the need to
introduce unnaturally  large hierarchies between  the scalar-potential
quartic couplings,  or between the VEVs  of the $\Phi$  and $S$ fields
which may  reintroduce an additional  hierarchy problem. 

In the  present study, we have  taken the view that  the generation of
the electroweak  scale $M_{\rm EW}$ is  the result of  the breaking of
the scale  invariance of the Higgs  sector of the  MSISM.  Instead, we
have tacitly assumed that quantum gravity effects are small and do not
destabilize     the     gauge     hierarchy.     As     was     argued
in~\cite{MNmar2007,Bardeen},  for  example,  the  latter  may  be  the
consequence of  a conformally UV  complete theory of  quantum gravity,
which we  are currently lacking.  However, a  necessary ingredient for
such a theory  to succeed appears to be the  absence of any additional
scale  between $M_{\rm  EW}$ and  $M_{\rm Planck}$.   It  is therefore
important  that the quartic  couplings remain  perturbative up  to the
Planck  scale, without  the  presence  of a  Landau  pole which  could
introduce an  additional unwanted higher scale in  the theory, through
non-perturbative  effects  that  could  dynamically  break  the  scale
invariance and so destabilize the gauge hierarchy.

We have  investigated the phenomenological implications  of the Type-I
and  Type-II MSISM, in  particular, whether  they realize  explicit or
spontaneous  CP  violation, neutrino  masses  or  predict dark  matter
candidates.   The key features  of the  different scenarios  have been
summarized in Table~\ref{tab:taxonomy}.   To naturally account for the
very small light-neutrino masses through the seesaw mechanism, we have
extended the Type-II MSISM  with right-handed neutrinos.  Our analysis
shows that the  right-handed neutrino mass scale $m_N$  cannot be much
higher than the  TeV scale and so heavy  Majorana neutrinos might lead
to  observable like-sign  dilepton effects  at the  LHC. On  the other
hand, the  addition of right-handed neutrinos  generically renders all
scalar  fields  unstable  and  so  prevents them  from  acting  as  DM
particles. However, we have shown that this problem could be solved by
promoting a  parity symmetry  present in the  scalar potential  of the
model to  the neutrino Yukawa  sector and to the  complete Lagrangian.
One of the scenarios satisfying this criterion is the Type-II MSISM of
maximal SCPV.

There are  several issues  which are beyond  the scope of  the present
paper, but  need to  be studied in  greater detail.   Specifically, it
would  be interesting  to  determine the  precise  constraints on  the
parameter  space  derived  from  the predicted  DM  relic  abundances.
Similarly, additional  constraints may be  derived from considerations
of  the  baryon asymmetry  in  the  Universe.   Finally, it  would  be
interesting  to  investigate, whether  the  presence  of  some of  the
quasi-flat  directions  in  the   MSISM  could  also  serve  to  drive
cosmo\-logical inflation.   These are some  of the issues  that remain
open within the MSISM, which we aim to address in the near future.


\subsection*{Acknowledgements}
L. A-N. thanks the participants of the ``Workshop on Multi-Higgs
Models'' in Lisbon, Portugal ($16^{\mathrm{th}} - 18^{\mathrm{th}}$
September 2009) for useful discussions. 

\newpage
\appendix

\setcounter{equation}{0}
\section{The Yukawa and Gauge Sectors of the MSISM}\label{App:SM}

Here we  briefly discuss the  Yukawa and electroweak gauge  sectors of
the MSISM, which closely resemble the SM.  This brief exposition will
enable us  to set  up the notation  and determine  the gauge-dependent
masses  and couplings that  enter our  calculations for  the effective
potential,  the  anomalous  dimensions  and  the  electroweak  oblique
parameters.

The gauge-invariant  part of the Lagrangian describing  the Yukawa and
electroweak gauge sectors is given by
\begin{eqnarray}
	\label{eqn:SMlagrangian}
\mathcal{L}_{\mathrm{inv}} & = &  -\:
\frac{1}{4} G^{a}_{\mu \nu} G^{a, \mu \nu}\: -\:
\frac{1}{4} F^{i}_{\mu \nu} F^{i, \mu \nu}\: - \frac{1}{4} B_{\mu
  \nu} B^{\mu \nu}\: \nonumber\\
& &  +\: \bar{\psi}i \gamma^{\mu}D_{\mu} \psi\: +\:
(D^{\mu} \Phi)^{\dagger} (D_{\mu} \Phi)\: +\: 
(\partial_\mu S^*)(\partial^\mu S) \nonumber\\
&&  -\ \Big( \,   
{\bf h}^{u}_{ij}\bar{Q}_{iL} \tilde{\Phi} u_{jR}\: +\: 
{\bf h}^{d}_{ij}\bar{Q}_{iL} \Phi d_{jR}\: +\:  
{\bf h}^{e}_{ij}\bar{L}_{iL} \Phi e_{jR}\ +\ {\rm H.c.}\Big) \; ,  
\end{eqnarray} 
where     $G^{a}_{\mu    \nu}     =     \partial_{\mu}G^{a}_{\nu}    -
\partial_{\nu}G^{a}_{\mu}  +  g_{s}  f^{abc}  G^{b}_{\mu}G^{c}_{\nu}$,
$F^{i}_{\mu \nu}=\partial_{\mu}A^{i}_{\nu}-\partial_{\nu}A^{i}_{\mu} +
g      \varepsilon^{ijk}     A^{j}_{\mu}A^{k}_{\nu}$      and     $B_{\mu
  \nu}=\partial_{\mu}B_{\nu}-\partial_{\nu}B_{\mu}$   are   the  field
strength  tensors  of  the  SU(3)$_c$, SU(2)$_L$  and  U(1)$_Y$  gauge
fields,   $G^{a}_{\mu}$  (with  $a=1,\dots,8$),   $A^{i}_{\mu}$  (with
$i=1,2,3$) and $B_{\mu}$, respectively.  Correspondingly, $g_{s}$, $g$
and $g'$ are the SU(3)$_c$, SU(2)$_L$ and U(1)$_Y$ gauge couplings and
$D_{\mu}$  is   the  covariant   derivative  defined  as   $D_{\mu}  =
\partial_{\mu}  -   ig_{s}  \frac{\lambda^{a}}{2}  G^{a}_{\mu}   -  ig
\frac{\tau^{i}}{2}  A^{i}_{\mu}  -  i\frac{Y}{2}  g'  B_{\mu}$,  where
$\lambda^{a}$ ($\tau^i$) are the  usual Gell-Mann (Pauli) matrices and
$Y$ is the $U(1)_{Y}$ weak hypercharge of the various fields,
\begin{eqnarray}
&& Y(\Phi)= 1 \, , \qquad Y(S) = 0 \, , \qquad Y(L_{L})=-1\, , \qquad
  Y(e_{R})=-2 \, , 
  \nonumber\\ 
&& Y(Q_{L})=\frac{1}{3} \ , \qquad Y(u_{R})= \frac{4}{3}\ , \qquad Y(d_{R})
  = - \frac{2}{3}\ . 
\end{eqnarray}
In~(\ref{eqn:SMlagrangian}),  we  have  used  $\psi$  to  collectively
represent all the fermions of the model,
\begin{equation}
Q_{iL}= \left( \begin{array}{c}
u_{i}\\
d_{i}
\end{array}\right)_{L}, \qquad u_{iR} \, , \qquad d_{iR} \, , \qquad 
L_{iL}= \left( \begin{array}{c}
\nu^{0}_{i}\\
e_{i}
\end{array}\right)_{L}, \qquad e_{iR}\ ,
\end{equation}
where the  subscripts $L$  and $R$ denote  the left-  and right-handed
chiralities of the fermion. Each type of fermion has three generations
represented by $i=1,2,3$, i.e. $e_{i} = (e, \mu, \tau)$.  The matrices
${\bf h}^{u,d,e}_{ij}$ contain the Yukawa couplings for the SM up- and
down-type  quarks  and  charged   leptons.   Finally,  we  denote  the
hypercharge   conjugate  field   of  the   Higgs  doublet   $\Phi$  as
$\tilde{\Phi}=i \tau^{2} \Phi^{*}$.

A convenient gauge-fixing scheme to remove the tree-level mixing terms
between  the Goldstone  and gauge  bosons  is the  $R_{\xi}$ class  of
gauges.   Adopting this  scheme and  decomposing linearly  the neutral
component of  $\Phi$ about  its one-loop induced  VEV, as  $v_{\phi} +
\phi$, we  may write the  gauge-fixing and the  induced Faddeev--Popov
Lagrangians as~follows:
\begin{eqnarray} 
	\label{eqn:gfandfp}
\mathcal{L}_{\rm GF} & = & -\frac{1}{2 \xi} \Big[ ( \partial_{\mu} G^{a
    \mu})^{2} +  ( \partial_{\mu} A^{i
    \mu})^{2} + (\partial_{\mu} B^{\mu})^{2} \Big]  -
\frac{i}{2\sqrt{2}} g v_{\phi}  (G^{-} - G^{+})
\partial^{\mu}A^{1}_{\mu}  \nonumber\\ 
& & -  \frac{1}{2\sqrt{2}}  g v_\phi  (G^{-} + G^{+})
\partial^{\mu}A^{2}_{\mu} + \frac{1}{2} g v_\phi G
\partial^{\mu}A^{3}_{\mu} - \frac{1}{2} g' v_\phi  G
\partial^{\mu}B_{\mu}  \nonumber\\ 
& & - \tilde{m}_{G^{\pm}}^{2} G^{+} G^{-} - \frac{1}{2}
\tilde{m}_{G}^{2} G^{2} \; , \nonumber\\ 
\mathcal{L}_{\rm FP} & = & - \bar{\eta}^{a} \partial_{\mu}(
\partial^{\mu}\delta^{ac} - g_{s} f^{abc} G^{b \mu}) \eta^{c} +
\omega_{i}^{\dagger} m^{f}_{ij} \omega_{j} + 
\omega_{i}^{\dagger} m^{f}_{i} \chi  + \chi^{\dagger} m^{f}_{i} 
\omega_{i} + \chi^{\dagger} m^{f} \chi  \; ,
\end{eqnarray}
where  $\eta^{a}$ ($a=1,...,8$),  $\omega_{i}$  ($i =  1,  2, 3$)  and
$\chi$  are  the  SU(3)$_c$,  SU(2)$_L$  and  U(1)$_Y$  ghost  fields,
respectively, and
\begin{eqnarray}
m^{f}_{ii} & = & - \partial^{\mu}\partial_{\mu} - \frac{1}{4}g^{2} \xi
v_\phi \phi - \frac{1}{4} g^{2} \xi v_\phi^{2} \; , \quad 
m^{f}_{12} = - m^{f}_{21} = g \partial_{\mu}
A^{3,\mu} - \frac{1}{4}g^{2} \xi v_\phi  G \; ,
\nonumber\\ 
m^{f}_{13} & = & - m^{f}_{31} = - g \partial_{\mu} A^{2,\mu} -
\frac{1}{4\sqrt{2}}g^{2} \xi v_\phi  (G^{-} + G^{+}) \; ,
\nonumber\\ 
m^{f}_{23} & = & -m^{f}_{32} = g \partial_{\mu} A^{1,\mu} +
\frac{i}{4\sqrt{2}}g^{2} \xi v_\phi  (G^{-} - G^{+}) \; ,
\nonumber\\ 
m^{f}_{1} & = & - \frac{1}{4\sqrt{2}} g g' \xi v_\phi
(G^{-} + G^{+}) \; , \quad m^{f}_{2} = \frac{i}{4\sqrt{2}} g g' \xi
v_\phi (G^{-} - G^{+}) \; , \nonumber\\ 
m^{f}_{3} & = & \frac{1}{4} g g' \xi v_\phi \phi +
\frac{1}{4}g g' \xi v_\phi^{2} \; , \quad m^{f} = -
\partial^{\mu}\partial_{\mu} -  \frac{1}{4}g^{'2} \xi v_\phi \phi 
- \frac{1}{4} g^{'2} \xi v_\phi^{2} \; .   
\end{eqnarray}
The   would-be   Goldstone    bosons   obtain   gauge-dependent   mass
contributions  due to  the gauge  fixing term  $\mathcal{L}_{\rm GF}$,
given by
\begin{equation} 
	\label{eqn:gmasses}
m_{G^{\pm}}^{2} = \frac{1}{4} g^{2} \xi v^2_\phi\; , \qquad  \quad 
m_{G}^{2} = \frac{1}{4} (g^{2} + g^{' 2}) \xi v_\phi^{2} \; .
\end{equation}
Similarly, the ghosts also  gain gauge-dependent mass eigenvalues from
$\mathcal{L}_{\rm GF}$, i.e.
 \begin{equation}
 	\label{eqn:ghostmasses}
m_{\omega_{\pm}}^{2}  = \frac{1}{4}  g^{2}  \xi v^2_\phi  \; ,  \qquad
m_{\omega_{Z}}^{2} = \frac{1}{4} (g^{2} + g^{' 2}) \xi v_\phi^{2} \; ,
\qquad m_{\omega_{A}}^{2} = 0 \; , \qquad m_{\eta^{a}}^{2} = 0 \; ,
\end{equation}	
where $\omega_{\pm} = \frac{1}{\sqrt{2}} (\omega_{1} \mp i \omega_{2})
$, $\omega_{Z} = \frac{1}{\sqrt{g^{2} + g^{'2}}}( g \omega_{3} - g'
\chi)$  
and  $\omega_{A} = \frac{1}{\sqrt{g^{2} + g^{'2}}}(g' \omega_{3} + g \chi)$.

We  should note that  after EWSSB,  all $v_\phi$-dependent  masses and
couplings     affect      the     one-loop     effective     potential
$V^{\mathrm{1-loop}}_{\mathrm{eff}}$  along  the  flat direction,  but
they do  not influence the  one-loop anomalous dimensions  and $\beta$
functions, which may be computed in the symmetric phase of the theory.
In the same context, we  also note that the $v_{\phi}$-dependent terms
contribute to  the electroweak oblique  parameters, $S$, $T$  and $U$,
which are conventionally calculated in the Feynman-'t Hooft gauge $\xi
= 1$.


\setcounter{equation}{0}
\section{The One-Loop Effective Potential of the MSISM}\label{App:EffPot}

Here we calculate the one-loop  effective potential of the MSISM.  To
this end, we use the functional expression \cite{Jackiw, Zinn-Justin}:
\begin{equation} 
	\label{V1loop}
V^{\mathrm{1-loop}}_{\mathrm{eff}} =  -C_{s} \frac{i \hbar}{2}  \Big(
\mathrm{Tr} \ln H_{\varphi_{1} \varphi_{2}}(\varphi_{c}) -
\mathrm{Tr} \ln H_{\varphi_{1} \varphi_{2}}(0) \Big) \; ,   
\end{equation}
where $H_{\varphi_{1} \varphi_{2}}$  is the second derivative of the
classical action $S = \int d^4 x {\cal L}$, i.e. 
\begin{equation}
	\label{eqn:H}
H_{\varphi_{1} \varphi_{2}}(\varphi_{c})  =   \frac{\delta^{2}
  S}{\delta  \varphi_{1}   (x_1)\delta  \varphi_{2} (x_{2})}
\bigg|_{\varphi = \varphi_{c}} \; . 
\end{equation}
In the above,  $\varphi$ collectively denotes each of  the fields,
\begin{displaymath}
\{ \Phi, S, A^{i}_{\mu}, B_{\mu}, \omega_{\pm}, \omega_{Z},
\omega_{A}, \eta^{a}, u_{i}, d_{i}, e_{i}, \nu_{i}, N_{i} \} 
\end{displaymath}
where $\varphi_{c}$ is the classical field defined as the VEV of the
operator $\varphi$ in the presence of the source $J(x)$ 
and  $C_{s}  = +1~(-1)$  for fields   obeying   the   Bose--Einstein
(Fermi--Dirac)   statistics.  Moreover, the trace $\mathrm{Tr}$ in
(\ref{V1loop}) acts over all space and internal degrees of freedom.
For our purposes, a more convenient representation of (\ref{V1loop})
is 
\begin{equation} 
	\label{V1loopeff}
V^{\mathrm{1-loop}}_{\mathrm{eff}} = - C_{s} \frac{i}{2}  \int^{1}_{0}
dx \mathrm{Tr} \bigg[ \frac{H_{\varphi_{1} \varphi_{2}}(\varphi_{c}) -
    H_{\varphi_{1} \varphi_{2}}(0)}{x \left( H_{\varphi_{1}
      \varphi_{2}}(\varphi_{c}) -  H_{\varphi_{1} \varphi_{2}}(0)
    \right) + H_{\varphi_{1} \varphi_{2}}(0)}  \bigg] \; . 
\end{equation} 
In  momentum space  of $n  = 4  - 2 \varepsilon$  dimension,  this
last expression becomes 
\begin{equation} 
	\label{V1loopk}
V^{\mathrm{1-loop}}_{\mathrm{eff}} = - C_{s} \frac{i}{2} \int^{1}_{0}
dx \int \frac{d^n k}{(2\pi)^n} \mathrm{tr} \bigg[ \frac{H_{\varphi_{1}
      \varphi_{2}}(\varphi_{c}) -  H_{\varphi_{1} \varphi_{2}}(0)}{x
    \left( H_{\varphi_{1} \varphi_{2}}(\varphi_{c}) -  H_{\varphi_{1}
      \varphi_{2}}(0) \right) + H_{\varphi_{1} \varphi_{2}}(0)}
  \bigg] 
\end{equation} 
and  $\mathrm{tr}$ now  symbolizes the  trace only  over  the internal
degrees of  freedom, e.g.~over the polarizations of  the gauge fields,
the spinor components of the fermions or the Yukawa coupling matrices.

The one-loop effective  potential of the MSISM can  now be calculated
by applying (\ref{V1loopk}) to the scalars, gauge bosons (GB), ghosts,
charged fermions (CF) and neutrinos (N) individually, i.e.
\begin{eqnarray}
 	\label{Vefftotal}
 V^{\mathrm{1-loop}}_{\mathrm{eff}}  & = &
 V^{\mathrm{1-loop}}_{\mathrm{eff}}(\mathrm{Scalar}) +
 V^{\mathrm{1-loop}}_{\mathrm{eff}}(\mathrm{GB}) +
 V^{\mathrm{1-loop}}_{\mathrm{eff}}(\mathrm{Ghost}) \nonumber\\ 
 & & + V^{\mathrm{1-loop}}_{\mathrm{eff}}(\mathrm{CF}) +
 V^{\mathrm{1-loop}}_{\mathrm{eff}}(\mathrm{N}) \; .    
\end{eqnarray} 

For  the  scalar  contribution,  this is  a  non-trivial derivation,
since $H_{\varphi_{1}  \varphi_{2}}(\varphi_{c})$ as defined in
(\ref{eqn:H}) is the $6\times 6$ matrix:
\begin{equation} 
\left( \begin{array}{c c c c}
H_{\Phi^{\dagger} \Phi}  & H_{\Phi^{\dagger} \Phi^{\dagger}}  &
H_{\Phi^{\dagger} S}  & H_{\Phi^{\dagger} S^{*}} \\ 
H_{\Phi \Phi}  & H_{\Phi \Phi^{\dagger}}  & H_{\Phi S}  & H_{\Phi S^{*}} \\
H_{S \Phi}  & H_{S \Phi^{\dagger}}  & H_{S S}  & H_{S S^{*}} \\
H_{S^{*} \Phi}  & H_{S^{*} \Phi^{\dagger}}  & H_{S^{*} S}  & H_{S^{*} S^{*}} 
\end{array}\right ) \; .
\end{equation}
Observe    that     $H_{\Phi^{\dagger}    \Phi},\    H_{\Phi^{\dagger}
  \Phi^{\dagger}} ,\ H_{\Phi \Phi} $ and $H_{\Phi \Phi^{\dagger}}$ are
$2  \times 2$  matrices,  $H_{S  S},\ H_{S  S^{*}},  H_{S^{*} S}$  and
$H_{S^{*}  S^{*}}$ are  complex  numbers, and  the remaining  entries,
e.g.~$H_{\Phi  S},\ H_{\Phi S^{*}}$  etc, are  two-dimensional complex
vectors.  This  internal matrix structure  needs be treated  with care
and  must  be  preserved  when  determining  the  matrix,  $[x  \left(
  H_{\varphi_{1}     \varphi_{2}}(\varphi_{c})     -    H_{\varphi_{1}
    \varphi_{2}}(0)  \right) +  H_{\varphi_{1} \varphi_{2}}(0)]^{-1}$.
Taking this fact into account, the scalar contribution is found to be
\begin{eqnarray}
	\label{eqn:scalar1loopeff}
 V^{\mathrm{1-loop}}_{\mathrm{eff}}(\mathrm{Scalar)}  & = &
 \frac{1}{64 \pi^{2}} \bigg[\, 2M_{G^{\pm}}^{4} \bigg(-
   \frac{1}{\varepsilon}-\frac{3}{2} + \ln
   \frac{M_{G^{\pm}}^{2}}{\bar{\mu}^{2}} \bigg)\ +\ M_{G}^{4} \bigg(-
   \frac{1}{\varepsilon}-\frac{3}{2} + \ln
   \frac{M_{G}^{2}}{\bar{\mu}^{2}} \bigg) \nonumber\\ 
 & & +\ \sum_{i=1}^{3} M_{H_i}^{4} \bigg(
   -\frac{1}{\varepsilon}-\frac{3}{2} + \ln
   \frac{M_{H_i}^{2}}{\bar{\mu}^{2}} \bigg)\, \bigg] \; ,
\end{eqnarray}
where $\ln\bar{\mu}^{2} =-\gamma +\ln 4\pi\mu^{2}$, $\gamma \approx 0.5772$
is   the   Euler--Mascheroni   constant   and  $\mu$   is   't-Hooft's
renormalization scale.  The Goldstone mass terms in the above equation
are given by
\begin{equation}
M_{G}^{2} = M_{G^{\pm}}^{2} =  \lambda_{1} \Phi^{\dagger} \Phi +
\lambda_{3} S^*S  + \lambda_{4} S^{2} + \lambda_{4}^{*} S^{* 2}\; .
\end{equation} 
These  mass   terms  vanish  along  the  flat   direction  because  of
(\ref{eqn:diffwrtphi}).  However,  after EWSSB they  obtain additional
$\xi$-dependent   contributions  through   the   gauge  fixing   terms
[cf.~(\ref{eqn:gmasses})].

The   masses   $M_{H_{1,2,3}}^{2}$   appearing   in~(\ref{eqn:scalar1loopeff})
correspond to the eigenvalues of the matrix
\begin{equation} 
	\label{eqn:123}
{\cal M}^2_S\ =\ \left( \begin{array}{c c c}
M_{\phi}^{2} & M_{\phi \sigma} & M_{\phi J} \\
M_{\phi \sigma}  & M_{\sigma}^{2} &  M_{\sigma J} \\
M_{\phi J}   & M_{\sigma J}  & M_{J}^{2} 
\end{array}\right )\; ,
\end{equation}
where
\begin{eqnarray}  
	\label{eqn:scalarmassesfor123}
M_{\phi}^{2} & = & \frac{3}{2}\lambda_{1} \phi^{2} + \frac{1}{2}
(\lambda_{3} + \lambda_{4} + \lambda_{4}^{*}) \sigma^{2} +
i(\lambda_{4} - \lambda_{4}^{*})\sigma J +  \frac{1}{2} (\lambda_{3} -
\lambda_{4} - \lambda_{4}^{*}) J^{2} \; ,\nonumber\\ 
M_{\sigma}^{2} & = & \frac{1}{2}(\lambda_{3} + \lambda_{4} +
\lambda_{4}^{*}) \phi^{2} + \frac{3}{2}(\lambda_{2} + 2\lambda_{5} + 2
\lambda_{5}^{*} + \lambda_{6} + \lambda_{6}^{*}) \sigma^{2} 
\nonumber\\ 
& & + 3i(\lambda_{5} - \lambda_{5}^{*} + \lambda_{6} -
\lambda_{6}^{*}) \sigma J + \frac{1}{2}(\lambda_{2} - 3\lambda_{6} -
3\lambda_{6}^{*}) J^{2} \; , \nonumber\\ 
M_{J}^{2} & = & \frac{1}{2}(\lambda_{3} - \lambda_{4} -
\lambda_{4}^{*}) \phi^{2} + \frac{1}{2}(\lambda_{2} - 3\lambda_{6} -
3\lambda_{6}^{*}) \sigma^{2} + 3i(\lambda_{5} - \lambda_{5}^{*} -
\lambda_{6} + \lambda_{6}^{*}) \sigma J \nonumber\\ 
& & + \frac{3}{2}(\lambda_{2} - 2\lambda_{5} - 2 \lambda_{5}^{*} +
\lambda_{6} + \lambda_{6}^{*}) J^{2} \; , \nonumber\\ 
M_{\phi \sigma} & = & \phi \Big[ (\lambda_{3} +  \lambda_{4} +
\lambda_{4}^{*}) \sigma + i ( \lambda_{4} -  \lambda_{4}^{*}) J\Big] \; ,
\nonumber\\ 
M_{\sigma J} & = & i \Bigg[ \frac{1}{2}(\lambda_{4} - \lambda_{4}^{*}) \phi^{2}  
+ \frac{3}{2}(\lambda_{5} - \lambda_{5}^{*} + \lambda_{6} - \lambda_{6}^{*}) 
\sigma^{2} -i (\lambda_{2}  - 3 \lambda_{6} - 3 \lambda_{6}^{*}) \sigma J
 \nonumber\\ 
& & + \frac{3}{2}(\lambda_{5} - \lambda_{5}^{*} - \lambda_{6} + 
\lambda_{6}^{*}) J^{2}  \Bigg] \; , \nonumber\\ 
M_{\phi J} & = & \phi \Big[ i  (\lambda_{4} -  \lambda_{4}^{*})
\sigma +  (\lambda_{3} -  \lambda_{4} -  \lambda_{4}^{*}) J   \Big] \; . 
\end{eqnarray}    
Note that $M^{2}_{\phi,  \sigma, J}$ reduce to the  squared mass terms
for the $\phi$,  $\sigma$ and $J$ fields, respectively,  if all mixing
terms $M_{\phi  \sigma, \phi J,  \sigma J}$ between the  scalar fields
vanish along  a given flat  direction.  In~addition, we  should remark
here that  one of the  eigenvalues of the matrix  (\ref{eqn:123}) will
always be zero along a minimal flat direction, since it corresponds to
the pseudo-Goldstone boson $h$ of scale invariance.

We  now  turn  our   attention  to  the  gauge-boson  contribution  in
(\ref{Vefftotal}), which  has been calculated in  the $R_{\xi}$ gauge.
The gauge-boson contribution reads:
\begin{eqnarray}
	\label{eqn:gaugeboson1loopeff}
V^{\mathrm{1-loop}}_{\mathrm{eff}}(\mathrm{GB)}  & = & \frac{1}{64
  \pi^{2}} \bigg[\, 6M_{W}^{4} \bigg(-
  \frac{1}{\varepsilon}-\frac{5}{6} + \ln
  \frac{M_{W}^{2}}{\bar{\mu}^{2}} \bigg)\   +\ 2 \xi^{2} M_{W}^{4}
  \bigg(- \frac{1}{\varepsilon}-\frac{3}{2} + \ln \frac{ \xi
    M_{W}^{2}}{\bar{\mu}^{2}} \bigg) \nonumber\\ 
& & + \ 3M_{Z}^{4} \bigg(- \frac{1}{\varepsilon}-\frac{5}{6} + \ln
  \frac{M_{Z}^{2}}{\bar{\mu}^{2}} \bigg)\ +\ \xi^{2} M_{Z}^{4} \bigg(-
  \frac{1}{\varepsilon}-\frac{3}{2} + \ln \frac{\xi
    M_{Z}^{2}}{\bar{\mu}^{2}} \bigg)\,  \bigg]\; , 
\end{eqnarray}
where
\begin{equation} 
	\label{eqn:WZmasses}
M_{W}^{2}  =  \frac{g^{2}}{2} \Phi^{\dagger} \Phi\ \; ,
\qquad M_{Z}^{2}  =  \frac{g^{2}+g'^{2}}{2}  \Phi^{\dagger} \Phi \; . 
\end{equation}
In the same class of $R_{\xi}$ gauges, the ghost contribution is given after EWSSB
by
\begin{equation}
	\label{eqn:ghost1loopeff}
V^{\mathrm{1-loop}}_{\mathrm{eff}}(\mathrm{Ghost)}\  =\ -\; \frac{2}{64
  \pi^{2}} \bigg[\, 2 M_{\omega_{\pm}}^{4} \bigg(-
  \frac{1}{\varepsilon}-\frac{3}{2} + \ln \frac{ M_{\omega_{\pm}}^{2}}
  {\bar{\mu}^{2}} \bigg)\ +\ M_{\omega_{Z}}^{4} \bigg(-
  \frac{1}{\varepsilon}-\frac{3}{2} + \ln \frac{M_{\omega_{Z}}^{2}}
  {\bar{\mu}^{2}} \bigg)\,  \bigg]\; ,
\end{equation} 
where  $M^2_{\omega_{\pm}}  = \xi  M^2_W$  and  $M^2_{\omega_Z} =  \xi
M^2_Z$ are the field-dependent ghost masses.

Next, we calculate the charged fermion contribution to the effective
potential~(\ref{Vefftotal}). This is given by
\begin{eqnarray}
	\label{eqn:chargedfermion1loopeff}
V^{\mathrm{1-loop}}_{\mathrm{eff}}(\mathrm{CF)}  & = & -\;\frac{4}{64
  \pi^{2}} \bigg[\, 3\sum_{i=1}^{3} M^{4}_{ui} \bigg(-
  \frac{1}{\varepsilon}- 1 + \ln \frac{M_{ui}^{2}}{\bar{\mu}^{2}}
  \bigg)\  +\ 3\sum_{i=1}^{3} M_{di}^{4} \bigg(- \frac{1}{\varepsilon}
  - 1 + \ln \frac{M_{di}^{2}}{\bar{\mu}^{2}} \bigg)  \nonumber\\ 
& & +\ \sum_{i=1}^{3} M_{ei}^{4} \bigg(- \frac{1}{\varepsilon}- 1 +
  \ln \frac{M_{ei}^{2}}{\bar{\mu}^{2}} \bigg)\,    \bigg]\; , 
\end{eqnarray}
where $M^2_{f  i}$ ($f = u,  \, d, \,  e$) are the eigenvalues  of the
background  $\Phi$-dependent  squared  mass  matrix for  the  $f$-type
fermion:  $({\bf h}^{f\dagger}{\bf  h}^{f})\,  \Phi^\dagger\Phi$. Note
the factor  3 in  front of the  up- and down-type  quark contributions
which counts the SU(3)$_{c}$ colour degrees of freedom.

If the MSISM is extended with right-handed neutrinos, these will give
rise  to   additional  quantum  effects  on   the  one-loop  effective
potential~(\ref{Vefftotal}).  The   contribution  of  the   light  and
heavy Majorana neutrinos to the effective potential is given by
\begin{eqnarray}
	\label{eqn:neutrino1loopeff}
V^{\mathrm{1-loop}}_{\mathrm{eff}}(\mathrm{N)} & = &  -\ \frac{2}{64
  \pi^{2}}\; \bigg\{\, \mathrm{Tr}\,\bigg[ ({\bf M_{\nu}
    M_{\nu}^{\dagger}})^{2} 
\bigg( - \frac{1}{\varepsilon}- 1 + \ln \frac{{\bf M_{\nu} 
    M_{\nu}^{\dagger}}}{\bar{\mu}^{2}} \bigg) \bigg] \nonumber\\ 
& & +\ \mathrm{Tr}\,\bigg[ ({\bf M}_N {\bf M}_N^{\dagger})^{2} 
\bigg( - \frac{1}{\varepsilon}- 1 + \ln \frac{{\bf M}_N
    {\bf M}_N^{\dagger}}{\bar{\mu}^{2}} \bigg) \bigg]\,  \bigg\}\; ,
\end{eqnarray} 
where  ${\bf M_{\nu}}$  is  the background  $\Phi$- and  $S$-dependent
light-neutrino mass matrix,
\begin{equation}
{\bf M_{\nu}} = (\Phi\Phi^T)\, {\bf h^{\nu}}\, {\bf M}_N^{-1}\, {\bf
  h}^{\nu T}\; ,   
\end{equation}
and ${\bf M}_N$ is  the respective $S$-dependent heavy-neutrino  mass
matrix:
\begin{equation}
{\bf M}_N\ =\ {\bf h}^N S\: +\: {\bf \tilde{h}}^{N\dagger} S^{*} \; .
\end{equation}

Finally,  an important  remark is  in order.   The  one-loop effective
potential  $V^{\mathrm{1-loop}}_{\mathrm{eff}}$  is  in general  gauge
dependent  through   (\ref{eqn:gaugeboson1loopeff})  and  after  EWSSB
through  (\ref{eqn:ghost1loopeff}) and  the  Goldstone $\xi$-dependent
mass  terms in  (\ref{eqn:scalar1loopeff})  as well.   However, it  is
known  that  the effective  potential  becomes gauge-independent  when
evaluated at local extrema~\cite{Nielsen,Lisa}.  Within the context of
perturbation  theory,  the  one-loop  effective  potential  should  be
$\xi$-independent,  if  it  is   evaluated  along  a  stationary  flat
direction~\cite{Garbrecht}.   This  is  exactly  the case  of  the  GW
approach  to  the  effective  potential~(\ref{eqn:general1looppotAB}).
Therefore,  as  a  consistency   check,  we  have  verified  that  the
$\xi$-dependent terms due to  gauge, Goldstone and ghost contributions
cancel against each other in the effective potential~(\ref{Vefftotal})
when evaluated along a stationary flat direction.


\setcounter{equation}{0}
\section{One-Loop Anomalous Dimensions 
and {\boldmath $\beta$}-Functions}\label{App:betafns} 

In this section, we calculate the one-loop anomalous dimensions of the
fields and  the $\beta$ functions  of couplings in the  MSISM, within
the $\overline{\rm  {MS}}$ scheme of renormalization  in the $R_{\xi}$
class gauges.  Our calculation  is based on the so-called displacement
operator formalism, or $D$-formalism  in short, which was developed in
\cite{Binosi} as an  alternative approach to systematically performing
renormalization to  all orders in perturbation theory.   Since this is
not a common approach, we briefly review its basic features.

According  to  the   $D$-formalism,  the  renormalized  one-particle
irreducible  $n$-point correlation  functions,  denoted  hereafter
with a  script  $R$,  are related to the unrenormalized ones through: 
\begin{equation}\label{Dformalism}
\varphi_{R}^{n}\  \g^{R}_{\varphi^{n}}(\lambda_{R}, \xi_{R}; \mu)\  =\
e^{D} \Big( \varphi_{R}^{n}\ \g_{\varphi^{n}}(\lambda_{R}, \xi_{R};
\mu, \epsilon) \Big) \; , 
\end{equation}
where $D$ is the displacement operator that takes the form,
\begin{equation}
D\ =\ \delta \varphi \frac{ \partial}{ \partial \varphi_{R}}\: +\: \delta
\lambda \frac{ \partial}{ \partial \lambda_{R}}\:  +\: \delta \xi \frac{
  \partial}{ \partial \xi_{R}} \ , 
\end{equation}
where  $\varphi$  again  represents  all  the  fields  in  the  model,
$\lambda$ all  the coupling constants,  i.e. $\lambda_{i}, \ g  ,\ g',
\ g_{s},\  h^{f}_{ij}$, and  $\xi$ is the  gauge fixing  parameter. In
addition,    the   counterterm    renormalizations,   $\delta\varphi$,
$\delta\lambda$  etc, are  defined  as, $\delta  \varphi  = \varphi  -
\varphi_R = (Z^{1/2}_\varphi - 1) \varphi_R$, $\delta\lambda = \lambda
-\lambda_R = (Z_\lambda - 1) \lambda_R$ etc.

We  may  now  perform  a  loopwise expansion  of  the  operator  $e^D$
in (\ref{Dformalism}),  
\begin{equation}
e^{D}\ =\ 1\: +\: D^{(1)}\: +\: \Big( D^{(2)} +
\frac{1}{2}D^{(1)2}\Big)\: +\ \dots,
\end{equation}
where the superscript $(n)$ on $D$ denotes the loop order, i.e.
\begin{equation}
D^{(n)}\ =\ \delta \varphi^{(n)} \frac{ \partial}{ \partial \varphi_{R}}\
+\ \delta \lambda^{(n)} \frac{ \partial}{ \partial \lambda_{R}}\  +\
\delta \xi^{(n)} \frac{ \partial}{ \partial \xi_{R}} \ . 
\end{equation}
Correspondingly,   the   parameter   or  counterterm   shifts
$\delta \varphi^{(n)},  \delta \lambda^{(n)}$  and $\delta \xi^{(n)}$
are loopwise defined as 
\begin{equation}
\delta \varphi^{(n)}\ =\ Z_{\varphi}^{\frac{1}{2}(n)}\varphi_{R} \; ,
\qquad \delta \lambda^{(n)}\ =\ Z_{\lambda}^{(n)}\lambda_{R} \; ,  \qquad
\delta \xi^{(n)}\ =\ Z_{\xi}^{(n)}\xi_{R} \; . 
\end{equation}
Applying the $D$-formalism to one-loop, we have
\begin{equation}\label{D1loop}
\varphi_{R}^{n} \Gamma^{R(1)}_{\varphi^{n}}(\lambda_{R}, \xi_{R}; \mu)
= D^{(1)}\left( \varphi_{R}^{n}
\Gamma^{(0)}_{\varphi^{n}}(\lambda_{R}, \xi_{R}; \mu) \right) +
\varphi_{R}^{n} \Gamma^{(1)}_{\varphi^{n}}(\lambda_{R},  \xi_{R}; \mu,
\epsilon) \; . 
\end{equation} 
This  last equation  can be  used  to calculate  the wavefunction  and
coupling    constant    renormalizations,   $Z^{(1)}_{\varphi}$    and
$Z^{(1)}_{\lambda}$.   Having  thus  obtained $Z^{(1)}_{\varphi}$  and
$Z^{(1)}_{\lambda}$, we may  compute the one-loop anomalous dimensions
$\gamma_\varphi$ of the fields  and the $\beta_{\lambda}$ functions of
the couplings as follows:
\begin{eqnarray}
  \label{eqn:anomdimandbetaeqn}
\gamma_{\varphi}\ \!\!&\equiv&\!\! -\, \mu\, \frac{d\ln \varphi_R}{d\mu} \ =\
- \frac{1}{2}\ \lim_{\varepsilon \to 0}\; \sum_{\lambda_{i}}\
\varepsilon\, d_{\lambda_{i}} \lambda_{iR}\;
\frac{\partial}{\partial \lambda_{iR}}\,  Z^{(1)}_{\varphi}\; , 
\nonumber\\
\beta_{\lambda_{i}}\ \!\!&\equiv&\!\!  \mu\, 
\frac{d\lambda_{iR}}{d\mu} \ =\ 
\lambda_{iR}\; \lim_{\varepsilon \to 0}\; 
\sum_{\lambda_j}\, \varepsilon\, d_{\lambda_{j}}     
\lambda_{jR}\; \frac{\partial}{\partial \lambda_{jR}}\, 
 Z^{(1)}_{\lambda_{i}} \; , 
\end{eqnarray}
where $\varepsilon\, d_{\lambda}$  is the tree-level scaling dimension
of the  generic coupling $\lambda$ in  $n= 4-2\varepsilon$ dimensions,
with $d_{\lambda_i}  = 2$ for  the scalar quartic couplings,  $d_{g} =
d_{h} = 1$ for the gauge and  Yukawa couplings and $d_\xi = 0$ for the
gauge-fixing parameter.  It is useful to remark here that the one-loop
anomalous   dimensions  $\gamma_\varphi$   of  the   fields   and  the
$\beta_\lambda$ functions can be  calculated in the symmetric phase of
the theory.

Employing~(\ref{D1loop})  and   (\ref{eqn:anomdimandbetaeqn})  in  the
$\overline{\rm MS}$  scheme, we  may calculate the  one-loop anomalous
dimensions  and  $\beta$  functions  in  the  $R_{\xi}$  gauge.   More
explicitly, we obtain for the anomalous dimensions of the fields:
\begin{eqnarray} 
  \label{anomdim}
\gamma_{\Phi} & = & \frac{1}{(4\pi)^{2}} \left[\frac{1}{4}(\xi 
- 3)(3g^{2} +  g'^{2}) + T_{1}  \right] \; ,\nonumber\\[3mm]  
\gamma_{S} & = & \frac{1}{(4\pi)^{2}}\; \frac{1}{2}T_{2} \;,\nonumber\\[3mm]
\mbox{\boldmath$\gamma$}_{u_L} & = & \frac{1}{(4\pi)^{2}} \left[ \frac{1}{2}
  \Big( {\bf h}^{u} {\bf h}^{u \dagger}  
+ {\bf h}^{d} {\bf h}^{d \dagger} \Big) + \xi \Big( \frac{4}{3}g_{s}^{2} +
   \frac{3}{4} g^{2} + \frac{1}{36} g'^{2} \Big)\, {\bf 1}_3 \right]\;,   
\nonumber\\[3mm] 
\mbox{\boldmath$\gamma$}_{u_{R}} & = & \frac{1}{(4\pi)^{2}} \left[ {\bf h}^{u
  \dagger} {\bf h}^{u} 
+ \frac{4}{9}\xi \Big( 3g_{s}^{2} + g'^{2}\Big)\, {\bf 1}_3\,\right]\;,
\nonumber\\[3mm]
\mbox{\boldmath$\gamma$}_{d_{L}} & = & \frac{1}{(4\pi)^{2}} \left[ \frac{1}{2}
  \Big( {\bf h}^{u} {\bf h}^{u \dagger} 
+ {\bf h}^{d} {\bf h}^{d \dagger} \Big) + \xi \Big( \frac{4}{3}g_{s}^{2} +
   \frac{3}{4} g^{2} + \frac{1}{36} g'^{2} \Big)\, {\bf 1}_3\right]\;,
\nonumber\\[3mm]
\mbox{\boldmath$\gamma$}_{d_{R}} & = & \frac{1}{(4\pi)^{2}} \left[ {\bf h}^{d
  \dagger} {\bf h}^{d} 
+ \frac{1}{9}\xi \Big( 12g_{s}^{2} + g'^{2}\Big)\,{\bf 1}_3\right]\;,
\nonumber\\[3mm]
\mbox{\boldmath$\gamma$}_{\nu^0_L} & = & \frac{1}{(4\pi)^{2}} \left[ \frac{1}{2}
  \Big( {\bf h}^{e} {\bf h}^{e \dagger} +  {\bf h}^{\nu} {\bf h}^{\nu \dagger}
  \Big) + \frac{\xi}{4} \Big( 3g^{2} + g'^{2} \Big)\, {\bf 1}_3 \right] \; ,
\nonumber\\[3mm]  
\mbox{\boldmath$\gamma$}_{\nu^{0C}_L} & = &  \frac{1}{(4\pi)^{2}} 
  \left[\, \frac{1}{2}
  \Big( {\bf h}^{e *} {\bf h}^{e T} +  {\bf h}^{\nu *}
  {\bf h}^{\nu T} \Big) +
  \frac{\xi}{4} \Big( 3g^{2} + g'^{2} \Big)\, {\bf 1}_3 \right] \; ,
\nonumber\\[3mm]  
\mbox{\boldmath$\gamma$}_{\nu^0_R} & = &  \frac{1}{(4\pi)^{2}} \left( 
{\bf h}^{\nu \dagger} {\bf h}^{\nu} + \frac{1}{2} {\bf h}^{N \dagger} {\bf
  h}^{N} +
\frac{1}{2} {\bf \tilde{h}}^{N} {\bf \tilde{h}}^{N \dagger} \right)\;,
\nonumber\\[3mm]
\mbox{\boldmath$\gamma$}_{\nu^{0C}_R} & = & \frac{1}{(4\pi)^{2}}
\left( {\bf h}^{\nu T} 
  {\bf h}^{\nu *} +
  \frac{1}{2} {\bf h}^{N} {\bf h}^{N \dagger} + 
 \frac{1}{2} {\bf \tilde{h}}^{N \dagger} {\bf \tilde{h}}^{N}  \right) \; ,
\end{eqnarray} 
where $T_{1} = \mathrm{Tr}\big( 3  {\bf h}^{u} {\bf h}^{u \dagger} + 3
{\bf h}^{d}{\bf h}^{d \dagger}  + {\bf h}^{e}{\bf h}^{e\dagger} + {\bf
  h}^{\nu}{\bf  h}^{\nu \dagger}\big)$  and $T_{2}  = \mathrm{Tr}\big(
{\bf  h}^{N \dagger} {\bf  h}^{N} +  {\bf \tilde{h}}^{N  \dagger} {\bf
  \tilde{h}}^N\big)$.                    Notice                   that
$(\mbox{\boldmath$\gamma$}_{\nu^0_L})^*                               =
\mbox{\boldmath$\gamma$}_{\nu^{0C}_L}$                              and
$(\mbox{\boldmath$\gamma$}_{\nu^0_R})^*                               =
\mbox{\boldmath$\gamma$}_{\nu^{0C}_R}$,  where we  have used  $h^{N} =
h^{N T}$ and $\tilde{h}^{N} = \tilde{h}^{N T}$, which is a consequence
of  the  Majorana  constraint  on  the  left-handed  and  right-handed
neutrinos, $\nu^0_{iL}$ and $\nu^0_{iR}$.

Correspondingly, we start by listing the one-loop $\beta$ functions of
the scalar-potential quartic couplings:
\begin{eqnarray}
\beta_{\lambda_{1}} & = & \frac{1}{8 \pi^{2}} \bigg[\, 6
  \lambda_{1}^{2} + \lambda_{3}^{2} + 4 \lambda_{4}\lambda_{4}^{*} +
  \frac{3}{8}\bigg( 3g^{4} +  2 g^{2}g'^{2} +  g'^{4}
  \bigg) - T_{3} - \lambda_{1}\bigg( \frac{3}{2} \left(3g^{2} +
  g'^{2} \right) - 2T_{1}\bigg)\, \bigg] \; , \nonumber\\[3mm] 
\beta_{\lambda_{2}} & = & \frac{1}{8 \pi^{2}}\bigg[\, 5
  \lambda_{2}^{2} + 2 \lambda_{3}^{2} + 4 \lambda_{4}\lambda_{4}^{*} +
  54 \lambda_{5}\lambda_{5}^{*} + 36 \lambda_{6}\lambda_{6}^{*}
  - \mathrm{Tr}\Big({\bf h}^{N}{\bf h}^{N \dagger}{\bf h}^{N}{\bf
    h}^{N \dagger}\Big)  
   \nonumber\\ 
& & - 2 \mathrm{Tr}\Big({\bf \tilde{h}}^{N}{\bf \tilde{h}}^{N \dagger}
{\bf h}^{N \dagger}{\bf h}^{N}\Big) - 2
  \mathrm{Tr}\Big({\bf \tilde{h}}^{N \dagger}{\bf \tilde{h}}^{N}
{\bf h}^{N}{\bf h}^{N \dagger}\Big)
 -   \mathrm{Tr}\Big({\bf \tilde{h}}^{N \dagger}{\bf \tilde{h}}^N
{\bf \tilde{h}}^{N\dagger}{\bf \tilde{h}}^N\Big)  
+ \lambda_{2} T_{2}\, \bigg] \; , \nonumber\\[3mm]
\beta_{\lambda_{3}} & = & \frac{1}{8 \pi^{2}} \bigg[\, 3 \lambda_{1}
  \lambda_{3} + 2\lambda_{2} \lambda_{3} + 2 \lambda_{3}^{2} + 8
  \lambda_{4}\lambda_{4}^{*} + 6 \lambda_{4}\lambda_{5}^{*} + 6
  \lambda_{5}\lambda_{4}^{*} 
-2\mathrm{Tr}\Big({\bf h}^{N \dagger} {\bf h}^{N} {\bf h}^{\nu
  \dagger}{\bf h}^{\nu}\Big) \nonumber\\ 
& &  - 2\mathrm{Tr}\Big({\bf \tilde{h}}^N 
{\bf \tilde{h}}^{N \dagger} {\bf h}^{\nu \dagger} {\bf h}^{\nu}\Big) - 
\lambda_{3}\bigg(\frac{3}{4}\left(3g^{2} + g'^{2}\right) - T_{1} 
- \frac{1}{2} T_{2} \bigg)
  \, \bigg] \; ,  \\[3mm] 
\beta_{\lambda_{4}} & = & \frac{1}{8 \pi^{2}} \bigg[\, 3 \lambda_{1}
  \lambda_{4} + \lambda_{2} \lambda_{4} + 4 \lambda_{3}\lambda_{4} + 3
  \lambda_{3}\lambda_{5} + 6 \lambda_{4}^{*}\lambda_{6}  - 2
  \mathrm{Tr}\Big({\bf \tilde{h}}^N {\bf h}^{N} {\bf h}^{\nu \dagger}
         {\bf h}^{\nu}\Big)
  \nonumber\\  
& &  - \lambda_{4}\bigg(\frac{3}{4}\left(3g^{2} + g'^{2}\right) -
  T_{1} - \frac{1}{2}T_{2} \bigg)\, 
  \bigg] \; ,  \nonumber\\[3mm] 
\beta_{\lambda_{5}} & = & \frac{1}{8\pi^{2}} \bigg[\, 9 \lambda_{2}
  \lambda_{5} + 2 \lambda_{3} \lambda_{4} + 18
  \lambda_{5}^{*}\lambda_{6} - 
 \mathrm{Tr}\Big({\bf \tilde{h}}^{N \dagger} {\bf \tilde{h}}^N
{\bf h}^N {\bf \tilde{h}}^N\Big) - 
  \mathrm{Tr}\Big({\bf h}^{N}{\bf h}^{N \dagger} {\bf h}^{N}
  {\bf \tilde{h}}^N\Big)
 +  \lambda_{5} T_{2}\, \bigg] \; ,  \nonumber\\[3mm] 
\beta_{\lambda_{6}} & = &  \frac{1}{8 \pi^{2}} \bigg[\, 6 \lambda_{2}
  \lambda_{6} + 2\lambda_{4}^{2} + 9 \lambda_{5}^{2} -  
  \mathrm{Tr}\Big({\bf \tilde{h}}^N{\bf h}^N{\bf \tilde{h}}^N
  {\bf h}^{N}\Big) +
  \lambda_{6}T_{2}\,  \bigg] \; ,\nonumber
\end{eqnarray}
where $T_{3}  = \mathrm{Tr}\big( 6\,{\bf h}^u{\bf  h}^{u \dagger} {\bf
  h}^u{\bf  h}^{u  \dagger} +  6\,{\bf  h}^{d}{\bf h}^{d  \dagger}{\bf
  h}^{d}{\bf  h}^{d \dagger} +  2\,{\bf h}^{e}{\bf  h}^{e \dagger}{\bf
  h}^{e}{\bf   h}^{e   \dagger}   +   2\,{\bf   h}^{\nu}{\bf   h}^{\nu
  \dagger}{\bf  h}^{\nu}{\bf h}^{\nu  \dagger}\big)$.   Note that  the
one-loop $\beta$ functions of the complex conjugate quartic couplings,
i.e.~$\beta_{\lambda_{4,5,6}^{*}}$,         are        given        by
$\beta_{\lambda_{4,5,6}^{*}} = (\beta_{\lambda_{4,5,6}})^{*}$.

For  the one-loop $\beta$  functions of  the SU(3)$_c$,  SU(2)$_L$ and
U(1)$_Y$ gauge couplings, we use the well-established results:
\begin{equation} 
\beta_{g_{s}}\ =\  -\; \frac{1}{8 \pi^{2}}\; \frac{7}{2} g_{s}^{3}\;,\qquad
\beta_{g}\ =\ -\; \frac{1}{8 \pi^{2}}\; \frac{19}{12} g^{3} \; ,\qquad
\beta_{g'}\ =\  \frac{1}{8 \pi^{2} }\; \frac{41}{12} g'^{3} \; .
\end{equation}
Next, we present  the known one-loop $\beta$ functions  of the up-type and
down-type quark Yukawa couplings
\begin{eqnarray}
\mbox{\boldmath$\beta$}_{{\bf h}^{u}} & = & \frac{1}{8 \pi^{2}} 
\bigg[\, -\frac{17}{24}g'^{2} -  \frac{9}{8}g^{2} - 4 g_{s}^{2} +
\frac{1}{2}T_{1} + \frac{3}{4} \big( {\bf h}^{u} {\bf h}^{u \dagger} -
     {\bf h}^{d} {\bf h}^{d \dagger} \big)\, \bigg]\, {\bf h}^{u} \; ,
     \nonumber\\[3mm]  
\mbox{\boldmath$\beta$}_{{\bf h}^{d}} & = & \frac{1}{8 \pi^{2}} 
\bigg[\, -\frac{5}{24}g'^{2} -  \frac{9}{8}g^{2} - 4 g_{s}^{2} +
\frac{1}{2}T_{1} + \frac{3}{4} \left( {\bf h}^{d} {\bf h}^{d \dagger} -
     {\bf h}^{u} {\bf h}^{u \dagger} \right)\, \bigg]\, {\bf h}^{d} \; .
\end{eqnarray}
Finally, the one-loop $\beta$ functions of the light- and heavy-neutrino
Yukawa couplings are calculated to be
\begin{eqnarray} 
	\label{eqn:beta} 
\mbox{\boldmath$\beta$}_{{\bf  \tilde{h}}^N} & = & \frac{1}{8 \pi^{2}} 
\bigg[\, {\bf \tilde{h}}^N \bigg( \frac{5}{4} {\bf h}^{N}{\bf h}^{N \dagger} +
  \frac{1}{4} {\bf h^{\tilde{N} \dagger} h^{\tilde{N}}} + \frac{1}{2}
       {\bf h}^{\nu T}{\bf h}^{\nu *} \bigg)  \nonumber \\ 
& & +\  \bigg( \frac{5}{4} {\bf h}^{N \dagger} {\bf h}^{N} + 
 \frac{1}{4} {\bf \tilde{h}}^N {\bf \tilde{h}}^{N \dagger} 
+ \frac{1}{2} {\bf h}^{\nu \dagger }{\bf h}^{\nu} \bigg) 
{\bf \tilde{h}}^N\ +\ \frac{1}{4}  {\bf \tilde{h}}^N T_{2}\, \bigg]\; ,
   \nonumber\\[3mm]
\mbox{\boldmath$\beta$}_{{ \bf h}^N} & = & \frac{1}{8 \pi^{2}} 
\bigg[\, 
{\bf h}^N \bigg( \frac{5}{4} {\bf \tilde{h}}^{N}{\bf \tilde{h}}^{N\dagger} 
+ \frac{1}{4} {\bf h}^{N \dagger}{\bf  h}^{N} +
       \frac{1}{2} {\bf h^{\nu \dagger }h^{\nu}} \bigg)  \nonumber \\ 
& & +\  \bigg( \frac{5}{4} {\bf \tilde{h}}^{N \dagger}{\bf \tilde{h}}^{N} +
  \frac{1}{4} {\bf h}^{N} {\bf h}^{N \dagger} + 
\frac{1}{2} {\bf h}^{\nu T} {\bf h}^{\nu *} \bigg) {\bf h}^N\ + \
\frac{1}{4}  {\bf h}^N T_{2}\,  \bigg] \; , \nonumber\\[3mm]  
\mbox{\boldmath$\beta$}_{{ \bf h}^\nu} & = & \frac{1}{8 \pi^{2}} 
\bigg[\, {\bf h}^{\nu} \bigg( -\frac{3}{8}g'^{2} -\frac{9}{8}g^{2} +
       \frac{1}{2}T_{1} \bigg)\ +\ \frac{3}{4}\bigg( {\bf h}^{\nu} 
{\bf h}^{\nu \dagger} - {\bf h}^{e} {\bf h}^{e \dagger} \bigg) 
{\bf h}^{\nu}       \nonumber \\ 
& & +\ \frac{1}{4}  {\bf h}^{\nu} 
\bigg( {\bf h}^{N \dagger} {\bf h}^{N} +  
{\bf \tilde{h}}^{N} {\bf \tilde{h}}^{N \dagger} \bigg)\,  \bigg] \; .
\end{eqnarray}

The one-loop anomalous dimensions and $\beta$ functions can be used to
verify the  renormalizability of $V^{\mathrm{1-loop}}_{\mathrm{eff}}$.
To    be    specific,    the    potential    $V=V^{\mathrm{tree}}    +
V^{\mathrm{1-loop}}_{\mathrm{eff}}$   should  be   UV   finite  after
renormalization.  In the so-called $\overline{\rm MS}$ renormalization
scheme~\cite{MSbar},  the one-loop UV counter-terms for  the  fields and
coupling constants are explicitly given by
\begin{equation}
\delta\varphi^{(1)} = Z_{\varphi}^{(1)\, 1/2} \varphi_{R} =
-\frac{1}{2}\left(\frac{1}{\varepsilon} - \gamma + \ln 4 \pi \right)
\gamma_{\varphi} \varphi_{R}\; , \quad 
\delta\lambda^{(1)} =  Z^{(1)}_{\lambda}
\lambda_{R} = \frac{1}{2} \left(\frac{1}{\varepsilon} - \gamma + \ln 4 \pi \right)
\beta_{\lambda}\; . 
\end{equation}

Taking  these relations  into  account, the  one-loop MSISM  effective
potential can  be renormalized in  the $\overline{\rm MS}$  scheme and
its complete analytic form is given by
 \begin{eqnarray}
 	\label{eqn:fullonelooprenormalisedpot}
V^{\mathrm{1-loop}}_{\mathrm{eff}}  & = & \frac{1}{64 \pi^{2}}\; \bigg\{\,
  2M_{G^{\pm}}^{4} \left( -\frac{3}{2} + \ln \frac{M_{G^{\pm}}^{2}}{\mu^{2}} \right) 
  + M_{G}^{4}\left(-\frac{3}{2} + \ln \frac{M_{G}^{2}}{\mu^{2}} \right) +
  \sum_{i=1}^{3} m_{H_i}^{4} \left(-\frac{3}{2} + \ln
  \frac{m_{H_i}^{2}}{\mu^{2}} \right)  \nonumber\\ 
 & & + 6M_{W}^{4} \left( -\frac{5}{6} + \ln \frac{M_{W}^{2}}{\mu^{2}}
  \right) +  3M_{Z}^{4} \left( -\frac{5}{6} + \ln
  \frac{M_{Z}^{2}}{\mu^{2}} \right) - 
   2 \xi^{2} M_{W}^{4} \left( -\frac{3}{2} + \ln \frac{ \xi
    M_{W}^{2}}{\mu^{2}} \right)\nonumber\\ 
& &-  \xi^{2} M_{Z}^{4} \left(
  -\frac{3}{2} + \ln \frac{\xi M_{Z}^{2}}{\mu^{2}} \right)
- 12  \sum_{i=1}^{3} M_{ui}^{4} \left( - 1 + 
 \ln \frac{M_{ui}^2}{\mu^{2}} \right)\\  
& &- 12  \sum_{i=1}^{3} M_{di}^4 
\left(  - 1 + \ln \frac{M_{di}^{2}}{\mu^{2}} \right)
 - 4  \sum_{i=1}^{3} M_{ei}^4 \left( - 1 + 
              \ln \frac{M_{ei}^{2}}{\mu^{2}} \right)\nonumber\\
& &
- 2 \mathrm{Tr}\bigg[ ({\bf M_{\nu} M_{\nu}^{\dagger}})^{2} \left( - 1 + \ln
  \frac{{\bf M_{\nu} M_{\nu}^{\dagger}}}{\mu^{2}} \right)\bigg]  - 2
\mathrm{Tr}\bigg[ ({\bf M}_N {\bf M}_N^{\dagger})^{2} \bigg( - 1 + \ln
  \frac{{\bf M}_N {\bf M}_N^{\dagger}}{\mu^{2}} \bigg)\bigg]\,
  \bigg\} \; ,\nonumber
\end{eqnarray}
where the mass  terms are defined in Appendices A  and B.  Notice that
along  a  stationary  flat   direction,  $\mu  \to  \Lambda$  and  the
$\xi$-dependent Goldstone-boson masses  $M_{G^\pm}$ and $M_{G}$, given
in    (\ref{eqn:gmasses}),   cancel   against    the   $\xi$-dependent
contributions  from the $W^\pm$  and $Z$  bosons and  their respective
ghost  fields.  Hence,  the complete  one-loop  renormalized effective
potential becomes gauge independent in this case.

\section{The Electroweak Oblique Parameters} \label{App:Obparams}

In order to calculate the  electroweak oblique parameters $S$, $T$ and
$U$, we  adopt the notation and  formalism developed in~\cite{Peskin}.
To this end,  we first review the definitions of the  $S$, $T$ and $U$
parameters  and present  their  basic relations  with the  gauge-boson
self-energies,  which we will  then use  to determine  the electroweak
oblique parameters in the MSISM.

In detail, the vacuum polarization amplitudes are defined as
\begin{equation}
i \Pi^{\mu \nu}_{XY} (q^{2})\ =\ i g^{\mu \nu} \Pi_{XY} (q^{2})\ +\
(q^{\mu} q^{\nu} \mathrm{terms})\; ,  
\end{equation}
where $XY = \{11, 22, 33, 3Q, QQ \}$ and
\begin{equation}
\Pi_{XY}(q^{2})\ =\ \Pi_{XY}(0)\ +\ q^{2} \Pi'_{XY}(q^{2})\; .
\end{equation}
These vacuum polarizations are related to the one-particle irreducible
self-energies of the $A$, $W^\pm$ and $Z$ gauge bosons through:
\begin{eqnarray}
	\label{eqn:1PIselfenergies} 
\Pi_{AA} & = & e^{2} \Pi_{QQ} \; , \quad  \quad \Pi_{WW}\,\ =\ \,
\frac{e^{2}}{\sin^{2} \theta_w} \Pi_{11} \; , \nonumber\\ 
\Pi_{ZA} & = & \frac{e^{2}}{\cos \theta_w \sin \theta_w} \left(
\Pi_{3Q} - \sin^2 \theta_w \Pi_{QQ}  \right) \; , \nonumber\\ 
\Pi_{ZZ} & = & \frac{e^{2}}{\cos^2 \theta_w \sin^{2} \theta_w}
\left( \Pi_{33} - 2 \sin^{2} \theta_w \Pi_{3Q} + \sin^{4}
\theta_w \Pi_{QQ}  \right) \; ,
\end{eqnarray}
where $e$  is the  electric charge and  $\theta_w$ is  the electroweak
mixing angle. One  can now solve the above  system of linear equations
for  the  vacuum polarization  amplitudes  $\Pi_{XY}$  and define  the
so-called  electroweak   oblique  parameters~\cite{Peskin}  in
terms of them as follows:
\begin{eqnarray}
\alpha_{\rm em}\, S & = & 4 e^{2} \Big [ \Pi'_{33} (0) - \Pi'_{3Q} (0) \Big
]\; , \nonumber\\ 
\alpha_{\rm em}\, T & = & \frac{e^{2}}{\sin^{2} \theta_w \cos^{2}
  \theta_w m_{Z}^{2}} \Big [ \Pi_{11} (0) - \Pi_{33} (0) \Big ]\; ,
\nonumber\\ 
\alpha_{\rm em}\, U & = & 4 e^{2} \Big [ \Pi'_{11} (0) - \Pi'_{33} (0) \Big ]\;, 
\end{eqnarray}
where  $\alpha_{\rm em}  = e^2/(4  \pi)$ is  the  electromagnetic fine
structure  constant.    Noting the $\sin^{2} \theta_{w}$ dependence of 
$\Pi_{33}$, $\Pi_{3Q}$ and $\Pi_{QQ}$ in $\Pi_{ZZ}$ 
(\ref{eqn:1PIselfenergies}), the  $S$, $T$ and $U$ parameters can be 
determined by calculating the $ZZ$ and $WW$  vacuum  polarization
amplitudes only. 

\begin{figure}
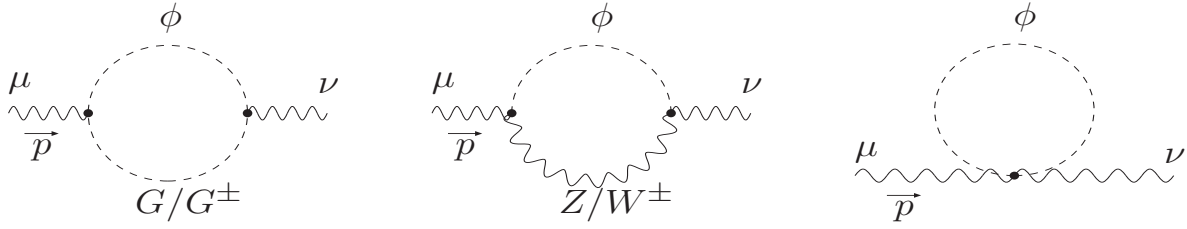

\begin{center}
\includegraphics[height=0.18\textwidth,width=0.27\textwidth]{Figures/scalarloop.epsi} 
\quad \ \ \ \ \
\includegraphics[height=0.18\textwidth,width=0.27\textwidth]{Figures/scalarvectorloop.epsi}
\quad \ \ \ \ \
\includegraphics[height=0.18\textwidth,width=0.27\textwidth]{Figures/4scalarloop.epsi}   
\end{center} 
\caption{\it   Feynman   diagrams   pertinent  to   the   scalar-boson
  contributions  to  the  eletroweak gauge-boson  vacuum  polarization
  amplitudes.}
\label{fig:vacpol}
\end{figure}

Our  interest is to  find the  difference in  the predictions  for the
electroweak  oblique parameters  in the  MSISM from  the corresponding
ones in the SM, i.e.~$\delta P = P_{\rm MSISM} - P_{\rm SM}$, where $P
= \{  S, T, U\}$. As  shown in Figure \ref{fig:vacpol},  the main loop
effect  beyond the  SM arises  from the  MSISM Higgs  scalars  $h$ and
$H_{1,2}$ that occur  in the $WW$ and $ZZ$  self-energies.  The sum of
these three  diagrams for  each one of  the three scalar  bosons, $h$,
$H_1$  and $H_2$,  is denoted  as $\widetilde{P}$.   Specifically, the
shifts  $\delta  P$ are  due  to the  Higgs  scalar  masses $m_h$  and
$m_{H_{1,2}}$, as well as their modified gauge couplings $g_{hVV}$ and
$g_{H_{1,2}VV}$ with  respect to the  SM coupling $g_{H_{\rm  SM}VV} =
1$, where $VV = \{ ZZ, WW\}$. Hence, the deviations of the electroweak
oblique parameters may be obtained by
\begin{eqnarray}
  \label{eqn:deltaP}
\delta P\ =\ g^2_{hVV} \widetilde{P}(m_{h})\: +\: g^2_{H_1VV}
\widetilde{P}(m_{H_{1}})\; +\: g^{2}_{H_2VV} \widetilde{P}(m_{H_{2}})\:
-\: \widetilde{P}(m_{H_{\rm SM}})\ .
\end{eqnarray}
Here,  the generic  function $\widetilde{P}(m)$  stands for  the functions
$\widetilde{S}(m)$, $\widetilde{T}(m)$, and  $\widetilde{U}(m)$, which are defined
as
\begin{eqnarray} 
  \label{eqn:tildeS}
\widetilde{S}(m) \!\!& = &\!\! \frac{1}{12 \pi}\;\bigg[ -
  \frac{1}{\epsilon}\: -\: \frac{1}{2}\ +\
  \frac{m^{4}(m^{2} - 3m_{Z}^{2})}{(m^{2} -
    m_{Z}^{2})^{3}}\;\ln \bigg(\frac{m^{2}}{\bar{\mu}^2}\bigg)\ +\  
\frac{m_{Z}^{4}(
    3m^{2} - m_{Z}^{2})}{(m^{2} - m_{Z}^{2})^{3}}\;
      \ln\bigg(\frac{m^2_Z}{\bar{\mu}^2}\bigg)
  \nonumber\\ 
\!\!&&\!\! -\ \frac{ 5m^{4} - 22m^{2}m_{Z}^{2} +
    5m_{Z}^{4}}{6\,(m^{2} - m_{Z}^{2})^{2}}\; \bigg] \; ,\\[3mm]
   \label{eqn:tildeT}
\widetilde{T}(m) \!\!& = &\!\! 
\frac{3}{16 \pi \sin^{2}\theta_w \cos^2\theta_w m^2_Z}\; 
\bigg[ \left( \frac{1}{\epsilon} + 1
  \right)(m_{Z}^{2} - m_{W}^{2})\ +\
  \frac{m^{2}m_{W}^{2}}{m^{2} -  m_{W}^{2}}\;
  \ln\bigg(\frac{m^{2}}{\bar{\mu}^2}\bigg) \nonumber\\ 
\!\!& &\!\! -\;
  \frac{m^{2}m_{Z}^{2}}{m^{2} - m_{Z}^{2}}
\ln\bigg(\frac{m^{2}}{\bar{\mu}^2}\bigg)\ -\
\frac{m_{W}^{4}}{m^{2} - m_{W}^{2}}
    \ln\bigg(\frac{m^{2}_W}{\bar{\mu}^2}\bigg)\ +\ 
\frac{m_{Z}^{4}}{m^{2} - m_{Z}^{2}}
\ln\bigg(\frac{m^{2}_Z}{\bar{\mu}^2}\bigg)\;\bigg] \; ,\qquad \\[3mm]
  \label{eq:tildeU}
\widetilde{U}(m) \!\!&=&\!\! \frac{1}{12 \pi}\bigg[\; 
\frac{m^{4}(m^{2}- 3m_{W}^{2})}{(m^{2} - m_{W}^{2})^{3}} 
\ln\bigg(\frac{m^{2}}{\bar{\mu}^2}\bigg)\ 
-\ \frac{m^{4}(m^{2} - 3m_{Z}^{2})}{(m^{2} - m_{Z}^{2})^{3}}
\ln\bigg(\frac{m^{2}}{\bar{\mu}^2}\bigg)\nonumber\\
\!\!& &\!\! 
+\ \frac{m_{W}^{4}(3m^{2} - m_{W}^{2})}{(m^{2} -
    m_{W}^{2})^{3}}
\ln\bigg(\frac{m^{2}_W}{\bar{\mu}^2}\bigg)\ 
+\ \frac{m_{Z}^{4}(m_{Z}^{2} -
    3m^{2})}{(m^{2} - m_{Z}^{2})^{3}}
\ln\bigg(\frac{m^{2}_Z}{\bar{\mu}^2}\bigg)\\
\!\!& &\!\! -\ \frac{ 5m^{4}
    - 22m^{2}m_{W}^{2} + 5m_{W}^{4}}{6\,(m^{2} -
    m_{W}^{2})^{2}}\ +\ \frac{ 5m^{4} - 22m^{2}m_{Z}^{2} +
    5m_{Z}^{4}}{6\,(m^{2} - m_{Z}^{2})^{2}}\; \bigg] \; . \nonumber 
\end{eqnarray}
In the above, we have  followed the standard convention and calculated
the electroweak  oblique parameters in  the Feynman-'t Hooft  $\xi =1$
gauge,  in which  $m_{G}  = m_{Z}$  and  $m_{G^{\pm}} =  m_{W^{\pm}}$.
Moreover,  it is important  to note  that $\delta  S$, $\delta  T$ and
$\delta U$ are UV finite and  independent of $\bar{\mu}$, as it can be
easily  checked  by means  of  the  coupling  sum rule:  $g^2_{hVV}  +
g^2_{H_1VV} + g^2_{H_2VV} = g^2_{H_{\rm SM}VV} = 1$.

The theoretical predictions for $\delta  S$, $\delta T$ and $\delta U$
in    the   MSISM    are   confronted    with    their   experimental
values~\cite{PDG}:
\begin{eqnarray}
\delta S_{\rm exp} & = & -0.10 \pm 0.10~(-0.08)\;,  \nonumber\\
\delta T_{\rm exp} & = & -0.08 \pm 0.11~(+0.09)\;,  \nonumber\\
\delta U_{\rm exp} & = & 0.15 \pm 0.11~(+0.01)\;, 
\end{eqnarray}
where the first uncertainty  is evaluated by assuming that $m_{H_{\rm SM}}
= 117$~GeV, while the second  one given in parenthesis should be added
to  the first  to  give  the uncertainty  for  assuming $m_{H_{\rm SM}}  =
300$~GeV.  Along  with the LEP2 95\%~CL limit  presented in Fig.~10(a)
of Ref.~\cite{LEP115GeV},  we also  adjust the experimental  limits on
$\delta S$, $\delta T$ and $\delta  U$ to give a corresponding 95\% CL
interval.  The  following limits have been  implemented throughout our
analysis:
\begin{eqnarray}
	\label{eqn:STUvalues}
-0.296\ < \!\!&\delta S_{\rm exp}&\!\! <\ 0.096 \;,  \nonumber\\
-0.296\ < \!\!& \delta T_{\rm exp} &\!\! <\ 0.136 \;,  \nonumber\\
-0.066\ < \!\!&\delta U_{\rm exp} &\!\! <\ 0.366\; .
\end{eqnarray}
For definiteness, we have  chosen here the Higgs-mass reference value,
$m^{\rm  ref}_{H_{\rm SM}}~=~117~{\rm GeV}$,  even though  the derived
constraints on  the electroweak oblique parameters  are independent of
the choice of $m^{\rm ref}_{H_{\rm SM}}$.

Finally, we should remark that  we have not included the contributions
of  the   light  and  heavy  Majorana   neutrinos,  $\nu_{1,2,3}$  and
$N_{1,2,3}$,   to   the   electroweak   oblique   parameters.    These
contributions  are suppressed  either by  the smallness  of  the light
neutrino masses  or because they  are proportional to  ${\rm Tr}\,({\bf
  h}^\nu\,{\bf h}^{\nu\,\dagger})^2$, i.e.~they  are suppressed by the
fourth  power   of  the   small  neutrino  Yukawa   couplings.   These
contributions can therefore be  safely neglected, when compared to the
dominant scalar-loop effects on the $S$, $T$ and $U$ parameters.


\end{document}